\documentclass[a4paper,english]{amsbook}
\usepackage{newcent}
\usepackage[T1]{fontenc}
\usepackage[latin1]{inputenc}
\makeindex
\usepackage{graphicx}
\usepackage{setspace}
\usepackage{amssymb}
\usepackage{epsfig}

\makeatletter

\providecommand{\LyX}{L\kern-.1667em\lower.25em\hbox{Y}\kern-.125emX\@}
\newcommand{\noun}[1]{\textsc{#1}}

 \numberwithin{section}{chapter}
 \theoremstyle{plain}    
 \newtheorem{thm}{Theorem}[section]
 \numberwithin{equation}{section} %% Comment out for sequentially-numbered
 \numberwithin{figure}{section} %% Comment out for sequentially-numbered
 \theoremstyle{plain}    
 \newtheorem{cor}[thm]{Corollary} %%Delete [thm] to re-start numbering
 \theoremstyle{definition}
 \newtheorem{defn}[thm]{Definition}
 \theoremstyle{definition}
  \newtheorem{example}[thm]{Example}
 \theoremstyle{plain}    
 \newtheorem{lem}[thm]{Lemma} %%Delete [thm] to re-start numbering
 \theoremstyle{remark}    
 \newtheorem{notation}[thm]{Notation} 
 \theoremstyle{remark}    
 \newtheorem{note}[thm]{Note} 
 \theoremstyle{plain}    
 \newtheorem{prop}[thm]{Proposition} %%Delete [thm] to re-start numbering
 \theoremstyle{remark}
 \newtheorem{rem}[thm]{Remark}
 \theoremstyle{remark}    
 \newtheorem{summary}[thm]{Summary} 

\usepackage{babel}
\makeatother
\begin{document}

\title{\emph{\noun{\[
\]
}} \noun{}}

\author{\Huge {TIME-OPTIMAL CONTROL OF SPIN SYSTEMS}\\[23mm]

%{\fontsize{20}{2} 

%\selectfont 

%Diplomarbeit}\\[17mm]

%\normalsize {angefertigt am}

%\Large {Mathematischen Institut}

%\Large {der Bayerischen Julius-Maximilians-Universit\"at 

%W\"urzburg}\\[23mm]

%\normalsize {vorgelegt von}

\Large {Jan Swoboda}\\[23mm]

\Large {January 2006}

}

\maketitle
\tableofcontents{}

\chapter*{Introduction}

This diploma thesis, written under the supervision of Prof. Uwe Helmke at W\"urzburg University, discusses some aspects of time-optimal control
theory of bilinear systems\[
(*)\quad \quad \dot{U}=\left(X_{0}+\sum _{i=1}^{m}v_{i}X_{i}\right)U,\]

$U\in G$, where $G$ is a Lie group, $X_{0},...,X_{m}$ are fixed
elements of its Lie algebra $\mathfrak{g}$, and $v_{i}:\left[0,T\right]\rightarrow \mathbb{R}$
are the control functions.

The resurgence of such systems in recent years has been caused by
applications in quantum computing and nuclear magnetic resonance spectroscopy,
which are related to the question of manipulating effectively ensembles
of coupled spin-particles. The dynamics of such spin systems are governed
by a Schrödinger equation which takes the form of equation $(*)$
with $G=SU\left(2^{n}\right)$.

In the focus of this work is the problem of time-optimal control of
such systems, i.e. the question of how to steer the system from a
given initial state $U(0)=U_{0}$ to a prescribed terminal state $U_{F}$
in least possible time. This is not quite a classical optimal control
problem, since the control variables $v_{i}$ can be chosen to be
arbitrarily large, so that there will be a whole subgroup $K$ of
$G$, all of whose points being reachable from identity within arbitrarily
small time $t$.

Non-linear optimal control problems like this play a crucial role
in control theory and a number of tools have been developed in order
to solve them, most notably the maximum principle of Pontrjagin. On
the other hand, under additional assumptions on the class of system
$(*)$, an explicit solution can be obtained using methods from the
theory of Lie groups and Lie algebras, such as the Cartan decomposition
and Riemannian symmetric spaces. This is one approach followed in
current research on this subject, cf. \cite{key-2}, \cite{key-18}
and \cite{key-15}.

The goal of this thesis is to take up the geometric ideas as formulated
in the paper \cite{key-18} and to present them in a setting which
is both mathematically rigorous and accessible without assuming too
many prerequisities.

The first chapter is aimed to give an overview on the relevant parts
of Lie Theory and Geometric Control Theory which serve as the framework
for the subsequent analysis of the time-optimal control problem associated
with system $(*)$.

In Chapter 2 we follow the geometric approach of \cite{key-18} with
its idea of replacing the original system on the Lie group $G$ by
a reduced system on a homogeneous space $G/K$. This idea is given
a precise formulation in terms of the equivalence theorem of Section
2.3. Although this result lacks, in contrast to Pontrjagin's maximum
principle, a recipe of how to compute time-optimal controls explicitly,
it nevertheless contributes towards a solution of the control problem.
This is mainly due to the following facts.

\begin{enumerate}
\item The equivalence theorem allows for a reduction of the dimension of
the state space of the control system.
\item The space of control paramaters of the reduced system is, in contrast
to that of the unreduced system, compact. This guarantees the existence
of controls which meet the maximality condition of Pontrjagin's maximum
principle.
\item The passage from $(*)$ to the reduced system is the first step towards
the complete solution of the time-optimal control problem in the special
case where the Lie algebra $\mathfrak{g}$ enjoys additional geometric
properties, see Section 2.5.
\end{enumerate}
As indicated in (2), the problem of time-optimal control of system
$(*)$ becomes, after replacing it suitably by a system with bounded
controls, approachable via the maximum principle. Our main result
in Section 2.4 is a computation of those trajectories of the reduced
systems that are, under certain additional assumptions, extremal in
the sense of the maximum principle. However, this yields only a large
family of candidates for a time-optimal solution of the control problem,
and it is not evident of how to determine amongst those the actually
optimal trajectories.

In Section 2.5 we are then specializing to the situation where the
Lie algebra $\mathfrak{g}$ has the additional property of being semisimple
and belonging to a symmetric Lie algebra pair $\left(\mathfrak{g},\mathfrak{k}\right)$
(cf. Definition 1.2.5). This makes the reduced time-optimal control
problem on the homogeneous space $G/K$ accessible to a geometric
solution. The reason for this is that now any point $\left[U_{F}\right]$
of the homogeneous space $G/K$ is contained in the projection $\left[A\right]\subseteq G/K$
of a suitable abelian subgroup $A$ of $G$, and one always can steer
the reduced system between any two points of $\left[A\right]$ along
a sequence of geodesics of $\left[A\right]$. An application of Kostant's
convexity theorem then shows that such a choice of controls is indeed
time-optimal. The original problem on the group $G$ is thus reduced
to a control problem on a so-called flat submanifold $\left[A\right]$
of $G/K$. The control problem reduced this way involves only commuting
vector fields, which makes it possible to solve it explicitly.\\
Optimal controls for the original control system $(*)$ may in a subsequent
step be obtained from those for the reduced system on $G/K$ by again
utilising the equivalence theorem as derived in Section 2.3.

The final chapter is devoted to a discussion of low-dimensional examples
of spin-systems such as one- and two-particle systems. These are well
suited for explicit computations, but are at the same time general
enough objects to illustrate the theory developed in the second chapter.

I am very grateful to my supervisor Professor Uwe Helmke for constantly
supporting me in writing this thesis. Also, I would like to thank
Martin Kleinsteuber for a number of helpful comments on this subject.
Finally, I am indebted to Dr. Gunther Dirr for all his commitment
in reading and discussing various aspects of this diploma thesis.

\chapter{Basic Notions from Lie Theory and Geometric Control}

\section{Lie Groups and Lie Algebras}

The problem of steering a quantum mechanical spin system we are interested
in can be formulated as a control problem on a Lie group, or a homogeneous
space. Its solution involves (amongst others) methods from the theory
of Lie groups, Lie algebras, and homogeneous spaces. In this section
I shall state only those definitions and theorems that will be used
later. I have nevertheless tried to make this exposition as self-contained
as possible. The results that will be mentioned are all standard.
They can be found in the book \cite{key-3} and will therefore be
stated without proof. 

\begin{defn}
A \emph{\index{Lie group}Lie group} $(G,\cdot )$ is a smooth manifold
$G$ endowed with the operations of group multiplication $\cdot $
and group inversion such that the map\begin{equation}
G\times G\longrightarrow G,\quad (g_{1},g_{2})\longmapsto g_{1}\cdot g_{2}^{-1}\label{eq:}\end{equation}
is smooth.
\end{defn}
\begin{example}
The general linear group $G=Gl_{n}\mathbb{R}$ of invertible, real
$(n\times n)$-matrices, and closed subgroups of this such as $Sl_{n}\mathbb{R}$
and $SO_{n}\mathbb{R}$. \\
The group $G=SU_{n}\subseteq Gl_{n}\mathbb{C}$ of unitary $(n\times n)$-matrices
of determinant $1$ will be the most important example to us.
\end{example}
\begin{defn}
A \emph{Lie algebra\index{Lie algebra}} is a (real or complex) vector
space $\mathfrak{g}$ together with a skew symmetric bilinear operation
$\left[\cdot ,\cdot \right]$ such that \emph{Jacobi's identity} holds:\begin{equation}
\left[X,\left[Y,Z\right]\right]+\left[Y,\left[Z,X\right]\right]+\left[Z,\left[X,Y\right]\right]=0\textrm{ for all }X,Y,Z\in \mathfrak{g}.\end{equation}

\end{defn}
The importance of Lie theory in many fields of mathematics and physics
arises from the fact that there is a natural linearization of both
the manifold structure (i.e. tangent spaces) and the conjugation map
$(g,h)\mapsto hgh^{-1}$ (giving the tangent spaces the structure
of Lie algebra) that allows to study nonlinear problems on the group
by translating them into linear problems on the Lie algebra level.
These two structures are closely related by the exponential map.

\begin{flushleft}In order to associate a Lie algebra structure to
a Lie group we make the following definition.\end{flushleft}

\begin{defn}
A vecor field $\chi $ on $G$ is called \emph{right-invariant} if
it satisfies\begin{equation}
\chi (g)=D_{\mathbf{1}}R_{g}\left(\chi \left(\mathbf{1}\right)\right)\textrm{ for all }g\in G\end{equation}
Here $R_{g}:h\mapsto hg$ denotes right-translation by $g$.
\end{defn}
\begin{lem}
Any right-invariant vector field $\chi \in \Gamma (TG)$ is smooth
and complete. The set of right-invariant vector fields is closed under
the Lie bracket $\left[\cdot ,\cdot \right]$ on $\Gamma (TG)$. Any
$X\in T_{\mathbf{1}}G$ can be extended uniquely to a right-invariant
vector field $\hat{X}$ with $\hat{X}(\mathbf{1})=X$. In particular,
the space of right-invariant vector fields has dimension equal to
$\dim G$.
\end{lem}
With these preparations in mind we are in position to endow the tangent
space at identity $\mathbf{1}$ of the Lie group $G$ with a Lie algebra
structure.

\begin{prop}
Set $\mathfrak{g}:=T_{\mathbf{1}}G$. Then $\mathfrak{g}$ is a Lie
algebra with bracket \begin{equation}
\left[X,Y\right]:=-\left[\hat{X},\hat{Y}\right](\mathbf{1}).\label{eq:}\end{equation}

\end{prop}
In view of Lemma 1.1.5 the following definition makes sense.

\begin{defn}
Let $X\in \mathfrak{g}$ and $\gamma _{X}$ be the integral curve
of $\hat{X}$ starting at $\gamma (0)=\mathbf{1}$. Then define the
\emph{exponential map}\index{exponential map}\begin{equation}
\exp :\mathfrak{g}\longrightarrow G\label{eq:}\end{equation}
by\begin{equation}
\exp (X):=\gamma _{X}(\mathbf{1}).\label{eq:}\end{equation}

\end{defn}
\begin{lem}
The so defined map $\exp $ is smooth, its differential at $X=0$
being $D_{0}\exp =\mathrm{id}_{\mathfrak{g}}$. In particular, $\exp $
is a diffeomorphism near $X=0$.
\end{lem}
\begin{example}
$G=Gl_{n}\mathbb{R}$ has Lie algebra $\mathfrak{g}=\mathbb{R}^{n\times n}$
and the exponential map is given by $\exp (X)=\sum _{k=0}^{\infty }\frac{1}{k!}X^{k}$.
\end{example}
There is a natural smooth Lie group action of $G$ on itself given
by\begin{equation}
\alpha :G\times G\longrightarrow G,\quad (g,h)\longmapsto ghg^{-1}\label{eq:}\end{equation}
with the identity as a fixed point. Thus differentiating $\alpha $
at the identity with respect to the second variable yields the smooth
group homomorphism\begin{equation}
\mathrm{Ad}:G\longrightarrow Gl(\mathfrak{g}),\quad g\longmapsto \mathrm{Ad}_{g}\label{eq:}\end{equation}
with\begin{equation}
\mathrm{Ad}_{g}(X)=\left.\frac{d}{dt}\alpha \left(g,\exp tX\right)\right|_{t=0}.\label{eq:}\end{equation}
We call this homomorphism \emph{adjoint representation of $G$} \emph{on
$\mathfrak{g}$}. Differentiating $\mathrm{Ad}$ at the identity $g=\mathbf{1}$
yields the linear map\begin{equation}
\mathrm{ad}:\mathfrak{g}\longrightarrow \mathrm{End}(X),\quad X\longmapsto \mathrm{ad}(X)\label{eq:}\end{equation}
with\begin{equation}
\mathrm{ad}X(Y)=\left.\frac{d}{dt}\mathrm{Ad}_{\exp tX}(Y)\right|_{t=0}.\label{eq:}\end{equation}
This map $\mathrm{ad}$ is a homomorphism of Lie algebras, i.e.\begin{equation}
\mathrm{ad}\left[X,Y\right]=\mathrm{ad}X\circ \mathrm{ad}Y-\mathrm{ad}Y\circ \mathrm{ad}X\label{eq:}\end{equation}
holds for all $X,Y\in \mathfrak{g}$. For this reason the map $\mathrm{ad}$
is called \emph{adjoint Lie algebra representation}.

\begin{thm}
For all $X,Y\in \mathfrak{g}$,\begin{equation}
\mathrm{ad}X(Y)=\left[X,Y\right].\label{eq:}\end{equation}

\end{thm}
For later use we make the following important definition.

\begin{defn}
On the Lie algebra $\mathfrak{g}$ define the \emph{Killing form}
$\kappa $ to be the bilinear form\begin{equation}
\left(X,Y\right)\longmapsto \left\langle X,Y\right\rangle =\mathrm{tr}\left(\mathrm{ad}X\circ \mathrm{ad}Y\right).\label{eq:}\end{equation}

\end{defn}
It is not difficult to see that the Killing form is symmetric and
$\mathrm{ad}$-invariant, i.e.\begin{equation}
\kappa \left(\left[X,Y\right],Z\right)=\kappa \left(X,\left[Y,Z\right]\right)\label{eq:}\end{equation}
holds for all $X,Y,Z\in \mathfrak{g}$.

\begin{example}
The Killing form on $\mathfrak{g}=\mathfrak{su}(n)$ is given by \begin{equation}
\kappa (X,Y)=2n\textrm{tr}(XY).\label{eq:}\end{equation}

\end{example}

\section{Homogeneous Spaces, Riemannian Symmetric Spaces, and Maximal Tori}

Throughout this section $G$ denotes a Lie group and $K\subseteq G$
a closed subgroup (which is known also to be a Lie group in the induced
topology). We first of all collect some facts concerning the space
of left cosets \begin{equation}
G/K:=\left\{ \left(gK\right|g\in G\right\} ,\label{eq:}\end{equation}
which we also call a \emph{$G$-homogeneous space\index{homogeneous space}.}

\begin{prop}
$G/K$ equipped with the quotient topology is a Hausdorff topological
space. It inherits from $G$ the structure of smooth manifold. This
manifold structure can be characterized as the unique differentiable
structure such that the canonical projection\begin{equation}
\pi :G\longrightarrow G/K\label{eq:}\end{equation}
is a smooth submersion. In the language of fibre bundles, the triple
$(\pi ,G,G/K)$ is a principal $K$-fibre bundle over $G/K$.
\end{prop}
\begin{proof}
\cite{key-29}, p. 32-33.
\end{proof}
There is a natural action of the Lie group $G$ on the homogeneous
space $G/K$:\begin{equation}
\varphi :G\times G/K\longrightarrow G/K,\quad (g,hK)\longmapsto ghK.\label{eq:}\end{equation}
This action is smooth and transitive on $G/K$. There is an important
converse to this observation:

\begin{thm}
\emph{(Theorem on transitive Lie group actions\index{theorem on transitive Lie group actions}).}
Let $\sigma :G\times M\rightarrow M$ be a smooth and transitive action
of $G$ on the manifold $M$, and $p\in M$ arbitrary. Then $\sigma $
is equivalent to the natural action $\varphi $ of $G$ on $G/\mathrm{Stab}_{p}\sigma $
in the sense that there is a diffeomorphism $\psi :M\rightarrow G/\mathrm{Stab}_{p}\sigma $
which satisfies\begin{equation}
\psi (\sigma (g,x))=\varphi (g,\psi (x))\label{eq:}\end{equation}
for all $g\in G$ and $x\in M$.
\end{thm}
\begin{proof}
\cite{key-29}, p. 33.
\end{proof}
We next give a criterion for the existence of a $G$- (left-) invariant
Riemannian metric on the homogeneous space $G/K$. Here a metric $\left\langle \cdot ,\cdot \right\rangle $
is called $G$\emph{-invariant\index{invariant metric}} if for all
$x\in G/K$, $h\in G$ and $X,Y\in T_{x}\left(G/K\right)$ \emph{}the
following holds:\begin{equation}
\left\langle X,Y\right\rangle _{x}=\left\langle D_{x}\varphi (h,\cdot )(X),D_{x}\varphi (h,\cdot )(Y)\right\rangle _{\varphi (h,x)},\label{eq:}\end{equation}
i.e. any diffeomorphism $\varphi (h,\cdot )$ of $G/K$ is an isometry
of $\left(G/K,\left\langle \cdot ,\cdot \right\rangle \right)$.

\begin{thm}
Let $G$ and $K$ as before, denote by $\mathfrak{g}$ and $\mathfrak{k}$
its Lie algebras, and let $\mathfrak{p}:=\mathfrak{g}/\mathfrak{k}=\left\{ \left[X\right]\left|X\in \mathfrak{g}\right.\right\} $
the quotient of the vector spaces $\mathfrak{g}$ and $\mathfrak{k}$.
Then there exists a $G$-invariant metric on the homogeneous space
$G/K$ if and only if the closure of the set\begin{equation}
\left\{ \left.\mathrm{Ad}_{k}:\mathfrak{p}\rightarrow \mathfrak{p},\left[X\right]\mapsto \left[\mathrm{Ad}_{k}X\right]\right|k\in K\right\} \label{eq:}\end{equation}
is compact in $\mathrm{End}(\mathfrak{p})$.
\end{thm}
\begin{proof}
\cite{key-4}, p. 67.
\end{proof}
\begin{example}
Theorem 1.2.3 can be applied in the situation of a compact Lie group
$G$. Consider the action $\sigma $ of $G\times G$ on $G$ which
is defined by\[
\sigma \left((g_{1},g_{2}),g\right):=g_{1}gg_{2}^{-1}.\]
$G\times G$ acts transitively on $G$ with stabilizer\[
K:=\mathrm{Stab}_{\sigma }(\mathbf{1})=\left\{ \left.(g,g)\right|g\in G\right\} \cong G.\]
Thus $G$ is $\left(G\times G\right)$-equivariantly (in the sense
of equation (1.2.4)) diffeomorphic to the homogeneous space $\left(G\times G\right)/K$.
Theorem 1.2.3 applies to this space because $K$ is compact and thus
has compact image under the continuous map $\mathrm{Ad}:K\rightarrow \mathrm{End}(\mathfrak{p}),\quad k\mapsto \mathrm{Ad}_{k}$.
Here $\mathfrak{p}$ denotes the quotient of $\mathfrak{g}\times \mathfrak{g}$
with $\mathfrak{k}=\left\{ \left.(X,X)\right|X\in \mathfrak{g}\right\} $.
So the manifold $G$ can be endowed with a metric $\left\langle \cdot ,\cdot \right\rangle $
which is invariant under the action $\sigma $. In particular, the
subgroups $G\times \mathbf{1}$ and $\mathbf{1}\times G$ act on $G$
by isometries. This means that the metric $\left\langle \cdot ,\cdot \right\rangle $
is invariant under both left and right translations, and for this
reason will be called bi-invariant. Another invariance property of
the metric $\left\langle \cdot ,\cdot \right\rangle $ is the following.
Since $K$ is the stabilizer subgroup of the identity element, we
see that for each $k\in K$ the map $D_{\mathbf{1}}\sigma (k)$, which
we define by\[
D_{\mathbf{1}}\sigma (k)(X):=\frac{d}{dt}\left.\sigma (k,\exp tX)\right|_{t=0}\]
for all $X\in \mathfrak{g}$, is an endomorphism of $\mathfrak{g}$.
By the definition of an invariant metric it follows that\[
\left\langle D_{\mathbf{1}}\sigma (k)(X),D_{\mathbf{1}}\sigma (k)(Y)\right\rangle _{\mathbf{1}}=\left\langle X,Y\right\rangle _{\mathbf{1}}\]
holds for all $X,Y\in \mathfrak{g}$ and $k\in K$. Now \[
D_{\mathbf{1}}\sigma (k)(X)=\frac{d}{dt}\left.\sigma (k,\exp tX)\right|_{t=0}=\mathrm{Ad}_{k}X,\]
 and therefore\[
\left\langle \mathrm{Ad}_{k}X,\mathrm{Ad}_{k}Y\right\rangle _{\mathbf{1}}=\left\langle X,Y\right\rangle _{\mathbf{1}}\]
for all $X,Y\in \mathfrak{g}$ and $k\in K$. Furthermore, let $Z\in \mathfrak{k}$
arbitrary. Then for all $t\in \mathbb{R}$,\[
\left\langle \mathrm{Ad}_{\exp tZ}X,\mathrm{Ad}_{\exp tZ}Y\right\rangle _{\mathbf{1}}=\left\langle X,Y\right\rangle _{\mathbf{1}}.\]
Differentiating this equation with respect to $t$ at $t=0$ and using
Theorem 1.1.10 yields\begin{equation}
\left\langle \left[X,Z\right],Y\right\rangle _{\mathbf{1}}=\left\langle X,\left[Z,Y\right]\right\rangle _{\mathbf{1}},\label{eq:}\end{equation}
i.e. the so-called $\mathrm{ad}$-invariance property of the metric
$\left\langle \cdot ,\cdot \right\rangle $.
\end{example}
\begin{defn}
Let $\mathfrak{g}$ a Lie algebra (over $\mathbb{R}$ or $\mathbb{C}$)
and $\mathfrak{k}\subseteq \mathfrak{g}$ a subalgebra. Then $\left(\mathfrak{g},\mathfrak{k}\right)$
is called a \emph{symmetric Lie algebra pair}\index{symmetric Lie algebra pair},
if there exists a Lie algebra automorphism $\theta :\mathfrak{g}\rightarrow \mathfrak{g}$,
which is involutive, i.e. $\theta ^{2}=\mathrm{id}$, and which has
$\mathfrak{k}$ as its $1$-eigenspace. Such an automorphism $\theta $
is called \emph{Cartan involution\index{Cartan involution}.}
\end{defn}
\begin{lem}
Let $\left(\mathfrak{g},\mathfrak{k}\right)$ and $\theta $ as in
the previous definition and denote by $\mathfrak{p}$ the $-1$-eigenspace
of $\theta $. Then $\mathfrak{g}=\mathfrak{k}\oplus \mathfrak{p}$.
This direct sum decomposition will in the following be named \emph{Cartan-like
decomposition}\index{Cartan-like decomposition}. It has the following
properties:

(i) (commutator relations)

$\bullet \qquad \left[\mathfrak{k},\mathfrak{k}\right]\subseteq \mathfrak{k},$

$\bullet \qquad \left[\mathfrak{k},\mathfrak{p}\right]\subseteq \mathfrak{p},$

$\bullet \qquad \left[\mathfrak{p},\mathfrak{p}\right]\subseteq \mathfrak{k}.$

(ii) With respect to the Killing form $\kappa $ on $\mathfrak{g}$,
we have that $\mathfrak{p}\subseteq \mathfrak{k}^{\perp }$ and $\mathfrak{k}\subseteq \mathfrak{p}^{\perp }$.
If moreover $\mathfrak{g}$ is a semisimple Lie algebra (cf. Definition
1.3.3), then $\mathfrak{p}=\mathfrak{k}^{\perp }$ and $\mathfrak{k}=\mathfrak{p}^{\perp }$,
i.e. $\mathfrak{g}$ is the orthogonal sum of $\mathfrak{k}$ and
$\mathfrak{p}$.
\end{lem}
\begin{proof}
(i) Let $X$ and $Y$ eigenvectors of $\theta $ with eigenvalues
$\lambda ,\mu \in \left\{ \pm 1\right\} $. $\theta $ is a Lie algebra
homomorphism, so that\[
\theta \left[X,Y\right]=\left[\theta (X),\theta (Y)\right]=\left[\lambda X,\mu Y\right]=\lambda \mu \left[X,Y\right],\]
which implies the result.\\
(ii) Let $X\in \mathfrak{k}$, $Y\in \mathfrak{p}$. Then, with respect
to a basis of $\mathfrak{g}$ adapted to the direct sum decomposition
$\mathfrak{g}=\mathfrak{k}\oplus \mathfrak{p}$, the endomorphisms
$\mathrm{ad}X$ and $\mathrm{ad}Y$ are (by (i)) represented by matrices\[
\mathrm{ad}X=\left(\begin{array}{cc}
 * & 0\\
 0 & *\end{array}
\right)\textrm{ and }\mathrm{ad}Y=\left(\begin{array}{cc}
 0 & *\\
 * & 0\end{array}
\right).\]
So\[
\mathrm{ad}X\circ \mathrm{ad}Y=\left(\begin{array}{cc}
 * & 0\\
 0 & *\end{array}
\right)\left(\begin{array}{cc}
 0 & *\\
 * & 0\end{array}
\right)=\left(\begin{array}{cc}
 0 & *\\
 * & 0\end{array}
\right),\]
which has trace equal to $0$, hence $\kappa (X,Y)=0$. This shows
$\mathfrak{p}\subseteq \mathfrak{k}^{\perp }$ and $\mathfrak{k}\subseteq \mathfrak{p}^{\perp }$.\\
Now let $\mathfrak{g}$ be semisimple. By Definition 1.3.3, this means
that the Killing form is non-degenerate on $\mathfrak{g}$. So for
any subspace $V$ of $\mathfrak{g}$ it follows that\[
\dim V+\dim V^{\perp }=\dim \mathfrak{g}.\]
In particular, $\dim \mathfrak{k}^{\perp }=\dim \mathfrak{p}$. Now
$\mathfrak{p}\subseteq \mathfrak{k}^{\perp }$, and therefore $\mathfrak{p}=\mathfrak{k}^{\perp }$.
The same conclusion applies to give $\mathfrak{k}=\mathfrak{p}^{\perp }$.
\end{proof}
Symmetric Lie algebra pairs are closely related to a special class
of homogeneous spaces $G/K$, called \emph{Riemannian symmetric spaces}.\index{Riemannian symmetric space}
Those can be characterized as Riemannian manifolds which are endowed
with a $G$-invariant metric $\left\langle \cdot ,\cdot \right\rangle $
such that there exists an isometry $\Phi $ which fixes the point
$K\in G/K$ and reverses all the geodesics through $K$. Any such
Riemannian symmetric space $\left(G/K,\left\langle \cdot ,\cdot \right\rangle \right)$
gives rise to a symmetric Lie algebra pair $\left(\mathfrak{g},\mathfrak{k}\right)$.
\\
There is a natural group homomorphism from $G$ to the Lie group of
isometries $\mathrm{Isom}\left(G/K\right)$ of $\left(G/K,\left\langle \cdot ,\cdot \right\rangle \right)$,
which is given by\begin{equation}
\varphi :g\longmapsto \varphi (g,\cdot ),\label{eq:}\end{equation}
with $\varphi (g,\cdot )$ as in equation (1.2.3). In the case where
the Lie group $G$ is connected and semisimple, one can show that
$\varphi $ is indeed an isomorphism between $G$ and the connected
component of identity of $\mathrm{Isom}\left(G/K\right)$, cf. \cite{key-39},
p. 143. The Cartan involution $\theta $ can then be viewed as an
{}``infinitesimal isometry'' which is related to $\Phi $ as follows:\begin{equation}
\theta (X)=\frac{d}{dt}\left.\varphi ^{-1}\left(\Phi \circ \varphi \left(\exp tX\right)\circ \Phi \right)\right|_{t=0},\label{eq:}\end{equation}
cf. \cite{key-9}, p. 227-228. Thus the expression $t\mapsto \varphi ^{-1}\left(\Phi \circ \varphi \left(\exp tX\right)\circ \Phi \right)$
can be regarded as an $1$-parameter subgroup of $G$. Equation (1.2.9)
then states that this $1$-parameter subgroup is generated by $\theta (X)$.\\
Symmetric Lie algebra pairs will be of importance in the sequel, since
the control problem we are going to discuss in Section 2.2 can be
reformulated as a problem on a homogeneous space $G/K$ (see Section
2.3) and then be solved, provided this space $G/K$ happens to be
symmetric (see Section 2.5).

\begin{example}
(cf. \cite{key-6}, Ch. X). For the following pairs $(G,K)$ of Lie
groups the homogeneous spaces $G/K$ are all symmetric. Pairs in the
left columns lead to noncompact examples, those in the middle columns
to compact ones.

\begin{tabular}{|c|c|c|c|c|c|}
\hline 
$\textrm{Type}$&
$G$&
$K$&
$G$&
$K$&
$\dim $\\
\hline
\hline 
A I&
$Sl_{n}\mathbb{R}$&
$SO(n)$&
$SU(n)$&
$SO(n)$&
$\frac{1}{2}(n-1)(n+2)$\\
\hline 
A II&
$SU^{*}(2n)$&
$Sp(n)$&
$SU(2n)$&
$Sp(n)$&
$(n-1)(2n+1)$\\
\hline 
A III&
$SU(p,q)$&
$S(U(p)\times U(q))$&
$SU(p+q)$&
$S(U(p)\times U(q))$&
$2pq$\\
\hline 
BD I&
$SO_{0}(p,q)$&
$SO(p)\times SO(q)$&
$SO(p+q)$&
$SO(p)\times SO(q)$&
$pq$\\
\hline 
D III&
$SO^{*}(2n)$&
$U(n)$&
$SO(2n)$&
$U(n)$&
$n(n-1)$\\
\hline 
C I&
$Sp_{n}\mathbb{R}$&
$U(n)$&
$Sp(n)$&
$U(n)$&
$n(n+1)$\\
\hline 
C II&
$Sp(p,q)$&
$Sp(p)\times Sp(q)$&
$Sp(p+q)$&
$Sp(p)\times Sp(q)$&
$4pq$\\
\hline
\end{tabular}

\begin{flushleft}These examples give rise to the following table of
symmetric Lie algebra pairs $\left(\mathfrak{g},\mathfrak{k}\right)$.
The action of the Cartan involution on $X\in \mathfrak{g}$ is described
under $\theta $, while the rightmost column shows the rank of the
algebra $\mathfrak{g}$, which will be defined in Theorem 1.2.8.\end{flushleft}

\begin{tabular}{|c|c|c|c|c|c|c|}
\hline 
$\textrm{Type}$&
$\mathfrak{g}$&
$\mathfrak{k}$&
$\mathfrak{g}$&
$\mathfrak{k}$&
$\theta $&
$\mathrm{rk}$\\
\hline
\hline 
A I&
$\mathfrak{sl}_{n}\mathbb{R}$&
$\mathfrak{so}(n)$&
$\mathfrak{su}(n)$&
$\mathfrak{so}(n)$&
$-X^{T}$&
$n-1$\\
\hline 
A II&
$\mathfrak{su}^{*}(2n)$&
$\mathfrak{sp}(n)$&
$\mathfrak{su}(2n)$&
$\mathfrak{sp}(n)$&
$-J_{n}X^{T}J_{n}^{-1}$&
$n-1$\\
\hline 
A III&
$\mathfrak{su}(p,q)$&
$\mathfrak{s}(\mathfrak{u}(p)\times \mathfrak{u}(q))$&
$\mathfrak{su}(p+q)$&
$\mathfrak{s}(\mathfrak{u}(p)\times \mathfrak{u}(q))$&
$I_{p,q}XI_{p,q}$&
$\min (p,q)$\\
\hline 
BD I&
$\mathfrak{so}(p,q)$&
$\mathfrak{so}(p)\times \mathfrak{so}(q)$&
$\mathfrak{so}(p+q)$&
$\mathfrak{so}(p)\times \mathfrak{so}(q)$&
$I_{p,q}XI_{p,q}$&
$\min (p,q)$\\
\hline 
D III&
$\mathfrak{so}^{*}(2n)$&
$\mathfrak{u}(n)$&
$\mathfrak{so}(2n)$&
$\mathfrak{u}(n)$&
$J_{n}XJ_{n}^{-1}$&
$\left[\frac{1}{2}n\right]$\\
\hline 
C I&
$\mathfrak{sp}_{n}\mathbb{R}$&
$\mathfrak{u}(n)$&
$\mathfrak{sp}(n)$&
$\mathfrak{u}(n)$&
$J_{n}XJ_{n}^{-1}$&
$n$\\
\hline 
C II&
$\mathfrak{sp}(p,q)$&
$\mathfrak{sp}(p)\times \mathfrak{sp}(q)$&
$\mathfrak{sp}(p+q)$&
$\mathfrak{sp}(p)\times \mathfrak{sp}(q)$&
$K_{p,q}XK_{p,q}$&
$\min (p,q)$\\
\hline
\end{tabular}\[
\]

\begin{flushleft}All those examples lead to \emph{irreducible} symmetric
spaces $G/K$. By this we mean that\end{flushleft}
\end{example}
\begin{itemize}
\item the Lie algebra $\mathfrak{g}$ is semisimple (cf. Definition 1.3.3)
and $\mathfrak{p}\subseteq \mathfrak{g}$ contains no ideal of $\mathfrak{g}$
other than the zero ideal, and
\item the Lie algebra representation\[
\mathfrak{k}\longrightarrow \mathrm{End}(\mathfrak{p}),\quad X\longmapsto \mathrm{ad}X\]
is irreducible.
\end{itemize}
In fact, the above list by E. Cartan exhausts all irreducible Riemannian
symmetric spaces (up to 12 compact and noncompact exceptional cases).

The notion of a \emph{maximal torus\index{maximal torus}} of a compact,
connected Lie group $G$ will be of considerable interest in the sequel.
By this we mean a Lie subgroup $T\subseteq G$ which is

\begin{enumerate}
\item abelian, 
\item compact,
\item maximal with respect to these properties, i.e. any subgroup $T'$
that satisfies (1) and (2) and contains $T$ already equals $T$. 
\end{enumerate}
It is easy to see that any maximal torus is isomorphic to a product
of copies of $\mathbb{S}^{1}$, hence the name.\\
A \emph{maximal abelian subalgebra\index{maximal abelian subalgebra}}
$\mathfrak{t}$ of a Lie algebra $\mathfrak{g}$ is defined to be
a subalgebra satisfying

\begin{enumerate}
\item $\left[\mathfrak{t},\mathfrak{t}\right]=0,$
\item any subalgebra $\mathfrak{t}'$ that contains $\mathfrak{t}$ and
satisfies (1) is equal to $\mathfrak{t}$.
\end{enumerate}
The salient facts concerning maximal tori of a \emph{compact} Lie
group, respectively maximal abelian Lie subalgebras, are collected
in the following theorem.

\begin{thm}
\emph{(Torus theorem\index{torus theorem}).} Let $G$ be a compact,
connected Lie group with Lie algebra $\mathfrak{g}$. Then the following
holds.

\begin{flushleft}(i) The equation $T=\exp \mathfrak{t}$ defines a
bijective correspondence between the maxi-\\
mal abelian subalgebras $\mathfrak{t}$ of $\mathfrak{g}$ and the
maximal tori $T$ of $G$. Every connected abelian subgroup of $G$
is contained in a maximal torus in $G$, and every abelian subalgebra
of $\mathfrak{g}$ is contained in a maximal abelian subalgebra of
$\mathfrak{g}$.\end{flushleft}

\begin{flushleft}(ii) All maximal tori in $G$ are conjugate to each
other, and $\mathrm{Ad}_{G}$ acts transitively on the set of maximal
abelian subalgebras of $\mathfrak{g}$. Each element of $G$ is conjugate
to an element of a given maximal torus, and $\mathrm{Ad}_{G}(\mathfrak{t})=\mathfrak{g}$
for any maximal abel-\\
ian subalgebra $\mathfrak{t}$ of $\mathfrak{g}$. In particular,
any two maximal abelian subalgebras have the same dimension. This
dimension is called the \emph{rank\index{rank of a compact Lie algebra}}
of the Lie algebra $\mathfrak{g}$ (the rightmost column of the second
table in Example 1.2.7).\end{flushleft}
\end{thm}
\begin{proof}
\cite{key-3}, Theorem 3.7.1 (iii), (iv).
\end{proof}

\section{Root Space Decomposition and Semisimple Lie Algebras}

\subsection{Root space decomposition of a compact Lie algebra}

Throughout this section let $\mathfrak{g}$ denote the Lie algebra
of a compact Lie group $G$. For simplicity we will refer to the Lie
algebra $\mathfrak{g}$ of such a Lie group as a \emph{compact Lie
algebra\index{compact Lie algebra}}, although there is an intrinsic
definition of compact Lie algebras in terms of the Killing form, which
differs from ours. For such a Lie algebra there exists the well-known
\emph{root space decomposition\index{root space decomposition}} into
a direct sum of subspaces which are simultaneously invariant under
the endomorphisms $\mathrm{ad}X$, where $X$ is an element of a maximal
abelian subalgebra $\mathfrak{t}$ of $\mathfrak{g}$. The properties
of such a decomposition will be widely used in the proof of the time-optimal
torus theorem, cf. Section 2.5.

To make the theorem on the Jordan normal form applicable it is convenient
to pass to the complexification $\mathfrak{g}_{\mathbb{C}}=\mathfrak{g}\otimes _{\mathbb{R}}\mathbb{C}$
of $\mathfrak{g}$ which we give the structure of complex Lie algebra
by defining its bracket as\begin{equation}
\left[A+\mathrm{i}B,A'+\mathrm{i}B'\right]:=\left[A,A'\right]-\left[B,B'\right]+\mathrm{i}\left(\left[A,B'\right]+\left[B,A'\right]\right)\label{eq:}\end{equation}
for $A,A',B,B'\in \mathfrak{g}.$ \\
There also is a unique linear extension of any $A\in \mathrm{End}(\mathfrak{g})$
to a complex linear endomorphism of $\mathfrak{g}_{\mathbb{C}}$,
which will again be denoted by $A$.\\
The existence of a root space decomposition is based on the following
lemma.

\begin{lem}
For each $X\in \mathfrak{g}$, the endomorphism $\mathrm{ad}X\in \mathrm{End}(\mathfrak{g}_{\mathbb{C}})$
is diagona-\\
lizable, with only purely imaginary eigenvalues.
\end{lem}
The proof, for which we refer the reader to \cite{key-3}, uses in
a crucial way the boundedness of the set $\mathrm{Ad}_{G}\subseteq \mathrm{End}(\mathfrak{g}_{\mathbb{C}})$
to conclude that the invariant subspaces $\mathfrak{g}_{j}=c_{j}I+N_{j}$
in the Jordan decomposition of $\mathrm{ad}X$ have nilpotent part
$N_{j}$ equal to $0$ and eigenvalues $c_{j}\in \mathrm{i}\mathbb{R}$.

\begin{thm}
\begin{flushleft}\emph{(Root space decomposition\index{theorem on the root space decomposition}
of $\mathfrak{g}_{\mathbb{C}}$).} Let $\mathfrak{t}$ be any abelian
subal-\\
gebra of $\mathfrak{g}$. Set $\alpha _{0}:=0\in \mathfrak{t}^{*}$.
Then there is a finite set $\Sigma =\left\{ \mathrm{i}\alpha _{1},...,\mathrm{i}\alpha _{m}\right\} $
of non-\\
zero real-linear forms $\mathfrak{t}\rightarrow \mathbb{R}$ and a
decomposition\begin{equation}
\mathfrak{g}_{\mathbb{C}}=\mathfrak{g}_{0}\oplus \bigoplus _{j=1}^{m}\mathfrak{g}_{\alpha _{j}}\label{eq:}\end{equation}
such that $\mathfrak{g}_{\alpha _{j}}\neq \left\{ 0\right\} $, and\begin{equation}
\mathrm{ad}X(Y)=\mathrm{i}\alpha _{j}(X)Y\textrm{ }\label{eq:}\end{equation}
holds for all $X\in \mathfrak{t}$, $Y\in \mathfrak{g}_{\alpha _{j}}$,
and $j=0,...,m$. Moreover, if $\alpha \in \left\{ 0\right\} \cup \Sigma $,
then $-\alpha \in \left\{ 0\right\} \cup \Sigma $, and $\mathfrak{g}_{-\alpha }=\overline{\mathfrak{g}_{\alpha }}$.
In particular, $\mathfrak{g}_{0}=\overline{\mathfrak{g}_{0}}$, so
that $\mathfrak{g}_{0}$ is of the form $\mathfrak{a}+\mathrm{i}\mathfrak{a}$
with $\mathfrak{a}=\mathfrak{g}_{0}\cap \mathfrak{g}$. Also, $\mathfrak{t}\subseteq \mathfrak{a}$.\\
The following commutator relations hold:\begin{equation}
\left[\mathfrak{g}_{\alpha _{i}},\mathfrak{g}_{\alpha _{j}}\right]\subseteq \left\{ \begin{array}{ll}
 \mathfrak{g}_{\alpha _{i}+\alpha _{j}}, & \textrm{if }\alpha _{i}+\alpha _{j}\in \left\{ 0\right\} \cup \Sigma ,\\
 \left\{ 0\right\} , & \textrm{otherwise}.\end{array}
\right.\label{eq:}\end{equation}
If $\mathfrak{t}$ is maximal abelian in $\mathfrak{g}$, then $\mathfrak{h}:=\mathfrak{t}+\mathrm{i}\mathfrak{t}=\mathfrak{g}_{0}$
and $\mathfrak{t}=\mathfrak{g}_{0}\cap \mathfrak{g}$. In that case
there are the decompositions\begin{equation}
\mathfrak{g}_{\mathbb{C}}=\mathfrak{h}\oplus \bigoplus _{j=1}^{m}\mathfrak{g}_{\alpha _{j}},\label{eq:}\end{equation}
and\begin{equation}
\mathfrak{g}=\mathfrak{t}\oplus \bigoplus _{\alpha _{j}\in \Sigma ^{+}}\left(\mathfrak{g}_{\alpha _{j}}\oplus \mathfrak{g}_{-\alpha _{j}}\right)\cap \mathfrak{g},\label{eq:}\end{equation}
where $\Sigma ^{+}$ is any subset of $\Sigma $ which satisfies\begin{equation}
\alpha _{j}\in \Sigma ^{+}\quad \Longleftrightarrow \quad -\alpha _{j}\notin \Sigma ^{+}\label{eq:}\end{equation}
for all $j=1,...,m$. The subspaces $\mathfrak{g}_{\alpha _{j}}$
are then called \emph{root spaces\index{root space}}, the linear
forms $\mathrm{i}\alpha _{j}$ are named \emph{roots\index{root}.}\end{flushleft}
\end{thm}
\begin{proof}
\cite{key-3}, p. 145-146.
\end{proof}
For any root $\mathrm{i}\alpha \in \Sigma $, the map $\alpha :\mathfrak{t}\rightarrow \mathbb{R}$
defines a non-zero, real-linear form. Its kernel $\ker \alpha $ therefore
defines a hyperplane in $\mathfrak{t}$, which is called the \emph{root
hyperplane\index{root hyperplane}} for $\alpha $. The connected
components of the set\begin{equation}
\mathfrak{t}\setminus \left(\bigcup _{\mathrm{i}\alpha \in \Sigma }\ker \alpha \right)\label{eq:}\end{equation}
are called \emph{Weyl chambers\index{Weyl chamber}}; they are open,
convex polyhedral cones in $\mathfrak{t}$.\\
We next define the \emph{Weyl group\index{Weyl group}} $W$ to be
the quotient \emph{\begin{equation}
W=N(\mathfrak{t})/\mathrm{Stab}(\mathfrak{t}),\label{eq:}\end{equation}
}with\begin{equation}
N(\mathfrak{t})=\left\{ \left.g\in G\right|\mathrm{Ad}_{g}\mathfrak{t}=\mathfrak{t}\right\} \label{eq:}\end{equation}
 the the normalizer of $\mathfrak{t}$ in $G$, and \begin{equation}
\mathrm{Stab}(\mathfrak{t})=\left\{ \left.g\in G\right|\mathrm{Ad}_{g}X=X\; \forall X\in \mathfrak{t}\right\} \label{eq:}\end{equation}
the pointwise stabilizer of $\mathfrak{t}$ in $G$. The group $W$
turns out to be finite. Furthermore, as a consequence of the Torus
Theorem 1.2.8, it can be shown that the isomorphism type of $W$ does
not depend on the choice of $\mathfrak{t}$. It is therefore justified
to call $W$ the Weyl group of the Lie algebra $\mathfrak{g}$.\\
There is the following action of $W=\left\{ \left.\left[g\right]\right|g\in N(\mathfrak{t})\right\} $
on $\mathfrak{t}$:\begin{equation}
W\times \mathfrak{t}\longrightarrow \mathfrak{t},\quad \left(\left[g\right],X\right)\longmapsto gXg^{-1}.\label{eq:}\end{equation}
The action of $W$ is transitive on the set of Weyl chambers. This
statement is part of the Weyl covering theorem\index{Weyl covering theorem},
cf. \cite{key-3}, p. 153. \\
At this point we conclude those general considerations on compact
Lie algebras and turn to compact semisimple algebras, where a refined
version of some of the statements made before can be given.

\subsection{Compact semisimple Lie algebras}

\begin{defn}
A Lie algebra $\mathfrak{g}$ (over $\mathbb{C}$ or over $\mathbb{R}$)
is called \emph{semisimple\index{semisimple Lie algebra}} if the
Killing form\begin{equation}
\kappa :\mathfrak{g}\times \mathfrak{g}\longrightarrow \mathfrak{g},\quad (X,Y)\longmapsto \textrm{tr}(\textrm{ad}X\circ \textrm{ad}Y)\label{eq:}\end{equation}
is nondegenerate on $\mathfrak{g}$.\\
The Lie algebra $\mathfrak{g}$ is called \emph{simple\index{simple Lie algebra}}
if it is not abelian and does not contain any ideals other than $\left\{ 0\right\} $
and $\mathfrak{g}$.
\end{defn}
\begin{rem}
The relation between simple and semisimple Lie algebras is such that
every semisimple Lie algebra splits uniquely (up to isomorphism) into
an orthogonal (with respect to the Killing form) sum of simple Lie
algebras, cf. \cite{key-8}, p. 23.
\end{rem}
\begin{example}
See Example 1.2.7. The Lie algebras $\mathfrak{g}$ which appear in
the second table are all simple. 

\begin{flushleft}For the remainder of this section let $\mathfrak{g}$
be a semisimple, compact Lie algebra over the reals. Furthermore,
let $\mathfrak{t}$ be a maximal abelian subalgebra of $\mathfrak{g}$,
and denote by $\Sigma $$\subseteq \mathfrak{t}^{*}$ the set of roots
in the root space decomposition of $\mathfrak{g}_{\mathbb{C}}$ with
respect to $\mathfrak{t}$. Since the Killing form is nondegenerate
on $\mathfrak{g}$ we may identify $\mathfrak{t}^{*}$ with $\mathfrak{t}$
in the following manner:\begin{equation}
\mathfrak{t}^{*}\ni \lambda \longmapsto X_{\lambda }\in \mathfrak{t}\textrm{ such that }\lambda (Y)=\kappa \left(X_{\lambda },Y\right)\textrm{ is satisfied for all }Y\in \mathfrak{g}.\label{eq:}\end{equation}
We will name $X_{\alpha }\in \mathfrak{t}$ a \emph{coroot\index{coroot}},
if $\mathrm{i}\alpha \in \Sigma $ is a root. The set of coroots is
denoted by $\Sigma ^{*}$. One observes that the coroots are perpendicular
to their respective root hyperplanes.\\
In addition to the results of Subsection 1.3.1, the following theorem
holds.\end{flushleft}
\end{example}
\begin{thm}
Let $\mathfrak{t}\subseteq \mathfrak{g}$ be a maximal abelian subalgebra
of the compact, semisimple Lie algebra $\mathfrak{g}$. Then:

\begin{flushleft}(i) There exists $X\in \mathfrak{t}$ such that $\mathfrak{t}=\ker (\mathrm{ad}X)$.
Conversely, the kernel of $\mathrm{ad}X$, $X\in \mathfrak{g}$ arbitrary,
is a maximal abelian subalgebra of $\mathfrak{g}$ if and only if
its dimen-\\
sion is the smallest one possible. Such an $X\in \mathfrak{g}$ is
also called \emph{a regular ele-}\\
\emph{ment\index{regular element}}. In that case the dimension of
$\ker \left(\mathrm{ad}X\right)$ coincides with the \emph{rank\index{rank of a compact Lie algebra}}
of the algebra $\mathfrak{g}$ as defined in 1.2.8.\end{flushleft}

\begin{flushleft}(ii) The set $\Sigma ^{*}$ of coroots spans $\mathfrak{t}$
as a vector space.\end{flushleft}

\begin{flushleft}(iii) The root spaces in a root space decomposition
of $\mathfrak{g}_{\mathbb{C}}$ with respect to $\mathfrak{t}$ are
all $1$-dimensional.\end{flushleft}
\end{thm}
\begin{proof}
Cf. \cite{key-8}, p. 80 for a proof of (i), and p. 39 for a proof
of (ii) and (iii). 
\end{proof}
\begin{example}
Consider the Lie algebra $\mathfrak{g}=\mathfrak{su}(n)$. 

\begin{flushleft}The complexification of $\mathfrak{g}$ is the simple
Lie algebra $\mathfrak{g}_{\mathbb{C}}=\mathfrak{sl}_{n}\mathbb{C}$.
Choose \[
\mathfrak{t}:=\left\{ \left.\mathrm{idiag}(\theta _{1},...,\theta _{n})\right|\theta _{j}\in \mathbb{R},\sum _{j}\theta _{j}=0\right\} \]
to serve as a maximal abelian subalgebra of $\mathfrak{g}$, and let
$\mathfrak{h}:=\mathfrak{t}+\mathrm{i}\mathfrak{t}$ be its complexification.
For $i\neq j$ set \[
E_{ij}:=(e_{rs})_{r,s=1,...,n}\textrm{ with }e_{rs}=\left\{ \begin{array}{ll}
 1, & \textrm{if}\: r=i,s=j,\\
 0, & \mathrm{otherwise}.\end{array}
\right.\]
Then the root space decomposition of $\mathfrak{g}_{\mathbb{C}}$
with respect to $\mathfrak{t}$ is given by\[
\mathfrak{g}_{\mathbb{C}}=\mathfrak{h}\oplus \bigoplus _{i\neq j}\mathbb{C}E_{ij}.\]
The corresponding roots are\[
\alpha _{ij}:\mathfrak{t}\longrightarrow \mathrm{i}\mathbb{R},\quad \mathrm{idiag}(\theta _{1},...,\theta _{n})\longmapsto \mathrm{i}(\theta _{i}-\theta _{j}).\]
 Indeed, a calculation yields\begin{eqnarray*}
\left[\left(\begin{array}{cccc}
 \mathrm{i}\theta _{1} &  &  & \\
  & \ddots  &  & \\
  &  & \ddots  & \\
  &  &  & \mathrm{i}\theta _{n}\end{array}
\right),\left(\begin{array}{cccc}
 0 &  &  & \\
  & \ddots  & 1 & \\
  &  & \ddots  & \\
  &  &  & 0\end{array}
\right)\right] & = & \mathrm{i}(\theta _{i}-\theta _{j})E_{ij}.
\end{eqnarray*}
The root spaces $\mathfrak{g}_{ij}=\mathbb{C}E_{ij}$ obey the commutator
relations\[
\left[\mathfrak{g}_{ij},\mathfrak{g}_{kl}\right]=\left\{ \begin{array}{ll}
 \mathfrak{g}_{il}, & \textrm{if}\: j=k,i\neq l,\\
 \mathfrak{g}_{kj}, & \textrm{if}\: i=l,j\neq k,\\
 0, & \mathrm{else}.\end{array}
\right.\]
>From the identification of $\mathfrak{h}^{*}$ with $\mathfrak{h}$
via the Killing form one obtains the co-\\
roots $X_{\alpha _{ij}}=\mathrm{idiag}(0,...,1,...,-1,...,0)\in \mathfrak{h}$
(here $1$ is the entry at the $i$-th, $-1$ at the $j$-th position).
They satisfy the relations\[
\frac{\kappa \left(X_{\alpha _{ij}},X_{\alpha _{kl}}\right)}{\left(\kappa \left(X_{\alpha _{ij}},X_{\alpha _{ij}}\right)\kappa \left(X_{\alpha _{kl}},X_{\alpha _{kl}}\right)\right)^{\frac{1}{2}}}=\left\{ \begin{array}{ll}
 0, & \textrm{if}\: \left\{ i,j\right\} \cap \left\{ k,l\right\} =\emptyset ,\\
 -1, & \textrm{if}\: i=l,j\neq k\textrm{ or }i\neq l,j=k,\\
 1, & \textrm{if}\: i=k,j\neq l\textrm{ or }i\neq k,j=l,\\
 -2, & \textrm{if}\: i=l,j=k,\\
 2, & \textrm{if}\: i=k,j=l,\end{array}
\right.\]
so that the coroots include angles equal to $0,\frac{\pi }{3},\frac{\pi }{2},\frac{2\pi }{3}$,
or $\pi $.\\
The Weyl group $W$ acts on $X\in \mathfrak{t}$ by permuting the
entries on the diagonal of $X$ and therefore turns out to be isomorphic
to the symmetric group $S(n)$.\\
On $\mathfrak{g}$ one defines the involution $\theta :X\longmapsto -X^{T}.$
Its $1$-eigenspace $\mathfrak{k}$ is equal to the subalgebra $\mathfrak{so}(n)$
of $\mathfrak{g}$, while its $-1$-eigenspace $\mathfrak{p}$ comprises
those matrices $X$ of $\mathfrak{g}$ which satisfy $X=X^{T}$. Therefore,
$(\mathfrak{g},\mathfrak{k})$ is a symmetric Lie algebra pair in
the sense of Definition 1.2.5. In fact, $\theta $ arises as the Cartan
involution asso-\\
ciated with the Riemannian symmetric space $SU(n)/SO(n)$, see Example
1.2.7. Notice also that \[
\mathfrak{k}=\sum _{i<j}\mathfrak{g}\cap \mathbb{C}(E_{ij}+\theta E_{ij}),\]
while\[
\mathfrak{n}:=\sum _{i<j}\mathfrak{g}\cap \mathbb{C}(E_{ij}-\theta E_{ij})\]
is complementary to $\mathfrak{h}$ in $\mathfrak{p}$. This is not
accidently and reflects a general correspondence between the Cartan-like
decomposition $\mathfrak{g}=\mathfrak{k}\oplus \mathfrak{p}$ and
the root space decomposition of $\mathfrak{g}_{\mathbb{C}}$ with
respect to a maximal abelian subalgebra $\mathfrak{h}\subseteq \mathfrak{p}$
of $\mathfrak{g}$, cf. \cite{key-6}, p. 336.\end{flushleft}
\end{example}
We next give a modified version of the Torus Theorem 1.2.8 (ii) in
the situation of a Lie group $G$ with semisimple Lie algebra $\mathfrak{g}$,
which will be of interest later on.

\begin{lem}
Let $G$ be a Lie group with semisimple Lie algebra $\mathfrak{g}$,
and $K$ a closed subgroup of $G$ such that its Lie algebra $\mathfrak{k}$
together with $\mathfrak{g}$ forms a symmetric Lie algebra pair.
Furthermore, let $\mathfrak{g}=\mathfrak{k}\oplus \mathfrak{p}$ be
the corresponding Cartan-like decomposition, and $\mathfrak{h}$ any
maximal abelian subalgebra of $\mathfrak{p}$. Then \begin{equation}
\mathfrak{p}=\bigcup _{k\in K}\mathrm{Ad}_{k}\mathfrak{h}.\label{eq:}\end{equation}

\end{lem}
\begin{proof}
\cite{key-6}, Chapter V, Lemma 6.3 (iii).
\end{proof}
To conclude this section we cite a theorem that relates the action
of the Weyl group $W$ on $\mathfrak{h}$ to the adjoint action of
$K$ on $\mathfrak{p}$.

\begin{thm}
\emph{(Kostant's convexity theorem\index{Kostant's convexity theorem}).}
In the setting of the previous Lemma 1.3.8, let $\Gamma :\mathfrak{p}\rightarrow \mathfrak{h}$
be the orthogonal projection with respect to the Killing form on $\mathfrak{g}$.
Then for any $X\in \mathfrak{h}$\emph{\begin{equation}
\Gamma (\textrm{Ad}_{K}X)=\mathfrak{c}(W\cdot X),\label{eq:}\end{equation}
}where $\mathfrak{c}$ denotes convex hull.
\end{thm}
\begin{proof}
\cite{key-26}.
\end{proof}
\begin{example}
Consider the symmetric Lie algebra pair $(\mathfrak{g},\mathfrak{k})$
with $\mathfrak{g}=\mathfrak{su}(2)$ and \[
\mathfrak{k}=\left\{ \left.\left(\begin{array}{cc}
 \mathrm{i}X & \\
  & -\mathrm{i}X\end{array}
\right)\right|X\in \mathbb{R}\right\} \subseteq \mathfrak{su}(2).\]
 Its Cartan-like decomposition is $\mathfrak{g}=\mathfrak{k}\oplus \mathfrak{p}$
with \[
\mathfrak{p}=\left\{ \left.\left(\begin{array}{cc}
 0 & z\\
 -\bar{z} & 0\end{array}
\right)\right|z\in \mathbb{C}\right\} .\]

\begin{flushleft}Fix $X_{0}=\left(\begin{array}{cc}
 0 & z_{0}\\
 -\bar{z_{0}} & 0\end{array}
\right)\in \mathfrak{p}$, $z_{0}\neq 0$, and set $\mathfrak{h}=\mathbb{R}X_{0}$. It is easily
seen that the Weyl orbit of $Y=\lambda X_{0}\in \mathfrak{h}$ is
$W\cdot Y=\left\{ \pm Y\right\} $ and the orbit of the adjoint action
of $K$ on $Y$ is\begin{eqnarray*}
\mathrm{Ad}_{K}Y & = & \left\{ \left.kYk^{-1}\right|k\in K\right\} \\
 & = & \left\{ \left.\left(\begin{array}{cc}
 e^{\mathrm{i}t} & 0\\
 0 & e^{-\mathrm{i}t}\end{array}
\right)\left(\begin{array}{cc}
 0 & \lambda z_{0}\\
 -\lambda \bar{z_{o}} & 0\end{array}
\right)\left(\begin{array}{cc}
 e^{-\mathrm{i}t} & 0\\
 0 & e^{\mathrm{i}t}\end{array}
\right)\right|t\in \mathbb{R}\right\} \\
 & = & \left\{ \left.\left(\begin{array}{cc}
 0 & \lambda e^{2\mathrm{i}t}z_{0}\\
 -\lambda e^{-2\mathrm{i}t}\bar{z_{o}} & 0\end{array}
\right)\right|t\in \mathbb{R}\right\} ,
\end{eqnarray*}
\end{flushleft}

\begin{flushleft}i.e. a circle $C\subseteq \mathfrak{p}$ centered
at the origin and passing through $Y$. The ortho-\\
gonal projection of $C$ on $\mathfrak{h}$ is the set $\left\{ \left.\alpha Y\right|-1\leq \alpha \leq 1\right\} $,
which in fact is the convex hull of $\left\{ \pm Y\right\} $.\end{flushleft}
\end{example}

\section{Some Definitions from Geometric Control Theory}

This section is aimed to introduce some basic definitions and results
from geometric control theory which serves as the appropriate framework
for the kind of problem to be considered later. Geometric control
theory is primarily interested in the investigation of controlled
dynamical systems on a manifold $M$, their behaviour being governed
by ODEs of the form\begin{equation}
\dot{x}=f(x(t),u(t)),\quad x(0)=x_{0},\label{eq:}\end{equation}
where the parameter $u$ (the {}``control function\index{control parameter}'')
is allowed to vary with time $t$ within a given parameter space $U\subseteq \mathbb{R}^{m}$
(the {}``control set\index{control set}'').\\
Given such a system (1.4.1) together with a control set $U$ one naturally
can ask the following questions.

\begin{enumerate}
\item Does there exists a control function $t\mapsto u(t)$ that transfers
the initial state $x_{0}$ of system (1.4.1) to a prescribed terminal
state $x_{F}=x(t_{F})$? Describe the set of all points in $M$ that
are reachable in this sense!
\item Proof the existence of time-optimal controls and give explicit construction
schemes for them.
\end{enumerate}
To make things precise we introduce some terminology.

\begin{defn}
A \emph{nonlinear control system\index{control system}} $\Sigma =(M,f_{u},U)$
is a triple consisting of a smooth manifold $M$, a parameter space
$U\subseteq \mathbb{R}^{m}$ and a family \begin{equation}
\left\{ \left.f_{u}\in \Gamma (TM)\right|u\in U\right\} \label{eq:}\end{equation}
 of vector fields on $M$. We will refer to $M$ as the \emph{state
space\index{state space}} of the control system, to $U$ as the \emph{space
of control parameters}\index{control parameter}, and to $u\in U$
as a \emph{control parameter}. A \emph{control} is a path $t\mapsto u(t)$
in the space of control parameters.\\
A curve $x:\left[0,T\right]\rightarrow M$ is called an \emph{integral
curve} for the control $u:\left[0,T\right]\rightarrow U,t\mapsto u(t)$
if it is absolutely continuous, and if \begin{equation}
\dot{x}(t)=f_{u(t)}(x(t))\label{eq:}\end{equation}
is satisfied for all $0<t<T$.
\end{defn}
\begin{notation}
We frequently write $f(x,u)$ rather than $f_{u}(x)$ for the value
of the vector field $f_{u}$ at the point $x$.
\end{notation}
To guarantee the existence of an integral curve as defined above,
we make the following standing assumptions:

\begin{itemize}
\item The map $u\mapsto f(x,u)$ is Lipschitzian for any fixed $x\in M$.
\item The vector field $f_{u}\in \Gamma (TM)$ is smooth for all $u\in U$.
\item The partial derivatives of the map $(x,u)\mapsto f(x,u)$ in directions
of $M$ are locally bounded in any point $(x_{0},u_{0})\in (M\times U)$.
\item The control $t\mapsto u(t)$ is measurable and locally bounded on
its interval of definition.
\end{itemize}
Under these assumptions, the existence and uniqueness of an integral
curve $t\mapsto x(t)$ with prescribed initial condition $x(t_{0})=x_{0}$
is guaranteed by the Caratheodory theorem for any control $t\mapsto u(t)$,
cf. \cite{key-24}, p. 28-29. 

\begin{defn}
\begin{flushleft}Let $\Sigma =(M,f_{u},U)$ be a control system, $x_{0}\in M$,
and $T\geq 0$. We define $R(x_{0},T)$ to be the set of all $x_{F}\in M$
with the property that there exists a control $u:[0,T]\rightarrow U$
which generates a trajectory $t\mapsto x(t)$ such that $x(0)=x_{0}$
and $x(T)=x_{F}$. We call $R(x_{0},T)$ the \emph{set of reachable
points from $x_{0}$ at time $t$}. \emph{}\\
Define furthermore the reachable set \emph{from $x_{0}$ within time
$T$} to be\begin{equation}
\mathbf{R}(x_{0},T)=\bigcup _{0\leq t\leq T}R(x_{0},t),\label{eq:}\end{equation}
and the \emph{reachable set\index{reachable set} for} $x_{0}$ to
be \emph{}\begin{equation}
\mathbf{R}(x_{0})=\bigcup _{0\leq T<\infty }\mathbf{R}(x_{0},T).\label{eq:}\end{equation}
The system $\Sigma $ is called \emph{controllable\index{controllability}},
if $\mathbf{R}(x_{0})=M$ holds. \end{flushleft}
\end{defn}
In the sequel we will pay attention to control problems on Lie groups
$G$ and homogeneous spaces $G/H$ only. We will therefore be confronted
with a special class of control systems.

\begin{defn}
A control system $\Sigma =(G,f_{u},U)$, $U\subseteq \mathbb{R}^{m}$,
on a Lie group $G$ is called \emph{affine right-invariant\index{affine right-invariant control system}}
if $\left\{ f_{u}\right\} _{u\in U}$ is a family of vector fields
on $G$ of the form\begin{equation}
f(g,u)=X_{0}(g)+\sum _{i=1}^{m}u_{i}X_{i}(g)\label{eq:}\end{equation}
with $X_{i}$, $i=0,...,m$, right-invariant and $u=(u_{1},...,u_{m})\in U\subseteq \mathbb{R}^{m}$. 
\end{defn}
Now what about questions (1) and (2) formulated above in the context
of Lie groups? There are very detailed investigations on those topics,
see e.g. the paper by V. Jurdjevic and H. J. Sussmann \cite{key-13}
where problem (1) is completely answered, the paper by D. Mittenhuber
\cite{key-19} for a treatise of question (2) as well as Jurdjevic's
book \cite{key-14}. The following is a survey of the results needed
to tackle the question of controllability of those quantum mechanical
systems we are finally interested in.

\begin{thm}
\emph{(Controllability of affine right-invariant systems on Lie groups).}
Let $\Sigma =(G,f_{u},U)$ with $f_{u}(g)=X_{0}(g)+\sum _{i=1}^{m}u_{i}X_{i}(g)$
be an affine right invariant system on the Lie group $G$. Denote
by $\mathfrak{g}$ the Lie algebra of $G$. Then the reachable set
$\mathbf{R}(\mathbf{1})$ is always a semi-group. If $\mathbf{R}(\mathbf{1})$
happens to be a group then it coincides with $\mathrm{S}(X_{0},...,X_{m})$,
the Lie subgroup of $G$ generated by the elements $\exp X_{0},...,\exp X_{m}\in G$.\\
Each of the following two conditions is sufficient for $\mathrm{S}(X_{0},...,X_{m})$
to be a Lie subgroup:

\begin{flushleft}(i) $X_{0}=0$ (absence of a drift term).\end{flushleft}

\begin{flushleft}(ii) $\mathrm{S}(X_{0},...,X_{m})$ is compact.\end{flushleft}

\begin{flushleft}If (ii) is satisfied, then there is a constant $T>0$
such that $\mathbf{R}(\mathbf{1})=\mathbf{R}(\mathbf{1},T)$.\\
Furthermore, under the additional assumption that $G$ is connected,
the following criterion on controllability holds:\begin{equation}
\Sigma \textrm{ }\mathrm{is}\textrm{ }\mathrm{controllable}\quad \Longleftrightarrow \quad \left\langle X_{0},...,X_{m}\right\rangle _{\mathrm{Lie}}=\mathfrak{g}.\label{eq:}\end{equation}
\end{flushleft}
\end{thm}
\begin{proof}
\cite{key-5}, Lemma 4.5 and Theorems 5.1, 6.5.
\end{proof}
A little bit more theory is needed to answer the remaining question
(2).

\section{Optimal Control and the Maximum Principle}

In this section we take up the discussion of time-optimal control
as formulated in question (2), Section 1.4.

\begin{defn}
Let $\Sigma =(M,f_{u},U)$ be a nonlinear control system and $\varphi :M\times U\rightarrow \mathbb{R}$
a continuous function. For a trajectory $t\mapsto \left(x(t),u(t)\right)$
of $\Sigma $ with initial point $x(t_{0})=a$ and terminal point
$x(t_{1})=b$ we define the \emph{cost of transfer\index{cost of transfer}
between $a$ and} $b$ to be \begin{equation}
\int _{t_{0}}^{t_{1}}\varphi (x(t),u(t))\, dt.\label{eq:}\end{equation}
A trajectory $t\mapsto (\bar{x}(t),\bar{u}(t))$ of $\Sigma $ that
transfers $a\in M$ to $b\in M$ is called \emph{optimal} if $\int _{t_{0}}^{t_{1}}\varphi (\bar{x}(t),\bar{u}(t))\, dt$
is minimal amongst all costs of transfer between $a$ and $b$.\\
In the special case $\varphi \equiv 1$ we refer to the corresponding
cost functional as \emph{time} and to the respective optimal trajectories
as being \emph{time-optimal}\index{time-optimal control}.
\end{defn}
It is convenient to implement the cost function $\varphi $ into the
given control system $\Sigma $ as follows. Set $\Sigma _{\mathrm{ext}}:=(\mathbb{R}\times M,\tilde{f}_{u},U)$,
where $\tilde{f}_{u}$ is the vector field on $\mathbb{R}\times M$
given by \begin{equation}
\tilde{f}_{u}(x_{0},x)=\left(\varphi (x,u),f_{u}(x)\right).\label{eq:}\end{equation}
We call $\Sigma _{\mathrm{ext}}$ the \emph{cost-extended system\index{cost extended system}
for} $(\Sigma ,\varphi )$.

\begin{flushleft}The geometric significance of the trajectories of
$\Sigma _{\mathrm{ext}}$ is that optimal trajecto-\\
ries $t\mapsto \bar{x}(t)$ of $\Sigma $ for the transfer of $a$
to $b$ arise as the projections on $M$ of those trajectories $t\mapsto (\bar{x}_{0}(t),\bar{x}(t))$
of $\Sigma _{\mathrm{ext}}$ that transfer $(0,a)$ to $(\bar{x}_{0}(T),b)$
with $\bar{x}_{0}(T)$ minimal. Such trajectories $t\mapsto (\bar{x}_{0}(t),\bar{x}(t))$
of the cost-extended system neces-\\
sarily have their terminal point $(\bar{x}_{0}(T),b)$ on the boundary
of $\mathbf{R}(0,a)$. We call this the \emph{extremality property}
(E)\index{extremal property (E)} of optimal trajectories for $\left(\Sigma ,\varphi \right)$.\end{flushleft}

\begin{flushleft}We shall now discuss a necessary condition for a
control function $u$ to gene-\\
rate a trajectory which enjoys the extremality property (E). This
will lead us to the well-known \emph{maximum principle of Pontrjagin}\index{Pontrjagin's maximum principle}.
To this aim we need to introduce some terminology from classical mechanics
(cf. e.g. \cite{key-32}). For simplicity we focus at first on the
case $M=\mathbb{R}^{n}$, and then extend the dis-\\
cussion to arbitrary smooth manifolds $M$.\end{flushleft}

\subsection{The case $M=\mathbb{R}^{n}$.}

In this situation we regard the manifold $N=\mathbb{R}^{n+1}\times \mathbb{R}^{n+1}$
as a state space on which a family $H(x,p,u)$ of so-called Hamiltonian
functions, parametrized by the elements $u$ of the space $U$ of
control parameters, is given. These Hamiltonian functions are defined
by \begin{equation}
H(\cdot ,u):N\longrightarrow \mathbb{R},\quad (x,p)\longmapsto \sum _{i=0}^{n}p_{i}\tilde{f}_{i}(x,u).\label{eq:}\end{equation}
We here consider the tangent vector $\tilde{f}(x,u)$ as an element
of $\mathbb{R}^{n+1}$.\\
Now any smooth function $H:N\rightarrow \mathbb{R}$ defines a \emph{Hamiltonian
vector field\index{Hamiltonian vector field}} $X_{H}$ on $N$, whose
coordinates are given by\begin{equation}
\left(\frac{\partial H}{\partial p_{0}},...,\frac{\partial H}{\partial p_{n}},-\frac{\partial H}{\partial x_{0}},...,-\frac{\partial H}{\partial x_{n}}\right).\label{eq:}\end{equation}
The Hamiltonian vector field $X_{H}$ can alternatively be described
via the canonical symplectic form $\omega $ on $N$. This is a closed
$2$-form which is defined at each point $(x,p)\in N$ by\begin{equation}
\omega (X,Y):=X_{x}Y_{p}-X_{p}Y_{x}\in \mathbb{R},\label{eq:}\end{equation}
for $X=(X_{x},X_{p}),Y=(Y_{x},Y_{p})\in T_{(x,p)}N\cong \mathbb{R}^{n+1}\times \mathbb{R}^{n+1}$.
The vector field $X_{H}$ can then be defined to be the unique vector
field on $N$ which satisfies\begin{equation}
dH(Y)=\omega (X_{H},Y)\label{eq:}\end{equation}
for every vector field $Y$ on $N$.\\
The Hamiltonian vector field $X_{u}:=X_{H(\cdot ,u)}$ associated
with the particular Hamiltonian function (1.5.3) is in coordinates
$\left(x_{0},...,x_{n},p_{0},...,p_{n}\right)$ of $N$ given by\begin{equation}
X_{u}=\left(\tilde{f}_{0}(x,u),...,\tilde{f}_{n}(x,u),-\sum _{j=0}^{n}p_{j}\frac{\partial \tilde{f}_{j}\left(x,u\right)}{\partial x_{0}},...,-\sum _{j=0}^{n}p_{j}\frac{\partial \tilde{f}_{j}\left(x,u\right)}{\partial x_{n}}\right).\label{eq:}\end{equation}

\begin{rem}
The function $\tilde{f}\left(\cdot ,u\right)=\left(\varphi (x,u),f_{u}(x)\right)$
does by definition not depend on the variable $x_{0}$. Thus the term
$-\sum _{j=0}^{n}p_{j}\frac{\partial \tilde{f}_{j}}{\partial x_{0}}\left(x,u\right)$
in equation (1.5.7) vanishes identically. So any integral curve $t\mapsto \left(x_{0}(t),...,x_{n}(t),p_{0}(t),...,p_{n}(t)\right)$
of the vector field $X_{u}$ has constant coordinate $p_{0}$. The
Hamiltonian function in a later formulation of Pontrjagin's maximum
principle (cf. Theorem 2.4.2) will for that reason depend on the variables
$x_{0},...,x_{n}$ and $p_{1},...,p_{n}$ only, while $p_{0}$ will
appear as a parameter. As long as we are in the situation of a Euclidian
state space $N=\mathbb{R}^{n}\times \mathbb{R}^{n}$, the relevant
Hamiltonian function reads\begin{equation}
H(x_{1},...,x_{n},p_{1},...,p_{n})=p_{0}\varphi (x_{1},...,x_{n},u)+\sum _{i=1}^{n}p_{i}x_{i},\label{eq:}\end{equation}
where $p_{0}$ is a constant.
\end{rem}
Now let $t\mapsto u(t)$ be a control function and denote by $t\mapsto \gamma (t):=\left(x(t),p(t)\right)$
the integral curve of the time-dependent Hamiltonian vector field
$X_{u(t)}$ with initial value $\left(x_{0},p_{0}\right)$. Equation
(1.5.7) implies that the projection of $\gamma $ on the first factor
of $N=\mathbb{R}^{n+1}\times \mathbb{R}^{n+1}$ is equal to the trajectory
$t\mapsto x(t)$ that arises from the control $t\mapsto u(t)$ of
the cost-extended system and has initial value $x_{0}$. The path
$t\mapsto \gamma (t)$ is for this reason called \emph{Hamiltonian
lift\index{Hamiltonian lift}} of the path $t\mapsto x(t)$. The existence
of such a Hamiltonian lift gives rise to the idea of expressing the
extremality condition (E) suitably as a condition on the time-dependent
Hamiltonian function $H(\cdot ,u(t))$. The following definition introduces
the correct extremality condition.

\begin{defn}
Let $t\mapsto u(t)$ be a control function. The \emph{extremal Hamiltonian\index{extremal Hamiltonian}}
$M\left(x(t),p(t)\right)$ associated with the integral curve $t\mapsto \left(x(t),p(t)\right)$
of the time-dependent Hamiltonian vector field $X_{u(t)}$ is defined
by\begin{equation}
M\left(x(t),p(t)\right):=\sup _{u\in U}H\left(x(t),p(t),u\right).\label{eq:}\end{equation}
The statement of Pontrjagin's maximum principle (PMP) is the following.
Assume that the control $t\mapsto u(t)$, $t\in \left[0,T\right]$
generates a trajectory $t\mapsto x(t)\in \mathbb{R}^{n+1}$ of the
cost-extended system $\Sigma _{\mathrm{ext}}$ which has the extremal
property (E). Then the Hamiltonian lift of $t\mapsto x(t)$ to the
path $t\mapsto \left(x(t),p(t)\right)$ satisfies the extremality
condition \begin{equation}
H\left(x(t),p(t),u(t)\right)=M\left(x(t),p(t)\right)\label{eq:}\end{equation}
almost everywhere on $\left[0,T\right]$.\\
We demonstrate how to make use of PMP in a concrete but typical situation
(see also \cite{key-24}, p. 191).
\end{defn}
\begin{example}
On $M=\mathbb{R}$ consider the system\begin{equation}
\ddot{x}_{1}=u,\quad \left|u\right|\leq 1\label{eq:}\end{equation}
or equivalently\begin{equation}
\left\{ \begin{array}{ll}
 x=(x_{1},x_{2})\in \mathbb{R}^{2}, & \left|u\right|\leq 1,\\
 \dot{x_{1}}=x_{2}, & \\
 \dot{x_{2}}=u. & \end{array}
\right.\label{eq:}\end{equation}
We wish to steer system (1.5.12) from $x(0)=x_{0}$ to $x(t_{F})=0$
such that $t_{F}$ is minimal. Thus we take $\varphi \equiv 1$ to
serve as a cost function. The family of admissible vector fields for
our problem is\[
f_{u}(x)=(x_{2},u).\]
The Hamiltonian function $H\left(\cdot ,u\right)$ as defined through
equation (1.5.8) reads in this example\[
H\left(x,p,u\right)=p_{0}+p_{1}x_{2}+p_{2}u,\quad p_{0}\textrm{ constant,}\]
and leads to the Hamiltonian vector field\[
\left\{ \begin{array}{l}
 \dot{x}=\frac{\partial H\left(\cdot ,u\right)}{\partial p},\\
 \dot{p}=-\frac{\partial H\left(\cdot ,u\right)}{\partial x},\end{array}
\right.\]
\begin{equation}
\Longleftrightarrow \left\{ \begin{array}{ll}
 \dot{x}_{1}=x_{2}, & \dot{x_{2}}=u,\\
 \dot{p}_{1}=0, & \dot{p}_{2}=-p_{1}.\end{array}
\right.\label{eq:}\end{equation}
 The Hamiltonian system (1.5.13) that belongs to a time-optimal control
$t\mapsto \tilde{u}(t)$ has by PMP a solution such that \[
H\left(x(t),p(t),\tilde{u}(t)\right)=\max _{\left|u\right|\leq 1}H\left(x(t),p(t),u\right)=\max _{\left|u\right|\leq 1}\left(p_{1}(t)x_{2}(t)+p_{2}(t)u\right).\]
>From this it is immediate that $\tilde{u}(t)=\textrm{sgn}\left(p_{2}(t)\right)$,
if $p_{2}(t)\neq 0$. Therefore, \[
\max _{\left|u\right|\leq 1}H\left(x(t),p(t),u\right)=p_{1}(t)x_{2}(t)+\left|p_{2}(t)\right|.\]
>From (1.5.13) it follows that $p_{2}(t)=\alpha +\beta t$. One now
solves the equations involving $x$ in (1.5.13) with $\tilde{u}(t)=\textrm{sgn}\left(p_{2}(t)\right)$
to obtain the extremal trajectories of system (1.5.12). The result
is that for any initial value $x(0)=(x_{1}(0),x_{2}(0))$ there is
exactly one time-optimal trajectory $t\mapsto x(t)$. This is obtained
from choosing the control $u$ to be $u\equiv +1$ as long as\[
x_{1}>\frac{x_{2}^{2}}{2},\quad x_{2}<0\quad \textrm{or}\quad x_{1}>-\frac{x_{2}^{2}}{2},\quad x_{2}>0\]
is satisfied, and switching to $u\equiv -1$ if\[
x_{1}=\frac{x_{2}^{2}}{2},\quad x_{2}<0\quad \textrm{or}\quad x_{1}=-\frac{x_{2}^{2}}{2},\quad x_{2}>0,\]
or otherwise choosing $u\equiv -1$ as long as\[
x_{1}<\frac{x_{2}^{2}}{2},\quad x_{2}<0\quad \textrm{or}\quad x_{1}<-\frac{x_{2}^{2}}{2},\quad x_{2}>0\]
holds, and then switching to $u\equiv +1$.
\end{example}

\subsection{The general case.}

We now discuss how the previous considerations carry over to the case
of an arbitrary smooth manifold $M$. The state space is now taken
to be\begin{equation}
N:=T^{*}\left(\mathbb{R}\times M\right),\label{eq:}\end{equation}
the cotangent bundle of $\mathbb{R}\times M$. The manifold $N$ carries
in a canonical way a symplectic structure $\omega $ (i.e. $\omega $
is a non-degenerate closed $2$-form), which is defined to be \begin{equation}
\omega :=d\theta ,\label{eq:}\end{equation}
where the $1$-form $\theta $ is given by\begin{equation}
\theta _{\xi }(X):=\xi \left(D_{\xi }\pi (X)\right)\label{eq:}\end{equation}
for $\xi \in N$ and $X\in T_{\xi }N$. Here $\pi $ denotes canonical
projection from $N$ onto its base manifold $\mathbb{R}\times M$.\\
Using the symplectic form $\omega $ one can repeat the construction
of Hamiltonian vector fields, but now in a coordinate-free manner.
For any smooth function $H:N\rightarrow \mathbb{R}$ define the \emph{Hamiltonian
vector field\index{Hamiltonian vector field}} $X_{H}$ \emph{associated
with} $H$ to be the unique vector field on $N$ with the property
that\begin{equation}
dH_{\xi }(Y)=\omega _{\xi }\left(X_{H}(\xi ),Y\right)\label{eq:}\end{equation}
holds for all $\xi \in N$ and $Y\in T_{\xi }N$. One can show (cf.
\cite{key-32}) that in suitably defined local coordinates (so-called
\emph{Darboux coordinates}) the Hamiltonian vector field $X_{H}$
is of the same form as defined in (1.5.4) for the Euclidian case.\\
We again introduce a family $H\left(\cdot ,u\right):N\rightarrow \mathbb{R}$
of Hamiltonian functions, para-\\
metrized by the controls $u\in U$, as\begin{equation}
H\left(\xi ,u\right):=\xi \left(\tilde{f}_{u}\left(\pi (\xi )\right)\right).\label{eq:}\end{equation}
This definition can be shown to be consistent with that in (1.5.3),
and one also can prove that the trajectories $t\mapsto \xi (t)$ of
the Hamiltonian vector field $X_{u}:=X_{H\left(\cdot ,u\right)}$
are projected under $\pi $ to those of the vector field $\tilde{f}_{u}$
on $\mathbb{R}\times M$. In complete analogy to the linear case we
refer to the trajectories $t\mapsto \xi (t)$ as the \emph{Hamiltonian
lifts\index{Hamiltonian lift}} of the integral curves of $\tilde{f}_{u}$.\\
We finally adapt the extremality condition of the maximum principle
to the new situation of a general state space $N$.

\begin{defn}
Let $t\mapsto u(t)$ be a control function. The \emph{extremal Hamiltonian\index{extremal Hamiltonian}}
$M\left(\xi (t)\right)$ associated with the integral curve $t\mapsto \left(\xi (t)\right)$
of the time-dependent Hamiltonian vector field $X_{u(t)}$ is defined
by\begin{equation}
M\left(\xi (t)\right):=\sup _{u\in U}H\left(\xi (t),u\right).\label{eq:}\end{equation}
The statement of Pontrjagin's maximum principle on the relationship
between the extremality (E) of trajectories of the control system
$\Sigma $ and the extremality of their Hamiltonian lifts as formulated
in 1.5.1 remains valid also in the non-linear case. \\
For a proof and detailed discussion of PMP we refer the reader to
the books \cite{key-24} and \cite{key-14} and give here the precise
statement of the maximum principle for time-optimal control problems.
\end{defn}
\begin{thm}
Let $\Sigma =(M,f_{u},U)$ be a control system and $t\mapsto \tilde{u}(t)$,
$t\in [0,T]$, a time-optimal control. For each $u\in U$ define the
Hamilton function\begin{equation}
H(\cdot ,u):T^{*}M\longrightarrow \mathbb{R},\quad H(\xi ,u)=\xi \left(f_{u}\left(\pi (\xi )\right)\right)\label{eq:}\end{equation}
and denote by $X_{u}\in \Gamma (T^{*}M)$ the Hamiltonian vector field
for $H(\cdot ,u)$. Then any trajectory of $\dot{q}=f_{\tilde{u}}(q)$
in $M$ for the control function $\tilde{u}$ possesses a Hamiltonian
lift to a curve $t\mapsto \xi (t)$ in $T^{*}M$ with the property
that the extremality condition\begin{equation}
H(\xi (t),\tilde{u}(t))=M\left(\xi (t)\right)\label{eq:}\end{equation}
holds almost everywhere on $[0,T]$. 
\end{thm}
\begin{proof}
\cite{key-24}, Corollary 12.12.
\end{proof}

\section{Kronecker Product Formalism}

In this section we develop a formalism which allows for an elegant
description of linear transformations on the tensor product $V\otimes W$
of vector spaces $V$ and $W$. This formalism is well-suited for
calculations in quantum mechanical multi-particle systems.

\begin{notation}
In the sequel, all vector spaces are finite-dimensional over the field
$\mathbb{K}=\mathbb{R}$ or $\mathbb{K}=\mathbb{C}$. For short, we
will always write $V\otimes W$ for the $\mathbb{K}$-tensor product
of the vector spaces $V$ and $W$. If $V$ and $W$ carry the inner
product $\left\langle \cdot ,\cdot \right\rangle _{V}$ and respectively
$\left\langle \cdot ,\cdot \right\rangle _{W}$, then $V\otimes W$
will also be regarded an inner product space with the induced inner
product which is given by\begin{equation}
\left\langle v\otimes w,v'\otimes w'\right\rangle _{V\otimes W}=\left\langle v,v'\right\rangle _{V}\left\langle w,w'\right\rangle _{W}.\label{eq:}\end{equation}
We set $\mathrm{End}(V)$ for the vector space of $\mathbb{K}$-linear
endomorphisms of $V$. This is a $\mathbb{K}$-Lie algebra with bracket
$\left[A,B\right]=AB-BA$.
\end{notation}
\begin{defn}
Let $V$, $W$ vector spaces and $A\in \mathrm{End}(V)$, $B\in \mathrm{End}(W)$.
We define the \emph{Kronecker product\index{Kronecker product}} $A\otimes B\in \mathrm{End}(V\otimes W)$
by\begin{equation}
\left(A\otimes B\right)\left(v\otimes w\right)=Av\otimes Bw\label{eq:}\end{equation}
for $v\in V$, $w\in W$.\\
Set\begin{equation}
X:=\mathrm{Span}{}_{\mathbb{K}}\left\{ \left.A\otimes B\right|A\in \mathrm{End}(V),B\in \mathrm{End}(W)\right\} \subseteq \mathrm{End}(V\otimes W).\label{eq:}\end{equation}

\end{defn}
\begin{lem}
The Kronecker product has the following properties.

\begin{flushleft}(i) For all $A,B\in \mathrm{End}(V)$, $C,D\in \mathrm{End}(W)$
and $\lambda \in \mathbb{K}$,\begin{eqnarray*}
\left(A+B\right)\otimes C & = & A\otimes C+B\otimes C,\\
A\otimes (C+D) & = & A\otimes C+A\otimes D,\\
\left(\lambda A\right)\otimes B & = & \lambda \left(A\otimes B\right)=A\otimes \left(\lambda B\right),\\
\left(A\otimes C\right)\circ \left(B\otimes D\right) & = & \left(A\circ B\right)\otimes \left(C\circ D\right).
\end{eqnarray*}
\end{flushleft}

\begin{flushleft}(ii) If $\left\{ A_{i}\right\} _{i}$, $\left\{ B_{j}\right\} _{j}$
are bases of $\mathrm{End}(V)$, $\mathrm{End}(W)$, then $\left\{ A_{i}\otimes B_{j}\right\} _{(i,j)}$
is a basis of $X$. Moreover, $X=\mathrm{End}(V\otimes W)$.\end{flushleft}

\begin{flushleft}(iii) Let $\left\langle \cdot ,\cdot \right\rangle _{\mathrm{End}(V)}$
and $\left\langle \cdot ,\cdot \right\rangle _{\mathrm{End}(W)}$
be inner products on $\mathrm{End}(V)$ and on $\mathrm{End}(W)$.
Then an inner product $\left\langle \cdot ,\cdot \right\rangle $
on $\mathrm{End}(V\otimes W)$ is defined by linear continuation of
\begin{equation}
\left\langle A\otimes B,A'\otimes B'\right\rangle :=\left\langle A,A'\right\rangle _{\mathrm{End}(V)}\left\langle B,B'\right\rangle _{\mathrm{End}(W)}.\label{eq:}\end{equation}
If $\left\{ A_{i}\right\} _{i}$, $\left\{ B_{j}\right\} _{j}$ are
orthonormal bases of $\mathrm{End}(V)$ and $\mathrm{End}(W)$, then,
with res-\\
pect to the inner product as defined above, the set $\left\{ A_{i}\otimes B_{j}\right\} _{(i,j)}$
is an orthonor-\\
mal basis of $\mathrm{End}(V\otimes W)$. \end{flushleft}

\begin{flushleft}(iv) For all $A,A'\in \mathrm{End}(V)$ and $B,B'\in \mathrm{End}(W)$
the following formula holds:\begin{equation}
\left[A\otimes B,A'\otimes B'\right]=\left[A,A'\right]\otimes BB'+A'A\otimes \left[B,B'\right].\label{eq:}\end{equation}
(v) For all $A\in \mathrm{End}(V)$, $B\in \mathrm{End}(W)$\begin{equation}
\left(A\otimes B\right)^{H}=A^{H}\otimes B^{H}.\label{eq:}\end{equation}
\end{flushleft}

\begin{flushleft}(vi) With respect to ordered bases $\left\{ v_{i}\right\} _{i}$
of $V$,$\left\{ w_{j}\right\} _{j}$ of $W$, and $\left\{ v_{i}\otimes w_{j}\right\} _{(i,j)}$
of $V\otimes W$ (with the indices $(i,j)$ being ordered lexicographically),
the endomor-\\
phism $A\otimes B$ is represented by the matrix\begin{equation}
\left(r_{ik}s_{jl}\right)_{(i,j),(k,l)}\label{eq:}\end{equation}
if $A$ and $B$ are represented by matrices $\left(r_{ij}\right)_{(i,j)}$
and $\left(s_{kl}\right)_{(k,l)}$, respectively.\end{flushleft}

\begin{flushleft}(vii) For all $A,B\in \mathrm{End}(V)$,\begin{equation}
\mathrm{tr}(A\otimes B)=\mathrm{tr}(A)\mathrm{tr}(B).\label{eq:}\end{equation}
\end{flushleft}

\begin{flushleft}(viii) For all $A,B\in \mathrm{End}(V)$,\begin{equation}
A\otimes B=P\left(B\otimes A\right)P^{-1}\label{eq:}\end{equation}
with an involution $P\in Gl\left(V\otimes V\right)$.\end{flushleft}

\begin{flushleft}(ix) If $A\in Gl(V)$, $B\in Gl(W)$ then $A\otimes B\in Gl\left(V\otimes W\right)$
and has inverse \begin{equation}
\left(A\otimes B\right)^{-1}=A^{-1}\otimes B^{-1}.\label{eq:}\end{equation}
\end{flushleft}

\begin{flushleft}(x) Let $A,A'\in \mathrm{End}(V)$ and $B,B'\in \mathrm{End}(W)$
equivalent endomorphisms, i.e. $A=UA'U^{-1}$ and $B=VB'V^{-1}$ for
some $U\in Gl(V)$ and $V\in Gl(W)$. Then also $A\otimes B$ and
$A'\otimes B'$ are equivalent with\begin{equation}
A\otimes B=\left(U\otimes V\right)\left(A'\otimes B'\right)\left(U\otimes V\right)^{-1}.\label{eq:}\end{equation}
\end{flushleft}

\begin{flushleft}(xi) For all $A,B\in \mathrm{End}(V)$,\begin{equation}
\det \left(A\otimes B\right)=\left(\det A\det B\right)^{n},\label{eq:}\end{equation}
where $n=\dim V$.\end{flushleft}
\end{lem}\begin{proof}
(i) This follows from the bilinearity of the tensor product.\\
(ii) Let $\alpha _{ij}\in \mathbb{K}$ such that $\sum _{i,j}\alpha _{ij}\left(A_{i}\otimes B_{j}\right)=0$.
So for all $a,v\in V$, $b,w\in W$, and $z_{j}:=\left\langle B_{j}w,b\right\rangle _{W}$
it follows that\begin{eqnarray*}
0 & = & \left\langle \left(\sum _{i,j}\alpha _{ij}\left(A_{i}\otimes B_{j}\right)\right)\left(v\otimes w\right),a\otimes b\right\rangle \\
 & = & \left\langle \sum _{i,j}\alpha _{ij}\left(A_{i}v\otimes B_{j}w\right),a\otimes b\right\rangle \\
 & = & \sum _{i,j}\alpha _{ij}\left\langle A_{i}v,a\right\rangle _{V}\left\langle B_{j}w,b\right\rangle _{W}
\end{eqnarray*}
\begin{eqnarray*}
 & = & \sum _{i,j}\alpha _{ij}z_{j}\left\langle A_{i}v,a\right\rangle _{V}\\
 & = & \left\langle \left(\sum _{i,j}\alpha _{ij}z_{j}A_{i}\right)v,a\right\rangle _{V}.
\end{eqnarray*}
Therefore, because $\left\langle \cdot ,\cdot \right\rangle _{W}$
is non-degenerate,\[
\sum _{i,j}\alpha _{ij}z_{j}A_{i}=0.\]
Since $\left\{ A_{i}\right\} _{i}$ is linearly independent, we find
that for each $i$ \[
\sum _{j}a_{ij}z_{j}=0\quad \Longleftrightarrow \quad \sum _{j}a_{ij}\left\langle B_{j}w,b\right\rangle _{W}=0.\]
Because $w$ and $b$ are arbitrary, it follows that $\sum _{j}a_{ij}B_{j}=0$,
and finally, by linear independence of $\left\{ B_{j}\right\} _{j}$,
that $\alpha _{ij}=0$. Furthermore, the dimension of \[
\mathrm{End}(V)\otimes \mathrm{End}(W)=\mathrm{Span}{}_{\mathbb{K}}\left\{ \left.A\otimes B\right|A\in \mathrm{End}(V),B\in \mathrm{End}(W)\right\} \]
is\begin{eqnarray*}
\dim \mathrm{End}(V)\dim \mathrm{End}(W) & = & \left(\dim V\right)^{2}\left(\dim W\right)^{2}\\
 & = & \left(\dim \left(V\otimes W\right)\right)^{2}\\
 & = & \dim \mathrm{End}\left(V\otimes W\right).
\end{eqnarray*}
 So\[
\mathrm{End}(V)\otimes \mathrm{End}(W)=\mathrm{End}\left(V\otimes W\right).\]
(iii) The sesquilinearity of $\left\langle \cdot ,\cdot \right\rangle $
follows from that of $\left\langle \cdot ,\cdot \right\rangle _{\mathrm{End}(V)}$
and $\left\langle \cdot ,\cdot \right\rangle _{\mathrm{End}(W)}$
together with the bilinearity of the tensor product. The bilinear
form $\left\langle \cdot ,\cdot \right\rangle $ is positiv definite
because for all $A\in \mathrm{End}(V)$, $B\in \mathrm{End}(W)$ we
have that\[
\left\langle A\otimes B,A\otimes B\right\rangle =\left\langle A,A\right\rangle _{\mathrm{End}(V)}\left\langle B,B\right\rangle _{\mathrm{End}(W)}\geq 0,\]
and the last expression is equal to $0$ if and only if $A=0$ or
$B=0$, i.e. if and only if $A\otimes B=0$.\\
(iv) The Lie bracket of two matrices $(X_{ij})_{ij}$ and $(Y_{ij})_{ij}$
has matrix elements $\left[X,Y\right]_{rs}=\sum _{u}X_{ru}Y_{us}-Y_{ru}X_{us}$.
We apply this to the matrix representation of $\left[A\otimes B,A'\otimes B'\right]$
and find that\begin{eqnarray*}
\left[A\otimes B,A'\otimes B'\right]_{ij,kl} & = & \sum _{st}\left(A\otimes B\right)_{ij,st}\left(A'\otimes B'\right)_{st,kl}-\left(A'\otimes B'\right)_{ij,st}\left(A\otimes B\right)_{st,kl}\\
 & = & \sum _{st}A_{is}B_{jt}A_{sk}^{'}B_{tl}^{'}-A_{is}^{'}B_{jt}^{'}A_{sk}B_{tl}\\
 & = & \sum _{s}\left(\sum _{t}B_{jt}B_{tl}^{'}-B_{jt}^{'}B_{tl}\right)A_{is}A_{sk}^{'}\\
 &  & -\sum _{t}\left(\sum _{s}A_{is}^{'}A_{sk}-A_{is}A_{sk}^{'}\right)B_{jt}^{'}B_{tl}
\end{eqnarray*}
\begin{eqnarray*}
 & = & \sum _{s}\left[B,B'\right]_{jl}A_{is}A_{sk}^{'}-\sum _{t}\left[A^{'},A\right]_{ik}B_{jt}^{'}B_{tl}\\
 & = & \left[B,B'\right]_{jl}\sum _{s}A_{is}A_{sk}^{'}+\left[A,A^{'}\right]_{ik}\sum _{t}B_{jt}^{'}B_{tl}\\
 & = & \left[B,B'\right]_{jl}\left(AA^{'}\right)_{ik}+\left[A,A^{'}\right]_{ik}\left(B^{'}B\right)_{jl}
\end{eqnarray*}
 which proves the claim.\\
(v) For all $v,v'\in V$, $w,w'\in W$ we have that \begin{eqnarray*}
\left\langle v\otimes w,\left(A^{H}\otimes B^{H}\right)\left(v'\otimes w'\right)\right\rangle _{V\otimes W} & = & \left\langle v,A^{H}v'\right\rangle _{V}\left\langle w,B^{H}w'\right\rangle _{W}\\
 & = & \left\langle Av,v'\right\rangle _{V}\left\langle Bw,w'\right\rangle _{W}\\
 & = & \left\langle \left(A\otimes B\right)\left(v\otimes w\right),v'\otimes w'\right\rangle _{V\otimes W},
\end{eqnarray*}
hence $\left(A\otimes B\right)^{H}=A^{H}\otimes B^{H}$.\\
(vi) Let $\left\{ v_{i}\right\} _{i}$, $\left\{ w_{j}\right\} _{j}$
be ON-bases for $V$and $W$, respectively. Then $\left\{ v_{i}\otimes w_{j}\right\} _{ij}$
is an ON-basis for $V\otimes W$ and thus the matrix element $\left(A\otimes B\right)_{ij,kl}$
is given by\begin{eqnarray*}
\left(A\otimes B\right)_{ij,kl} & = & \left\langle v_{i}\otimes w_{j},\left(A\otimes B\right)\left(v_{k}\otimes w_{l}\right)\right\rangle _{V\otimes W}\\
 & = & \left\langle v_{i}\otimes w_{j},Av_{k}\otimes Bw_{l}\right\rangle _{V\otimes W}\\
 & = & \left\langle v_{i},Av_{k}\right\rangle _{V}\left\langle w_{j},Bw_{l}\right\rangle _{W}\\
 & = & A_{ik}B_{jl},
\end{eqnarray*}
(vii) This follows directly from the matrix representation of $A$,
$B$ and $A\otimes B$ as given in (vi):\begin{eqnarray*}
\textrm{tr}\left(A\otimes B\right) & = & \sum _{(ij),(ij)}\left(A\otimes B\right)_{ij,ij}\\
 & = & \sum _{i,j}A_{ii}B_{jj}\\
 & = & \left(\sum _{i}A_{ii}\right)\left(\sum _{j}B_{jj}\right)\\
 & = & \textrm{tr}(A)\textrm{tr}(B).
\end{eqnarray*}
(viii) A comparison of matrix elements of $A\otimes B$ and $B\otimes A$
shows that\[
\left(A\otimes B\right)_{ij,kl}=\left(B\otimes A\right)_{ji,lk}.\]
Thus a change of basis by a suitable transposition matrix $P$ transforms
$A\otimes B$ into $B\otimes A$.\\
(ix) For all $v\in V$, $w\in W$, \begin{eqnarray*}
\left(A^{-1}\otimes B^{-1}\right)\left(A\otimes B\right)\left(v\otimes w\right) & = & A^{-1}Av\otimes B^{-1}Bw\\
 & = & v\otimes w\\
 & = & \left(A\otimes B\right)\left(A^{-1}\otimes B^{-1}\right)\left(v\otimes w\right).
\end{eqnarray*}
(x) This becomes clear from\[
\left(U\otimes V\right)\left(A'\otimes B'\right)\left(U\otimes V\right)^{-1}=\left(UA'U^{-1}\right)\otimes \left(VB'V^{-1}\right)=A\otimes B.\]
(xi) The Kronecker product of a triangular matrix $A\in \mathrm{End}(V)$
and an arbitrary matrix $B\in \mathrm{End}(V)$ is the block triangular
matrix $A\otimes B$, the diagonal blocks consisting of the $\left(n\times n\right)$-matrices
$C_{ij}$ with\[
\left(C_{ij}\right)_{kl}=\left(A\otimes B\right)_{ik,jl}=A_{ij}B_{kl}.\]Thus $A\otimes B$ has determinant
\begin{eqnarray*}
\det \left(A\otimes B\right) & = & \prod _{i,j}\det C_{ij}\\
 & = & \prod _{i,j}\det \left(A_{ij}B_{kl}\right)_{kl}\\
 & = & \prod _{i,j}A_{ij}^{n}\det \left(B_{kl}\right)_{k,l}\\
 & = & \left(\det \left(B_{kl}\right)_{kl}\right)^{n}\prod _{i,j}A_{ij}^{n}\\
 & = & \left(\det \left(A_{ij}\right)_{ij}\det \left(B_{kl}\right)_{kl}\right)^{n},
\end{eqnarray*}
where in the last equation we have made use of the triangular form
of the matrix $(A_{ij})_{ij}$. Now any $A\in \mathrm{End}(V)$ is
conjugate to a triangular matrix $A'$, so by (xi) we find that $A\otimes B$
is conjugate to the block triangular matrix $A'\otimes B$. Thus\[
\det \left(A\otimes B\right)=\det \left(A'\otimes B\right)=\left(\det A'\det B\right)^{n}=\left(\det A\det B\right)^{n}.\]

\end{proof}
\begin{example}
Let $V=W=\mathbb{C}^{2}$, and let $\mathrm{End}(V)=\mathfrak{gl}_{2}\mathbb{C}$
be endowed with the inner product $\left\langle A,B\right\rangle =\frac{1}{2}\textrm{tr}\left(A^{H}B\right)$.
Then an orthonormal basis for $\mathfrak{gl}_{2}\mathbb{C}$ is given
by\begin{equation}
\mathbf{1}=\left(\begin{array}{cc}
 1 & 0\\
 0 & 1\end{array}
\right),I_{x}=\left(\begin{array}{cc}
 0 & 1\\
 -1 & 0\end{array}
\right),I_{y}=\left(\begin{array}{cc}
 0 & \mathrm{i}\\
 \mathrm{i} & 0\end{array}
\right),I_{z}=\left(\begin{array}{cc}
 \mathrm{i} & 0\\
 0 & -\mathrm{i}\end{array}
\right),\label{eq:}\end{equation}
and an orthonormal basis for the $\mathbb{R}$-vector space $\mathfrak{su}_{2}$
by $I_{x}$, $I_{y}$, $I_{z}$. According to the previous lemma,
an orthonormal basis for the $\mathbb{C}$-vector space $\mathrm{End}\left(\mathbb{R}^{2}\otimes \mathbb{R}^{2}\right)$
is made up by the set\[
\left\{ \mathbf{1}\otimes \mathbf{1},\mathbf{1}\otimes I_{x},\mathbf{1}\otimes I_{y},\mathbf{1}\otimes I_{z},I_{x}\otimes \mathbf{1},I_{x}\otimes I_{x},I_{x}\otimes I_{y},I_{x}\otimes I_{z},\right.\]
\[
\left.I_{y}\otimes \mathbf{1},I_{y}\otimes I_{x},I_{y}\otimes I_{y},I_{y}\otimes I_{z},I_{z}\otimes \mathbf{1},I_{z}\otimes I_{x},I_{z}\otimes I_{y},I_{z}\otimes I_{z}\right\} .\]
The special choice of the basis (1.6.13) is motivated from quantum
mechanics, where the matrices $I_{x}$, $I_{y}$, and $I_{z}$ are
called \emph{Pauli spin matrices}\index{Pauli spin matrix}. \\
Another aspect will become important later on. Consider the $\mathbb{R}$-linear
span of the set \begin{equation}
X:=\bigcup _{j=1}^{n}X_{j},\label{eq:}\end{equation}
where \[
X_{j}:=\left\{ \left.\mathrm{i}^{\varepsilon _{j}}\mathbf{1}\otimes ...\otimes I_{\alpha _{1}}\otimes ...\otimes I_{\alpha _{j}}\otimes ...\otimes \mathbf{1}\right|\alpha _{i}\in \left\{ x,y,z\right\} ,i=1,...,j\right\} ,\]
and\[
\varepsilon _{j}:=\left\{ \begin{array}{ll}
 1, & j\textrm{ even,}\\
 0,\textrm{ } & j\textrm{ odd.}\end{array}
\right.\]
So the set $X_{j}$ comprises (up to a sign $1$ or $\mathrm{i}$)
the $n$-fold tensor products of elements in $\left\{ \mathbf{1},I_{x},I_{y},I_{z}\right\} $
with exactly $j$ factors different from $\mathbf{1}$. By construction,
$X\subseteq \mathfrak{su}\left(2^{n}\right)$, because for each element
\[
Y=\mathrm{i}^{\varepsilon _{j}}\mathbf{1}\otimes ...\otimes I_{\alpha _{1}}\otimes ...\otimes I_{\alpha _{j}}\otimes ...\otimes \mathbf{1}\in X_{j},\]
the equation\[
Y+Y^{H}=\mathrm{i}^{\varepsilon _{j}}\mathbf{1}\otimes ...\otimes I_{\alpha _{1}}\otimes ...\otimes I_{\alpha _{j}}\otimes ...\otimes \mathbf{1}+\left(\mathrm{i}^{\varepsilon _{j}}\right)\mathbf{1}\otimes ...\otimes \left(-I_{\alpha _{1}}\right)\otimes ...\otimes \left(-I_{\alpha _{j}}\right)\otimes ...\otimes \mathbf{1}=0\]
holds by Lemma 1.6.3 (v). Since the span of $X$ has the maximal possible
dimension $4^{n}-1$, it follows that $X$ is a tensor product basis
of $\mathfrak{su}\left(2^{n}\right)$. This basis will be used for
further calculations in our discussion of concrete $n$-particle spin
systems, cf. Chapter 3.
\end{example}

\chapter{General Control Theory for Spin Systems}

\section{Quantum Mechanics of Spin Systems}

We here give a short overview of the basic principles of quantum mechanics,
and in particular describe the physics of spin systems whose control
properties are in the focus of this work. This exposition is by no
means complete but is intended to introduce all the terminology and
concepts needed in the subsequent sections. We refer the reader to
\cite{key-20} for an exhaustive treatment of the subject.

The premise of non-relativistic quantum mechanics is that the state
of physical objects like electrons, protons and neutrons as well as
larger systems of those like atoms and molecules is represented by
a \emph{wave-function\index{wave-function}} $\psi $. This function
$\psi $ carries all the information of the state of the system under
consideration. \\
The collection of the physical relevant wave-functions is given by
the \emph{state space}\index{state space}, a separable complex Hilbert
space $\mathcal{H}$. This space could e.g. be the space $L_{2}$
of square-integrable functions $\mathbb{R}^{3}\rightarrow \mathbb{C}$;
a wave-function $\psi \in L_{2}$ would then contain information of
where the particle is localized in three-space. To be a little bit
more concrete,\begin{equation}
\frac{\int _{Q}\psi (x)\psi ^{*}(x)\, dx}{\int _{\mathbb{R}^{3}}\psi (x)\psi ^{*}(x)\, dx}\label{eq:}\end{equation}
gives the probability of {}``finding'' the particle within a measurable
subset $Q$ of $\mathbb{R}^{3}$.\\
It is convenient to normalize the wave-function $\psi $ to have norm
\begin{equation}
\int _{\mathbb{R}^{3}}\psi (x)\psi ^{*}(x)\, dx=1;\label{eq:}\end{equation}
wave-functions which only differ by a non-zero scalar will be regarded
equivalent.\\
The time-evolution of a state $\psi $ is governed by \emph{Schrödinger's
equation}\index{Schrödinger's equation}\begin{equation}
\mathrm{i}\hbar \dot{\psi }=H\psi \label{eq:}\end{equation}
with $H$ a Hermitian operator, which is called the \emph{Hamilton
operator\index{Hamilton operator}} of the system and which might
also be time-dependent. It models the presence of a field acting on
the states and causing their dynamics.

As an example, the Hamilton operator for a single particle of mass
$m$ moving in a one-dimensional harmonic potential is given by\begin{equation}
H=-\frac{\Delta }{2m}+\alpha x^{2},\quad \alpha \in \mathbb{R}^{+},\label{eq:}\end{equation}
where $\Delta =\frac{\partial ^{2}}{\partial x^{2}}$ denotes the
Laplace operator. \\
One of the principles of quantum mechanics says that it is not possible
to observe the wave-function $\psi $ itself by performing an experiment
and thus to gain complete information about the system. What can be
observed is the spectrum of certain Hermitian operators called \emph{observables}\index{observable}.
These are for instance the operators $x$ (space), $-\mathrm{i}\hbar \nabla $
(momentum), $H$ (energy), $-\mathrm{i}\hbar x\times \nabla $ (angular
momentum), and others like e.g. {}``spin''.

Let $\psi _{j}$, $j\in \mathbb{N}$, be a complete set of orthonormalized
eigenvectors for the observable $A$ with eigenvalues $\lambda _{j}$
and assume the state $\psi $ at some fixed time $t_{0}$ to be given
by\begin{equation}
\psi =\sum _{j=1}^{\infty }\left\langle \psi ,\psi _{j}\right\rangle \psi _{j}.\label{eq:}\end{equation}
Then the measurement of $A$ at time $t_{0}$ will give the result
$\lambda _{j}$ with a certain probability, which simply is given
by the squared modulus $\left|\left\langle \psi ,\psi _{j}\right\rangle \right|^{2}$
of the coefficient of $\psi _{j}$ in above Fourier expansion. Thus
the expectation value of $A$ in the state $\psi (t_{0})$ is expressed
as\begin{equation}
\left\langle A\right\rangle =\sum _{j=1}^{\infty }\left|\left\langle \psi ,\psi _{j}\right\rangle \right|^{2}\lambda _{j}=\left\langle \psi ,A\psi \right\rangle .\label{eq:}\end{equation}
The process of measuring $A$ will change the state $\psi $ to $\psi =\psi _{j}$,
if the result of the observation was $\lambda _{j}$. It therefore
is not possible to perform at the state $\psi $ the exact measurement
of two or more non-commuting observables. This is the statement of
\emph{Heisenberg's uncertainty relation}\index{Heisenberg's uncertainty relation},
see \cite{key-20} for a quantitative discussion.\\
Define the \emph{time-evolution operator\index{time-evolution operator}}
$U$ to be the solution of the differential equation\begin{equation}
\mathrm{i}\hbar \dot{U}=HU,\quad U(0)=\mathbf{1}.\label{eq:}\end{equation}
This differential equation is again called Schrödinger equation. It
is easily seen that the dynamics of the state $\psi $ under the influence
of the Hamilton operator $H$ are given by\begin{equation}
\psi (t)=U(t)\psi (0)\textrm{ for all }t\geq 0.\label{eq:}\end{equation}
It is therefore sufficient to study $U$ in order to obtain a full
description of a given quantum mechanical system.

The discussion so far applies in particular to the \emph{spin\index{spin}}
of a quantum mechanical system, a phenomenon which is without analogue
in classical physics. The simpliest examples of quantum mechanical
systems containing spin are the fermions, or spin-$\frac{1}{2}$-particles,
like e.g. electrons, neutrons and protons. To carry spin in that cases
expresses the heuristic imagination that those particles possess an
angular momentum, which comes from a rotation around their own axis
and which is sensible towards a magnetic field (and only for that
reason is measurable).\\
The mathematical formulation of this phenomenon is as follows. Choose
$\mathcal{H}=\mathbb{C}^{2}$ to serve as the state space and let
$S=\left(S_{x},S_{y},S_{z}\right)$ denote the so-called operator
of \emph{total spin}\index{total spin}. Thus $S$ is a $3$-frame
of $(2\times 2)$-Hermitian operators $S_{x}$, $S_{y}$, $S_{z}$,
whose components will be specified below. The \emph{spin projection\index{spin projection}}
in direction of $e=\left(e_{x},e_{y},e_{z}\right)\in \mathbb{R}^{3}$,
$\left\Vert e\right\Vert =1$, is given by the Hermitian matrix\begin{equation}
S\cdot e:=\sum _{\alpha =x,y,z}e_{\alpha }S_{\alpha }\in \mathbb{C}^{2\times 2}.\label{eq:}\end{equation}
A measurement of $S\cdot e$ in the state $\psi \in \mathcal{H}$
has outcome $\pm \frac{1}{2}\hbar $. Let $\chi ^{\pm }$ be normalized
eigenstates for $S_{z}=S\cdot e_{z}$ with eigenvalues $\pm \frac{1}{2}\hbar $.
Then in the basis $\left\{ \chi ^{\pm }\right\} $ of $\mathcal{H}$
the matrix representation of the spin projection operators is as follows:\begin{equation}
S_{x}=\frac{\hbar }{2}\left(\begin{array}{cc}
 0 & 1\\
 1 & 0\end{array}
\right),\quad S_{y}=\frac{\hbar }{2}\left(\begin{array}{cc}
 0 & \mathrm{i}\\
 -\mathrm{i} & 0\end{array}
\right),\quad S_{z}=\frac{\hbar }{2}\left(\begin{array}{cc}
 1 & 0\\
 0 & -1\end{array}
\right).\label{eq:}\end{equation}
The dynamics of the wave function $\psi \in \mathcal{H}$ under the
influence of a magnetic field $B=\left(B_{x},B_{y},B_{z}\right)\in \mathbb{R}^{3}$
are described by the Hamiltonian operator\begin{equation}
H_{\mathrm{magn}}=\textrm{const}\, B\cdot S=\textrm{const}\left(B_{x}S_{x}+B_{y}S_{y}+B_{z}S_{z}\right)\in \mathbb{C}^{2\times 2}.\label{eq:}\end{equation}
The generalization to ensembles of $n$ spin-$\frac{1}{2}$-particles
is as follows. The \emph{total spin} in this situation is given by
the operator $S=\left(S_{x},S_{y},S_{z}\right)$, \begin{equation}
S_{\alpha }:=\bigotimes _{i=1}^{n}S_{i,\alpha }\in \mathbb{C}^{2^{n}\times 2^{n}},\quad \alpha \in \left\{ x,y,z\right\} ,\label{eq:}\end{equation}
with $S_{i,\alpha }$ the spin projection of the $i$-th particle
in direction of $e_{\alpha }$, and $\otimes $ denoting the Kronecker
product. The projection of $S$ in direction of $e=\left(e_{x},e_{y},e_{z}\right)$,
$\left\Vert e\right\Vert =1$, is the operator \begin{equation}
S\cdot e:=\bigotimes _{i=1}^{n}\left(S_{i}\cdot e\right)\in \mathbb{C}^{2^{n}\times 2^{n}}\label{eq:}\end{equation}
which is acting on the Hilbert space $\mathcal{H}=\left(\mathbb{C}^{2}\right)^{\otimes n}$.
Its eigenvalues are $-\frac{n\hbar }{2},-\frac{n\hbar }{2}+1,...,\frac{n\hbar }{2}-1,\frac{n\hbar }{2}$.
The eigenspace $\mathcal{E}_{m}$ to the eigenvalue $\frac{\hbar }{2}m$,
$m=-n,-n+2,...,n-2,n$, has dimension\begin{equation}
\dim \mathcal{E}_{m}=\left(\begin{array}{c}
 n\\
 \frac{n+m}{2}\end{array}
\right).\label{eq:}\end{equation}
If the $n$-particle system is exposed to a magnetic field $B=\left(B_{x},B_{y},B_{z}\right)$
which we assume to be equal to $B_{i}=\left(B_{ix},B_{iy},B_{iz}\right)$
at the locus of the $i$-th particle, then its spin will be described
by the Hamilton operator\begin{equation}
H=H_{d}+H_{\mathrm{magn}}\in \mathbb{C}^{2^{n}\times 2^{n}},\label{eq:}\end{equation}
with \begin{equation}
H_{\mathrm{magn}}=\textrm{const}\bigotimes _{i=1}^{n}B_{i}\cdot S_{i},\label{eq:}\end{equation}
the factor $B_{i}\cdot S_{i}$ as given by equation (2.1.11), and
an operator $H_{d}$, which is fixed and describes the coupling between
the spins of the individual particles. This Hamilton operator $H$
will typically be occuring in the discussion of the control properties
of Schrödinger's equation\begin{equation}
\mathrm{i}\hbar \dot{U}=HU,\label{eq:}\end{equation}
which is the content of the following sections.

\section{The Control Problem}

Given the Lie group $G=SU(2^{n})$ and the following family of Hermitian
operators $\left(\mathbb{C}^{2}\right)^{\otimes n}\rightarrow \left(\mathbb{C}^{2}\right)^{\otimes n}$:\begin{equation}
H(v_{1},...,v_{m})=H_{d}+\sum _{j=1}^{m}v_{j}H_{j},\quad v=(v_{1},...,v_{m})\in \mathbb{R}^{m}.\label{eq:}\end{equation}
Fix an element $U_{F}\in G$ and consider the right-invariant control
system on $G$ given by\begin{equation}
\dot{U}=-\mathrm{i}H(v)U,\quad U(0)=\mathbf{1}\label{eq:}\end{equation}
with $v$ acting as control variable. The question of interest to
us is whether it is possible to steer system (2.2.2) from the initial
state $U(0)$ to the final state $U_{F}$. If this is the case, what
will be the minimum amount of time to achieve this?

The motivation for treating that kind of problem in time-optimal control
arises from questions concerning the quantum mechanics of spin systems,
such as ensembles of electrons or neutrons. Indeed, the operator $H(v)$
acts as the Hamilton operator for systems of coupled spin particles
that are under the influence of an exterior magnetic field of fixed
direction and variable strength (modelled by the variable $v$). So
(2.2.2) is just Schrödinger's equation for the time-evolution operator
$U$ of such a system (with $\hbar $ set equal to $1$).\\
The desire to solve a control problem as formulated above came alongside
with the development of certain experiments in nuclear magnetic resonance
spectroscopy (NMR) and quantum computing. Here one needs to manipulate
ensembles of coupled nuclear spins and wishes to do so in least possible
time. See e.g. \cite{key-31} and \cite{key-28} for details on this
topic.

In the discussion to follow we are going to generalize this kind of
control problem from the specific case $G=SU(2^{n})$ to arbitrary
compact Lie groups.

\section{Equivalence Theorem}

Throughout this section $G$ denotes a compact Lie group with Lie
algebra $\mathfrak{g}$, while $K$ denotes a closed subgroup of $G$
with Lie algebra $\mathfrak{k}$. We are interested in the following
affine right-invariant control system on $G$:\begin{equation}
\dot{U}=\left(H_{d}+\sum _{j=1}^{m}v_{j}H_{j}\right)U,\quad U(0)=\mathbf{1},\label{eq:}\end{equation}
with $H_{d}\in \mathfrak{g}$ arbitrary but fixed, and $H_{1},...,H_{m}$
a fixed set of generators for the Lie algebra $\mathfrak{k}$.

\subsection{Heuristic considerations.}

Consider again the evolution equation of the affine right-invariant
control system (2.3.1), and let $U(0)=U_{0}\in G$ arbi-\\
trary. Note that in principle the control variables $v_{j}$, $j=1,...,m$,
may be chosen to be arbitrarily large in comparison to the norm of
the fixed drift Hamiltonian $H_{d}$. For such a choice of $v=(v_{1},...,v_{m})$
the control system (2.3.1) will behave roughly as\begin{equation}
\dot{U}=\left(\sum _{j=1}^{m}v_{j}H_{j}\right)U,\quad U(0)=U_{0}.\label{eq:}\end{equation}
 Note also that, if we choose $v$ to be constant, the solution of
ODE (2.3.2) will be \begin{equation}
U(t)=\exp \left(t\sum _{j=1}^{m}v_{j}H_{j}\right)U_{0}.\label{eq:}\end{equation}
>From our assumptions on $H_{j}$, $j=1,...,m,$ and Theorem 1.4.5
it follows that it is possible to steer system (2.3.2) to any point
$U$ in the coset $KU_{0}\subseteq G$, and, by choosing $\left|v\right|$
large, to achieve this in negligible time. Thus from the point of
view of time-optimal control, group elements contained in the same
coset $KU_{0}\subseteq G$ can be considered equivalent. The question
of interest therefore is to find appropriate control strategies to
steer system (2.3.1) from the identity coset $K$ to any other coset
$KU_{0}$ in least possible time. To this aim it turns out to be useful
to replace system (2.3.1) by another right-invariant system on the
group $G$, whose reachable sets coincide (modulo $K$) with that
of (2.3.1), but which has bounded controls so that the phenomenon
of arbitrary fast movement within a coset does no longer occur.\\
Which kind of right-invariant control system on $G$ will be the appropriate
one? To answer this question assume that (2.3.1) evolves from $U_{0}\in G$
under the influence of the control variable $v$, which we still assume
to be constant. So we set\begin{equation}
H_{0}:=\sum _{j=1}^{m}v_{j}H_{j},\label{eq:}\end{equation}
and consider on $G$ the ODE\begin{equation}
\dot{U}=\left(H_{d}+H_{0}\right)U,\quad U(0)=U_{0}.\label{eq:}\end{equation}
The solution of this ODE is given by\begin{equation}
U(t)=\exp \left(t\left(H_{d}+H_{0}\right)\right)U_{0}.\label{eq:}\end{equation}
We separate the flow $t\mapsto U(t)$ into two components $Q(t)$
and $P(t)$, i.e. $U(t)=Q(t)P(t)$. As we have noticed before, it
is possible to steer the original system (2.3.1) within a given coset
$Kg$ arbitrarily fast. For this reason, the factor $Q(t)$, which
merely describes motion within $K$, will be factored out. The allowed
directions for steering the new system on $G$ are therefore given
by the values of $\dot{P}(t)$. These are now obtained from the ansatz
$U(t)=Q(t)P(t)$ with $U(t)$ as in equation (2.3.6) and $Q(t)=\exp tH_{0}\in K$.
A calculation yields\begin{eqnarray*}
\dot{P}(t) & = & \dot{\widehat{Q^{-1}}}(t)U(t)+Q^{-1}(t)\dot{U}(t)\\
 & = & \exp \left(-tH_{0}\right)\left(-H_{0}\right)U\left(t\right)+\exp \left(-tH_{0}\right)\left(H_{d}+H_{0}\right)U\left(t\right)\\
 & = & \exp \left(-tH_{0}\right)H_{d}U\left(t\right)\\
 & = & \exp \left(-tH_{0}\right)H_{d}\exp \left(tH_{0}\right)\exp \left(-tH_{0}\right)U\left(t\right)\\
 & = & \left(\mathrm{Ad}_{\exp \left(-tH_{0}\right)}H_{d}\right)Q^{-1}(t)U(t)\\
 & = & \left(\mathrm{Ad}_{\exp \left(-tH_{0}\right)}H_{d}\right)P\left(t\right).
\end{eqnarray*}
This gives rise to the idea of replacing the original system (2.3.1)
by the following control system on $G$:\begin{equation}
\dot{P}=XP,\quad P(0)=\mathbf{1},\label{eq:}\end{equation}
where the control $X$ is taken from \begin{equation}
\textrm{Ad}_{K}H_{d}=\left\{ \left.kH_{d}k^{-1}\right|k\in K\right\} ,\label{eq:}\end{equation}
the $K$-adjoint orbit of $H_{d}$ in $\mathfrak{g}$. \\
In view of our initial considerations it would be desirable to interpret
system (2.3.7) as a control system on the space of right-cosets modulo
$K$. However, it turns out that the expression $XP$ can only be
given a precise meaning as a tangent vector of $G/K$, if $G/K$ is
taken to be the space of left-cosets, i.e. $G/K=\left\{ \left.gK\right|g\in G\right\} $.
Then $XP$ just means right-translation of the vector $X\in \mathfrak{g}$
by $P\in G/K$. On the other hand, the reachable sets $R(\mathbf{1},t)$
for both system (2.3.1) and system (2.3.7) are easily shown to be
$\textrm{Ad}_{K}$-invariant, so that the identity $KR(\mathbf{1},t)=R(\mathbf{1},t)K$
holds for all $t\geq 0$. This makes it plausible that (2.3.7) can
be used to define on the left-homogeneous space $G/K$ a control system,
which is equivalent to system (2.3.1) on the group $G$.\\
This idea will be given evidence in the subsequent section. \[
\]

\subsection{Equivalence Theorem}

\begin{defn}
The control system (2.3.1) will from now on be refered to as the \emph{unreduced
system}\index{unreduced system}. We furthermore define on $G$ the
\emph{adjoint system\index{adjoint system}} to be\begin{equation}
\dot{U}=XU,\quad U(0)=\mathbf{1}\quad X\in \mathrm{Ad}_{K}H_{d},\label{eq:}\end{equation}
and on $G/K$ the \emph{reduced system\index{reduced system}} to
be\begin{equation}
\dot{P}=XP,\quad P(0)=K,\quad X\in \mathrm{Ad}_{K}H_{d},\label{eq:}\end{equation}
where the expression $XP$ is explained as follows.\\
If $P=\pi (g)$, then\begin{equation}
XP:=D_{\mathbf{1}}(\pi \circ R_{g})(X)\in T_{gK}(G/K).\label{eq:}\end{equation}
This is well-defined: If we replace $g$ by $g'=gk$, $k\in K$, then
we find that\begin{eqnarray*}
D_{\mathbf{1}}(\pi \circ R_{g'})(X) & = & D_{\mathbf{1}}(\pi \circ R_{gk})(X)\\
 & = & D_{\mathbf{1}}(\pi \circ R_{k}\circ R_{g})(X)\\
 & = & D_{\mathbf{1}}(\pi \circ R_{g})(X)\\
 & = & XP,
\end{eqnarray*}
as $\pi \circ R_{k}=\pi $. 
\end{defn}
\begin{notation}
We label reachable and approximately reachable sets etc. for the unreduced,
adjoint and reduced systems by lower indices $1$,$2$ and $3$, respectively.
\end{notation}
\begin{note}
For convenience we add to the admissible vector fields of systems
$1-3$ the zero field. This does not change the reachable sets $\mathbf{R}_{i}(x,t)$
and $\mathbf{R}_{i}(x)$, $i=1,2,3$, nor does it have any effect
on the problem of finding time-optimal trajectories for those systems.
This assumption merely has the advantage that in the remainder we
need not distinguish between the sets $\mathbf{R}_{i}(x,t)$ and $R_{i}(x,t)$
and also might use the fact that the sets $R_{i}(x,t)$ are monotonely
increasing in $t$.
\end{note}
The remainder of this section is aimed to establish a theorem which
will show that all three of the systems defined above can be considered
equivalent. To be able to give a precise formulation of what {}``equivalence''
should be, we introduce some terminology.

\begin{defn}
Let $\Sigma =(M,f_{u},U)$ be a control system. Define the \emph{set}
$S(x,t_{0})$ \emph{of approximately reachable points\index{approximately reachable }}
from $x\in M$ within time $t_{0}\geq 0$ to be \begin{equation}
S(x,t_{0}):=\bigcap _{t>t_{0}}\overline{\mathbf{R}(x,t)}.\label{eq:}\end{equation}
Here $\mathbf{R}(x,t)$ refers to the reachable set from $x$ within
time $t$ as defined in 1.4.3. Thus a point $y\in M$ is contained
in $S(x,t_{0})$, if and only if for any neighbourhood $U$ of $y$
and any $\varepsilon >0$ there exists a point $z\in U\cap \mathbf{R}(x,t+\varepsilon )$.
Furthermore, we define the \emph{infimizing time\index{infimizing time}}
to steer $\Sigma $ from $x_{1}\in M$ to $x_{2}\in M$ to be\begin{equation}
t_{\textrm{inf}}\left(x_{1},x_{2}\right):=\left\{ \begin{array}{ll}
 \inf \left\{ \left.t\in \mathbb{R}\right|x_{2}\in S\left(x_{1},t\right)\right\} , & \textrm{if }x_{2}\in S\left(x_{1},t\right)\textrm{ for some }t\in \mathbb{R},\\
 \infty , & \textrm{otherwise}.\end{array}
\right.\label{eq:}\end{equation}

\end{defn}
\begin{rem}
As a consequence of the boundedness of the set $\mathrm{Ad}_{K}H_{d}$
it easily follows that \begin{equation}
S_{j}(\mathbf{1},t)=\overline{R_{j}(\mathbf{1},t)}\label{eq:}\end{equation}
holds for $j=2,3$ and for all $t\geq 0$, cf. Proposition 2.3.9.
On the other hand, the distinction between the closure of reachable
sets and approximately reachable sets in the case of system $1$ became
inevitable since here the set of controls is unbounded.
\end{rem}
The equivalence between the control systems of Definition 2.3.1 can
now be stated as follows.

\begin{thm}
\emph{(Equivalence theorem\index{equivalence theorem}).} For all
$t\geq 0$ the following holds:

(i) $S_{1}(\mathbf{1},t)=K\overline{R_{2}(\mathbf{1},t)}=\overline{R_{2}(\mathbf{1},t)}K$,

(ii) $\pi \left(S_{1}(\mathbf{1},t)\right)=\overline{R_{3}(K,t)}$,
where $\pi $ denotes canonical projection $G\rightarrow G/K$.
\end{thm}
The proof of the equivalence theorem is based on the subsequent propositions.

\begin{prop}
(i) The reachable sets for the adjoint system $2$ are $\mathrm{Ad}_{K}$-invariant,
i.e. the identities\begin{equation}
\mathrm{Ad}_{K}\left(R_{2}(\mathbf{1},t)\right)=R_{2}(\mathbf{1},t)\label{eq:}\end{equation}
and\begin{equation}
\mathrm{Ad}_{K}\left(\overline{R_{2}(\mathbf{1},t)}\right)=\overline{R_{2}(\mathbf{1},t)}\label{eq:}\end{equation}
hold for all $k\in K$ and $t\geq 0$.

(ii) Any trajectory $t\mapsto U(t)$, $t\in \left[0,t_{F}\right]$,
of system $2$ is mapped under $\pi $ to a trajectory $t\mapsto V(t):=\left(\pi \circ U\right)(t)$
of system $3$. Conversely, any trajectory $t\mapsto V(t)$ of system
$3$ can be lifted to a trajectory $t\mapsto U(t)$ of system $2$.
In particular,\begin{equation}
\pi \left(R_{2}(\mathbf{1},t)\right)=R_{3}(\mathbf{1},t)\label{eq:}\end{equation}
holds for all $t\geq 0$.
\end{prop}
\begin{proof}
(i) Let the control $t\mapsto X(t)$ of system $2$ generate the trajectory
$t\mapsto U(t)$. Then the control $t\mapsto \textrm{Ad}_{k}X(t)$,
$k\in K$, generates the trajectory $t\mapsto \textrm{Ad}_{k}U(t)$,
because\begin{eqnarray*}
\frac{d}{dt}\left.\textrm{Ad}_{k}U(t)\right|_{t=t_{0}} & = & k\dot{U}(t_{0})k^{-1}\\
 & = & kX(t_{0})U(t_{0})k^{-1}\\
 & = & \left(kX(t_{0})k^{-1}\right)\left(kU(t_{0})k^{-1}\right)\\
 & = & \textrm{Ad}_{k}X(t_{0})\cdot \textrm{Ad}_{k}U(t_{0}).
\end{eqnarray*}
This implies that the set $R_{2}(\mathbf{1},t)$ is $\textrm{Ad}_{K}$-invariant
for any $t\geq 0$. Since the map $\textrm{Ad}_{k}$ is a homeomorphism,
the same holds for $\overline{R_{2}(\mathbf{1},t)}$.

(ii) Let $X:\left[0,t_{F}\right]\rightarrow \mathrm{Ad}_{K}H_{d}$,
$t\mapsto X(t)$ be any control for the adjoint and for the reduced
system. Denote by $t\mapsto U_{2}(t)\in G$ and by $t\mapsto U_{3}(t)\in G/K$
the resulting trajectories. Then\begin{eqnarray*}
\frac{d}{dt}\left.\left(\pi \circ U_{2}(t)\right)\right|_{t=t_{0}} & = & D_{U_{2}(t_{0})}\pi \left(\dot{U}_{2}(t_{0})\right)\\
 & = & D_{U_{2}(t_{0})}\pi \left(X(t_{0})U_{2}(t_{0})\right)\\
 & = & D_{U_{2}(t_{0})}\pi \left(D_{\mathbf{1}}R_{U_{2}(t_{0})}\left(X(t_{0})\right)\right)\\
 & = & D_{\mathbf{1}}\left(\pi \circ R_{U_{2}(t_{0})}\right)\left(X(t_{0})\right)\\
 & = & X(t_{0})\left(\pi \circ U_{2}(t_{0})\right).
\end{eqnarray*}
This shows that both $U_{3}$ and $\pi \circ U_{2}$ satisfy ODE (2.3.10)
on $G/K$ together with the initial condition $P(0)=K$. Therefore
$U_{3}=\pi \circ U_{2}$ holds everywhere on $\left[0,t_{F}\right]$. 
\end{proof}
\begin{prop}
For all $t_{F}\geq 0$, \begin{equation}
R_{1}(\mathbf{1},t_{F})\subseteq KR_{2}(\mathbf{1},t_{F}).\label{eq:}\end{equation}

\end{prop}
\begin{proof}
Let $U_{F}\in R_{1}(\mathbf{1},t_{F})$ and $t\mapsto v(t)\in \mathbb{R}^{m}$
a control for system $1$ such that the corresponding trajectory $t\mapsto U(t)$
satisfies $U(t_{F})=U_{F}$. Now let $t\mapsto Q(t)\in K$ the solution
curve of the ODE\[
\dot{Q}=\left(\sum _{i=1}^{m}v_{i}H_{i}\right)Q,\quad Q(0)=\mathbf{1},\]
and $t\mapsto P(t)\in G$ the solution curve of the ODE\[
\dot{P}=\left(Q^{-1}H_{d}Q\right)P,\quad P(0)=\mathbf{1}.\]
On $[0,t_{F}]$ consider the map $t\mapsto V(t):=Q(t)P(t)$. It satisfies
$V(0)=\mathbf{1}$ and\begin{eqnarray*}
\dot{V}(t) & = & \dot{Q}(t)P(t)+Q(t)\dot{P}(t)\\
 & = & \left(\sum _{i=1}^{m}v_{i}(t)H_{i}\right)Q(t)P(t)+Q(t)\left(Q^{-1}(t)H_{d}Q(t)\right)P(t)\\
 & = & \left(\sum _{i=1}^{m}v_{i}(t)H_{i}+H_{d}\right)Q(t)P(t),
\end{eqnarray*}
which shows, by the uniqueness part of the Caratheodory theorem, that
$V$ coincides with $U$ on $\left[0,t_{F}\right]$. Now let system
$2$ evolve according to the control law $t\mapsto Q^{-1}(t)H_{d}Q(t)$
for $t\in \left[0,t_{F}\right]$. Then \[
P(t_{F})=Q^{-1}(t_{F})V(t_{F})=Q^{-1}(t_{F})U(t_{F})=Q^{-1}(t_{F})U_{F}\in KU_{F}.\]
This shows $U_{F}\in KP(t_{F})\subseteq KR_{2}(\mathbf{1},t_{F})$,
as claimed.
\end{proof}
\begin{prop}
The set $\overline{R_{2}(x,t)}$ is compact for all $x\in G$ and
$t\geq 0$. Moreover, the following holds:\begin{equation}
\overline{R_{2}(x,t)}=\bigcap _{n=1}^{\infty }\overline{R_{2}\left(x,t+\frac{1}{n}\right)}.\label{eq:}\end{equation}

\end{prop}
\begin{proof}
Let $\left\langle \cdot ,\cdot \right\rangle $ be any right-invariant
metric on $G$. The set $R_{2}(x,t)$ is bounded, because the set
$\mathrm{Ad}_{K}H_{d}\subseteq \mathfrak{g}$ of controls is bounded
by a constant $M$ (in the norm induced by the scalar product $\left\langle \cdot ,\cdot \right\rangle _{\mathbf{1}}$
on $\mathfrak{g}=T_{\mathbf{1}}G$) . So, by the right-invariance
of the metric $\left\langle \cdot ,\cdot \right\rangle $, we have
that $\left\Vert f_{u}(z)\right\Vert =\left\Vert u\right\Vert \leq M$
for all $u\in \mathrm{Ad}_{K}H_{d}$ and $z\in G$. Hence the distance
$d(x,y)$ between $x$ and any $y\in R_{2}(x,t)$ can be estimated
as follows:\[
d(x,y)\leq \int _{0}^{t}\left\Vert u(s)\right\Vert \, ds\leq \int _{0}^{t}M\, ds=Mt.\]
Therefore, $R_{2}(x,t)$ is bounded, and $\overline{R_{2}(x,t)}$
is compact.\\
To prove equation (2.3.19), we first observe that $\overline{R_{2}(x,t)}$
is contained in any of the sets $\overline{R_{2}\left(x,t+\frac{1}{n}\right)}$,
$n\in \mathbb{N}$, thanks to the convention made in 2.3.3. So $\overline{R_{2}(x,t)}\subseteq \bigcap _{n=1}^{\infty }\overline{R_{2}\left(x,t+\frac{1}{n}\right)}$.
Now assume that there exists \[
y\in \left(\bigcap _{n=1}^{\infty }\overline{R_{2}\left(x,t+\frac{1}{n}\right)}\right)\setminus \overline{R_{2}(x,t)}.\]
Then $y$ has distance $d>0$ from the compact set $\overline{R_{2}(x,t)}$.
>From this and the boundedness of the controls it follows that the
infimizing time needed to steer system 2 from $\overline{R_{2}(x,t)}$
to $y$ is positive, i.e.\[
\inf \left\{ \left.\varepsilon >0\right|y\in \overline{R_{2}\left(x,t+\varepsilon \right)}\right\} >0,\]
in contradiction to $y\in \bigcap _{n=1}^{\infty }\overline{R_{2}\left(x,t+\frac{1}{n}\right)}$.
\end{proof}
\begin{prop}
For all $t\geq 0$ the following holds:\begin{equation}
R_{2}(\mathbf{1},t)\subseteq K\overline{R_{1}(\mathbf{1},t)}.\label{eq:}\end{equation}

\end{prop}
\begin{proof}
Let $x_{0}\in R_{2}(\mathbf{1},t)$. By definition there exists a
control $X:\left[0,t_{F}\right]\rightarrow \mathrm{Ad}_{K}H_{d}$
such that the resulting trajectory $t\mapsto P(t)$ of system $2$
satisfies $P\left(t_{F}\right)=x_{0}$. Theorem 1.2.2 allows us to
identify the smooth manifold $\mathrm{Ad}_{K}H_{d}$ with the homogeneous
space $K/\textrm{Stab}_{\textrm{K}}H_{d}$. From this identification
it becomes clear that the path $t\mapsto X(t)\in K/\textrm{Stab}_{\textrm{K}}H_{d}$
can be lifted to a path $t\mapsto Q^{-1}(t)\in K$ with $Q(0)=\mathbf{1}$
and the same regularity properties as $t\mapsto X(t)$. We therefore
have \begin{equation}
X(t)=Q^{-1}(t)H_{d}Q(t)\label{eq:}\end{equation}
on $\left[0,t_{F}\right]$. Notice that the path $t\mapsto Q(t)$
need not occur as a trajectory of the control system\begin{equation}
\dot{Q}=\left(\sum _{i=1}^{m}v_{i}H_{i}\right)Q,\quad Q(0)=\mathbf{1},\quad v=(v_{1},...,v_{m})\in \mathbb{R}^{m}\label{eq:}\end{equation}
on $K$. But as it is pointed out in \cite{key-18} and proved in
\cite{key-5}, there exists a sequence of control functions $t\mapsto v^{n}(t)\in \mathbb{R}^{m}$,
$t\in \left[0,t_{F}\right]$, such that the resulting sequence $\left(Q^{n}\right)_{n}$
of trajectories for system (2.3.22) converges in $L_{1}$ against
$Q$. Now define for $t\in \left[0,t_{F}\right]$\[
X^{n}(t):=\left(Q^{n}\right)^{-1}(t)H_{d}Q^{n}(t)\in \mathrm{Ad}_{K}H_{d}.\]
Then, by equation (2.3.21) and the definition of $Q^{n}$, the sequence
$\left(X^{n}\right)_{n}$ converges in $L_{1}$ against $X$. Furthermore,
let $P^{n}$ the solution of the ODE\[
\dot{P}^{n}=X^{n}P^{n},\quad P^{n}(0)=\mathbf{1}.\]
The convergence of $\left(X^{n}\right)_{n}$ against $X$ implies
\[
\lim _{n\rightarrow \infty }P^{n}\left(t_{F}\right)=P\left(t_{F}\right),\]
cf. \cite{key-24}, p. 41-42. We finally set $U^{n}:=Q^{n}P^{n}$.
Since the function $Q^{n}$ satisfies the ODE (2.3.22), it follows
from the same calculation as in the proof of Proposition 2.3.8 that
$U^{n}$ solves the ODE\[
\dot{U}^{n}=\left(H_{d}+\sum _{i=1}^{m}v_{i}^{n}H_{i}\right)U^{n},\quad U^{n}(0)=\mathbf{1}\]
on $\left[0,t_{F}\right]$ and therefore is a trajectory of system
$1$. So we have found that\begin{eqnarray*}
x_{0} & = & P\left(t_{F}\right)\\
 & = & \lim _{n\rightarrow \infty }P^{n}\left(t_{F}\right)\\
 & = & \lim _{n\rightarrow \infty }\left(Q^{n}\right)^{-1}\left(t_{F}\right)U^{n}\left(t_{F}\right)\\
 & = & Q^{-1}\left(t_{F}\right)\lim _{n\rightarrow \infty }U^{n}\left(t_{F}\right)\\
 & \in  & K\overline{R_{1}(\mathbf{1},t_{F}}),
\end{eqnarray*}
as claimed.
\end{proof}
\begin{prop}
$K\subseteq S_{1}(\mathbf{1},0).$ As a consequence, if $z\in S_{1}(\mathbf{1},t)$
for some $t\geq 0$, then $zk\in S_{1}(\mathbf{1},t)$ holds for all
$k\in K$.
\end{prop}
\begin{proof}
By assumption, the elements $H_{1},...,H_{m}\in \mathfrak{k}$ generate
$\mathfrak{k}$ as a Lie algebra. This implies that the system\begin{equation}
\dot{W}=\left(\sum _{i=1}^{m}v_{i}H_{i}\right)W,\quad W(0)=\mathbf{1}\label{eq:}\end{equation}
is controllable as a system on $K$, cf. Theorem 1.4.5. Moreover,
by the same theorem, there exists a constant $T>0$ such that $\mathbf{R}(\mathbf{1})=\mathbf{R}(\mathbf{1},T)$.
Since the norm of the operator $\sum _{i=1}^{m}v_{i}H_{i}$ may be
chosen to be arbitrarily large, this equation holds for any constant
$T>0$. \\
>From now on let $T>0$ and $W_{F}\in K$ be arbitrary but fixed, and
choose a control $t\mapsto \left(v_{1}(t),...,v_{m}(t)\right)$, $t\in \left[0,t_{F}\right]$,
$t_{F}<T$, such that the resulting trajectory $t\mapsto W(t)$ satisfies
$W(t_{F})=W_{F}$. Then for all $n\in \mathbb{N}$ the trajectory
$t\mapsto W^{n}(t)$ of (2.3.23), which results from the control $t\mapsto v^{n}(t):=nv(t)$,
satisfies $W^{n}\left(\frac{1}{n}t_{F}\right)=W_{F}$. Now consider
the ODE\[
\dot{U}^{n}=\left(H_{d}+n\sum _{i=1}^{m}v_{i}(t)H_{i}\right)U^{n},\quad U^{n}(0)=\mathbf{1},\]
and let $\varepsilon >0$ be arbitrary. Then for $n$ sufficiently
large it follows that\[
\left\Vert U^{n}\left(\frac{1}{n}t_{F}\right)-W_{F}\right\Vert =\left\Vert U^{n}\left(\frac{1}{n}t_{F}\right)-W^{n}\left(\frac{1}{n}t_{F}\right)\right\Vert <\varepsilon ,\]
cf. \cite{key-38}, p. 122. Thus \[
W_{F}\in \overline{R_{1}(\mathbf{1},t_{F})}\subseteq \overline{R_{1}(\mathbf{1},T)}\]
holds for all $T>0$. By definition, this implies $W_{F}\in S_{1}(\mathbf{1},0)$.
Since $W_{F}\in K$ was chosen to be arbitrarily, it follows that
$K\subseteq S_{1}(\mathbf{1},0)$. \\
The addendum that $zk\in S_{1}(\mathbf{1},t)$, if $z\in S_{1}(\mathbf{1},t)$
and $k\in K$, is an immediate consequence of the right-invariance
of system $1$. Namely, if we can steer system $1$ into any neighbourhood
$Z_{\varepsilon }$ of $z$ at time $t+\varepsilon $ and into any
neighbourhood $K_{\varepsilon }$ of $k$ at time $\varepsilon $,
than the system can likewise be steered at time $t+2\varepsilon $
into an arbitrary small neighbourhood $Z_{\varepsilon }K_{\varepsilon }$
of $zk$.
\end{proof}
We now turn to the proof of the Equivalence Theorem 2.3.4.

\begin{proof}
(i) From Proposition 2.3.7 (i) we obtain\[
\bigcup _{k\in K}k\overline{R_{2}(\mathbf{1},t)}=\bigcup _{k\in K}k\overline{R_{2}(\mathbf{1},t)}k^{-1}k=\bigcup _{k\in K}\overline{R_{2}(\mathbf{1},t)}k,\]
so that the identity\[
K\overline{R_{2}(\mathbf{1},t)}=\overline{R_{2}(\mathbf{1},t)}K\]
holds for all $t\geq 0$.\\
(A) $S_{1}(\mathbf{1},t)\subseteq K\overline{R_{2}(\mathbf{1},t)}$.\\
Let $x\in S_{1}(\mathbf{1},t)$. By definition of $S_{1}(\mathbf{1},t)$
there exists a sequence $\left(y_{n}\right)_{n}$ in $G$ with $y_{n}\in R_{1}\left(\mathbf{1},t+\frac{1}{n}\right)\cap U_{\frac{1}{n}}(x)$
and $\lim _{n\rightarrow \infty }y_{n}=x$. Here $U_{\varepsilon }(x)$
denotes the open ball around $x$ of radius $\varepsilon $. For all
$n\in \mathbb{N}$ there exists, by Proposition 2.3.8, $z_{n}\in R_{2}\left(\mathbf{1},t+\frac{1}{n}\right)$
and $k_{n}\in K$ such that $y_{n}=k_{n}z_{n}$. Since all $z_{n}$
are contained in the compact set $\overline{R_{2}(\mathbf{1},t+1)}$
and $K$ also is a compact set, we find a subsequence of $\left(z_{n},k_{n}\right)_{n}$
(which we label again by $n$) and which satisfies\[
\lim _{n\rightarrow \infty }\left(z_{n},k_{n}\right)=\left(z_{0},k_{0}\right)\in \overline{R_{2}(\mathbf{1},t+1)}\times K.\]
By definition of $z_{n}$ we have that $z_{0}\in \overline{\mathbf{R}_{2}(\mathbf{1},t+\varepsilon )}$
for all $\varepsilon >0$. This implies together with Proposition
2.3.9 that\[
z_{0}\in \bigcap _{n=1}^{\infty }\overline{R_{2}\left(\mathbf{1},t+\frac{1}{n}\right)}=\overline{R_{2}\left(\mathbf{1},t\right)}.\]
It then follows that\begin{eqnarray*}
x & = & \lim _{n\rightarrow \infty }y_{n}\\
 & = & \lim _{n\rightarrow \infty }\left(k_{n}z_{n}\right)\\
 & = & \lim _{n\rightarrow \infty }k_{n}\lim _{n\rightarrow \infty }z_{n}\\
 & = & k_{0}z_{0}\\
 & \in  & K\overline{R_{2}(\mathbf{1},t)}.
\end{eqnarray*}
(B) $K\overline{R_{2}(\mathbf{1},t)}\subseteq S_{1}(\mathbf{1},t)$.\\
Let $k\in K$, $x\in R_{2}(\mathbf{1},t)$. Then by Proposition 2.3.10,
$x=k'y$ for some $k'\in K$ and $y\in \overline{R_{1}(\mathbf{1},t)}$.
It follows that\[
kx\in K\overline{R_{1}(\mathbf{1},t)}\subseteq KS_{1}(\mathbf{1},t).\]
Now by Proposition 2.3.11, $K\subseteq S_{1}(\mathbf{1},0)$, hence
$KS_{1}(\mathbf{1},t)=S_{1}(\mathbf{1},t)$. This shows $kx\in S_{1}(\mathbf{1},t)$
and implies $KR_{2}(\mathbf{1},t)\subseteq S_{1}(\mathbf{1},t)$.
Furthermore, $K$ and also $S_{1}(\mathbf{1},t)$ is closed, so that
\[
\overline{KR_{2}(\mathbf{1},t)}=K\overline{R_{2}(\mathbf{1},t)}\subseteq S_{1}(\mathbf{1},t).\]
 (ii) From Proposition 2.3.7 (ii) and the fact that the images of
compact sets under the continuous map $\pi $ are compact, it follows
that \[
\overline{R_{3}(K,t_{F})}=\overline{\pi \left(R_{2}(\mathbf{1},t_{F})\right)}=\pi \left(\overline{R_{2}(\mathbf{1},t_{F})}\right).\]
 Combining this with part (i) of the proof we find that\[
\overline{R_{3}(K,t_{F})}=\pi \left(\overline{R_{2}(\mathbf{1},t_{F})}\right)=\pi \left(\overline{R_{2}(\mathbf{1},t_{F})}K\right)=\pi \left(S_{1}(\mathbf{1},t_{F})\right).\]
This concludes the proof of the equivalence theorem.
\end{proof}
\begin{cor}
For all $x\in G$,\begin{equation}
t_{\mathrm{inf,}1}\left(\mathbf{1},x\right)=t_{\mathrm{inf,}3}\left(K,\pi (x)\right).\label{eq:}\end{equation}

\end{cor}
\begin{proof}
Statement (ii) of the equivalence theorem implies that for all $p\in G/K$
there exists $y\in \pi ^{-1}(p)$ such that $t_{\textrm{inf,}3}\left(K,p\right)=t_{\textrm{inf,1}}\left(\mathbf{1},y\right)$.
Now for any $x=yk\in \pi ^{-1}(p)$ we have by Proposition 2.3.11
that $t_{\textrm{inf,1}}\left(\mathbf{1},x\right)=t_{\textrm{inf,1}}\left(\mathbf{1},y\right)$.
So $t_{\textrm{inf,1}}{}\left(\mathbf{1},x\right)=t_{\textrm{inf,}3}{}\left(K,\pi (x)\right)$
holds for all $x\in G$, as claimed.
\end{proof}
\begin{cor}
Assume the set\begin{equation}
W:=\bigcup _{\lambda \in \left[0,1\right]}\lambda \mathrm{Ad}_{K}H_{d}\subseteq \mathfrak{g}\label{eq:}\end{equation}
to be convex. Then the equivalence theorem can be restated as follows:

(i) $S_{1}(\mathbf{1},t)=KR_{2}(\mathbf{1},t)=R_{2}(\mathbf{1},t)K$,

(ii) $\pi \left(S_{1}(\mathbf{1},t)\right)=R_{3}(K,t)$.
\end{cor}
\begin{proof}
Replacing the set $\mathrm{Ad}_{K}H_{d}$ of control parameters for
system $2$ by the set $W$ will not change the reachable sets $R_{2}(\mathbf{1},t)$
since the trajectories that can occur then are just reparametrisations
of the trajectories one already had before for system $2$. Also there
will be no effect on the infimizing, respectively minimizing times
because the tangents $\dot{P}(t)$ of the now occuring trajectories
are of equal or smaller length than before, since $\lambda \in \left[0,1\right]$.
The same holds for the reduced system $2$. Because the set $W$ is
compact and due to our assumption convex, we may apply Filippov's
Theorem (cf. \cite{key-24}, Theorem 10.3) to obtain the compactness
of the sets $R_{2}(\mathbf{1},t)$ and $R_{3}(K,t)$. Thus in the
statement of the equivalence theorem, the expression $\overline{R_{2}(\mathbf{1},t)}$
can be replaced by $R_{2}(\mathbf{1},t)$, and $\overline{R_{3}(K,t)}$
can be replaced by $R_{3}(K,t)$.
\end{proof}
\begin{rem}
We do not know if there is a criterion of how to decide in a concrete
situation (where a subgroup $K\subseteq G$ and a vector $H_{d}\in \mathfrak{g}$
are given), whether the set $W$ of the previous corollary is convex.
This union of adjoint orbits turns out to be convex for instance in
the example of Section 3.2. But one can also find low-dimensional
examples, where $W$ is not a convex set.
\end{rem}

\section{The Maximum Principle for Compact Lie Groups }

In Section 1.5 the maximum principle of Pontrjagin (PMP) has been
discussed as a tool for determining extremal trajectories in a given
optimal control situation. The application of PMP involves the optimization
of functionals on the set $U$ of controls, which take the form\begin{equation}
u\longmapsto H(x,u)\textrm{,}\label{eq:}\end{equation}
where $x$ is an arbitrary but fixed point of the phase space $T^{*}M$,
while \begin{equation}
H(\cdot ,u):T^{*}M\longrightarrow \mathbb{R}\label{eq:}\end{equation}
is the Hamiltonian function associated with the optimal control problem.
An immediate application of the maximum principle to our original
(unreduced) control system $1$ yields in general no further information
on the optimality of a given control function, since in this case
the space $U$ of control parameters is unbounded. So in general there
need not be a control $u\in U$ which maximizes the functional (2.4.1).
This is one of the main differences to the adjoint system $2$. Here
the space $\mathrm{Ad}_{K}H_{d}\subseteq \mathfrak{g}$ of control
parameters is compact, so that the above functional always attains
its maximum. Thus the passage from system $1$ to system $2$ via
the equivalence theorem makes the time-optimal control problem of
Section 2.2 accessible to an application of Pontrjagin's maximum principle.
We take this observation \emph{}as an occasion to discuss PMP on Lie
groups in some detail.

Our discussion will be specialized to optimal control of right-invariant
systems\index{right-invariant system}. By this we mean a control
system $\Sigma =(G,f_{u},U)$, where the set $U$ of control parameters
is contained in the Lie algebra $\mathfrak{g}$ of $G$, and the admissible
vector fields $f_{u}$, $u\in U$, are the right-invariant extensions
of $u$, see Lemma 1.1.5. The results we are now going to discuss
apply in particular to the adjoint system (2.3.9). The reason why
right-invariant systems allow for a significant simpler formulation
of Pontrjagin's maximum principle is due to the following facts.

\begin{itemize}
\item The right-invariant vector fields $f_{u}$, $u\in U$, can be considered
to be contained in the finite dimensional Lie algebra $\mathfrak{g}$,
not just as elements of the infinite-dimensional algebra $\Gamma (TG)$
of arbitrary vector fields on $G$.
\item There is a canonical isomorphism between $T^{*}G$ and $G\times \mathfrak{g}^{*}$,
which allows to describe Hamiltonian functions and Hamiltonian vector
fields by globally defined coordinates.
\item Moreover, if $G$ is compact, it is possible to make use of the existence
of an $\textrm{ad}$-invariant inner product on $\mathfrak{g}$, which
allows to further identify $T^{*}G$ with $G\times \mathfrak{g}$.
An advantage of this identification is that Hamiltonian functions
and Hamiltonian vector fields often become easier to describe in coordinates
of $\mathfrak{g}$ and $G$ rather than in coordinates of $T^{*}G$,
respectively of $G$ and $\mathfrak{g}^{*}$.
\end{itemize}
\begin{flushleft}Our discussion follows the book \cite{key-24} by
Agrachev and Sachkov. Similar results can also be found in Mittenhuber's
paper \cite{key-19}.\end{flushleft}

\begin{prop}
Let $G$ be a Lie group with Lie algebra $\mathfrak{g}$. Denote by
$T^{*}G$ the cotangent bundle of $G$. Then the map \begin{equation}
\Phi :G\times \mathfrak{g}^{*}\longrightarrow T^{*}G,\quad (g,\lambda )\longmapsto (D_{g}R_{g^{-1}})^{*}(\lambda )\label{eq:}\end{equation}
is a vector bundle isomorphism. Here $(D_{g}R_{g^{-1}})^{*}:\mathfrak{g}^{*}\rightarrow \left(T_{g}G\right)^{*}$
denotes the dual of the linear map $D_{g}R_{g^{-1}}:T_{g}G\rightarrow \mathfrak{g}$.
\end{prop}
\begin{proof}
\cite{key-19}, p. 187.
\end{proof}
For the remainder of this section we restrict our discussion to the
case of a compact Lie group $G$. Such groups can be endowed with
an $\mathrm{ad}$-invariant Riemannian metric $\left\langle \cdot ,\cdot \right\rangle $,
cf. Example 1.2.4. This can now be used to first identify $\mathfrak{g}^{*}$
with $\mathfrak{g}$ via the scalar product $\left\langle \cdot ,\cdot \right\rangle _{\mathbf{1}}$.
Combining this identification with the map $\Phi $ of Proposition
2.4.1 then yields an isomorphism between the vector bundles $G\times \mathfrak{g}$
and $T^{*}G$. The $\mathrm{ad}$-invariance of the metric is certainly
not necessary for the existence of a vector bundle isomorphism $T^{*}G\cong G\times \mathfrak{g}$.
Such can be defined by using any Riemannian metric on $G$, and thus
also exists for noncompact Lie groups $G$. The point is that the
equations defining the Hamiltonian vector fields which occur in the
general statement of the maximum principle, cf. equation (1.5.16),
become particularly simple when choosing a trivialization via an $\mathrm{ad}$-invariant
metric. This point of view is substanziated in the following theorem. 

\begin{thm}
Let $\Sigma =(G,f_{u},U)$, $U\subseteq \mathfrak{g}$, be a right-invariant
control system on a compact Lie group $G$. Moreover, let $\varphi $
be a cost function, which does not depend on the position variable
$g\in G$, i.e. $\varphi :\mathfrak{g}\rightarrow \mathbb{R}$. Use
the above identification $T^{*}G\cong G\times \mathfrak{g}$ to define
the Hamiltonian function\begin{equation}
H^{\nu }(\cdot ,u):G\times \mathfrak{g}\longrightarrow \mathbb{R},\quad (g,X)\longmapsto \left\langle X,u\right\rangle _{\mathbf{1}}+\nu \varphi (u),\label{eq:}\end{equation}
where $u\in U$ and $\nu \in \left\{ -1,0\right\} $. Then the Hamiltonian
vector field associated with $H^{\nu }(\cdot ,u)$ reads \begin{equation}
\left\{ \begin{array}{l}
 \dot{g}=\frac{\partial H^{\nu }(\cdot ,u)}{\partial X}g,\\
 \dot{X}=\left[X,\frac{\partial H^{\nu }(\cdot ,u)}{\partial X}\right].\end{array}
\right.\label{eq:}\end{equation}
The Hamiltonian lift $t\mapsto \left(g(t),X(t)\right)\in G\times \mathfrak{g}$
of any optimal trajectory $t\mapsto g(t)$ for the optimal control
problem $\left(\Sigma ,\varphi \right)$ which results from a control
$t\mapsto \tilde{u}(t)$, $t\in \left[0,T\right]$, satisfies the
extremality condition\begin{equation}
H^{\nu }\left(g(t),X(t),\tilde{u}(t)\right)=\sup _{u\in U}H^{\nu }\left(g(t),X(t),u\right)\label{eq:}\end{equation}
almost everywhere on $\left[0,T\right]$.
\end{thm}
\begin{proof}
\cite{key-24}, Theorem 12.10, and p. 264. 
\end{proof}
\begin{rem}
See Remark 1.5.2 for the role of the parameter $\nu \in \mathbb{R}$
in the Hamilton function $H^{\nu }(\cdot ,u)$ of the previous theorem.
It can be argued that only the cases $\nu =-1$ (so-called normal
case) and $\nu =0$ (so-called abnormal case) need to be distinguished,
as Hamilton functions with $\nu >0$ lead to trajectories that maximize
the cost functional, while those with $\nu <0$ can be replaced by
$H^{\nu =-1}$ after rescaling the cost function $\varphi $, cf.
\cite{key-24}, p. 180. 
\end{rem}
\begin{example}
Let $\Sigma =\left\{ G,f_{u},U\right\} $ be a right-invariant control
system on the compact Lie group $G$, where $U:=\left\{ \left.u\in \mathfrak{g}\right|\left\langle u,u\right\rangle _{\mathbf{1}}=1\right\} $.
We are interested in time-optimal trajectories of $\Sigma $, and
therefore set $\varphi \equiv 1$. The Hamiltonian function (2.4.4)
in this case reads\begin{equation}
H^{\nu }(\cdot ,u):G\times \mathfrak{g}\longrightarrow \mathbb{R},\quad (g,X)\longmapsto \left\langle X,u\right\rangle _{\mathbf{1}}+\nu ,\label{eq:}\end{equation}
and the corresponding Hamiltonian vector field is\begin{equation}
\left\{ \begin{array}{l}
 \dot{g}=ug,\\
 \dot{X}=\left[X,u\right].\end{array}
\right.\label{eq:}\end{equation}
The maximality condition of PMP implies that $u(t)$ and $X(t)$ have
to be parallel for almost all $t$ in order to maximize the term $\left\langle X,u\right\rangle _{\mathbf{1}}$
in (2.4.7). But then, according to the second equation in (2.4.8),
$\dot{X}\equiv 0$, so that $X$ is constant. Then also the control
function $u$ is constant. So the first equation of (2.4.8) reads
$\dot{g}=Zg$ with some $Z\in U$, which is independent of $t$. This
equation can be integrated and yields the solution $g(t)=\exp (tZ)g_{0}$.
So the fastest way to steer system $\Sigma $ is along the integral
curves of right-invariant vector fields. \\
Since $G$ has been endowed with a bi-invariant metric, our time-optimal
problem can be, by the choice of $U$, considered a length-optimal
problem. We thus have as a result that a lenght-minimizing curve necessarily
is of the form $t\mapsto \exp (tZ)g_{0}$. Indeed, it can be shown
that any geodesic of the Riemannian manifold $\left(G,\left\langle \cdot ,\cdot \right\rangle \right)$
is of that form.
\end{example}
Theorem 2.4.2 can be applied to the problem of finding time-optimal
controls for the adjoint system \begin{equation}
\dot{U}=XU,\quad U(0)=\mathbf{1},\quad X\in \mathrm{Ad}_{K}H_{d}.\label{eq:}\end{equation}
 The associated Hamilton function and Hamilton vector fields to this
control problem are as stated in Theorem 2.4.2, namely (with $\varphi \equiv 1$
as cost function)\begin{equation}
H(\cdot ,u):G\times \mathfrak{g}\longrightarrow \mathbb{R},\quad (g,X)\longmapsto \left\langle X,u\right\rangle _{\mathbf{1}}+\upsilon \label{eq:}\end{equation}
and\begin{equation}
\left\{ \begin{array}{l}
 \dot{g}=ug,\\
 \dot{X}=\left[X,u\right].\end{array}
\right.\label{eq:}\end{equation}
The maximality condition of PMP for system (2.4.9) therefore reads\begin{equation}
H_{\max }(g,X)=\max _{u\in \mathrm{Ad}_{K}H_{d}}\left\langle X,u\right\rangle _{\mathbf{1}}.\label{eq:}\end{equation}
We next derive a necessary condition for $u\in \mathrm{Ad}_{K}H_{d}$
to satisfy (2.4.12).

\begin{prop}
Fix $X\in \mathfrak{g}$. Then $u_{0}=\mathrm{Ad}_{k_{0}}H_{d}\in \mathrm{Ad}_{K}H_{d}$
is a local maximum of the function $u\mapsto \left\langle X,u\right\rangle _{\mathbf{1}}$,
if the following holds.

(i) $\left\langle \left[X,u_{0}\right],Z\right\rangle _{\mathbf{1}}=0\textrm{ for all }Z\in \mathfrak{k}$,

(ii) $\left\langle \left[Z,X\right],\left[u_{0},Z\right]\right\rangle _{\mathbf{1}}<0\textrm{ for all }Z\in \mathfrak{k}$.

Condition (i) is necessary for (2.4.12) to hold.
\end{prop}
\begin{proof}
If (2.4.12) holds, then for all $Z\in \mathfrak{k}$ \begin{eqnarray*}
0 & = & \frac{d}{dt}\left.\left\langle \mathrm{Ad}_{\exp tZ}u_{0},X\right\rangle _{\mathbf{1}}\right|_{t=0}\\
 & = & \left\langle \mathrm{ad}Z(u_{0}),X\right\rangle _{\mathbf{1}}\\
 & = & \left\langle \left[Z,u_{0}\right],X\right\rangle _{\mathbf{1}}\\
 & = & -\left\langle \left[X,u_{0}\right],Z\right\rangle _{\mathbf{1}},
\end{eqnarray*}
which gives the necessarity of condition (i). Now the function $k\mapsto f(k):=\left\langle \mathrm{Ad}_{k}H_{d},X\right\rangle _{\mathbf{1}}$
has a local maximum in $k_{0}$ if condition (i) together with\[
\left.\frac{d^{2}}{dt^{2}}f\left(\exp tZ\cdot k_{0}\right)\right|_{t=0}<0\textrm{ for all }Z\in \mathfrak{k}\]
holds. But \begin{eqnarray*}
f\left(\exp tZ\cdot k_{0}\right) & = & \left\langle \mathrm{Ad}_{\exp tZ\cdot k_{0}}H_{d},X\right\rangle _{\mathbf{1}}\\
 & = & \left\langle e^{t\mathrm{ad}Z}(u_{0}),X\right\rangle _{\mathbf{1}}\\
 & = & \left\langle u_{0},X\right\rangle _{\mathbf{1}}+t\left\langle \left[Z,u_{0}\right],X\right\rangle _{\mathbf{1}}+\frac{1}{2}t^{2}\left\langle \left[Z,\left[Z,u_{0}\right]\right],X\right\rangle _{\mathbf{1}}+O\left(t^{3}\right),
\end{eqnarray*}
so condition $\left.\frac{d^{2}}{dt^{2}}f\left(\exp tZ\cdot k_{0}\right)\right|_{t=0}<0$
is equivalent to (ii).
\end{proof}
Our next goal is to derive a family of solutions of ODE (2.4.11).

\begin{thm}
Let $H_{d}\in \mathfrak{k}^{\perp }$, $A\in \mathrm{Ad}_{K}H_{d}$,
and $C\in \mathfrak{k}$. Then for\begin{equation}
u(t):=\mathrm{Ad}_{\exp \left(-Ct\right)}A,\label{eq:}\end{equation}
a solution of ODE (2.4.11) is given by\begin{equation}
\left\{ \begin{array}{l}
 g(t)=\exp (-Ct)\exp (Ct+At),\\
 X(t)=-u(t)+C.\end{array}
\right.\label{eq:}\end{equation}
The corresponding Hamilton function is $H(g,X,u)=\left\langle X,u\right\rangle _{\mathbf{1}}$.
This Hamilton function also satisfies the maximality condition (i)
of the previous proposition. Hence $t\mapsto g(t)$ is an extremal
trajectory of the time-optimal control problem associated with system
(2.4.9).
\end{thm}
\begin{proof}
We first notice that $u(t)\in \mathrm{Ad}_{K}H_{d}$ holds for all
$t$ by the choice of $C\in \mathfrak{k}$. A differentiation with
respect to $t$ now yields\begin{eqnarray*}
\dot{g}(t) & = & -C\exp (-Ct)\exp (Ct+At)+\exp (-Ct)(C+A)\exp (Ct+At)\\
 & = & \exp (-Ct)A\exp (Ct+At)\\
 & = & \exp (-Ct)A\exp (Ct)\exp (-Ct)\exp (Ct+At)\\
 & = & u(t)g(t)
\end{eqnarray*}
and\begin{eqnarray*}
\dot{X}(t) & = & -\dot{u}(t)\\
 & = & C\mathrm{Ad}_{\exp \left(-Ct\right)}A-\left(\mathrm{Ad}_{\exp \left(-Ct\right)}A\right)C\\
 & = & \left[C,u(t)\right]\\
 & = & \left[X(t),u(t)\right],
\end{eqnarray*}
as claimed. The fact that Hamilton system (2.4.11) arises from the
Hamilton function $H(\cdot ,u)$ is part of Theorem 2.4.2. It remains
to show that for $X=X(t)$, $u=u(t)$, and for all $Z\in \mathfrak{k}$
the equation\[
0=\left\langle \left[X,u\right],Z\right\rangle _{\mathbf{1}}\]
holds. Since \[
-\dot{u}=\left[X,u\right]\]
 and $\dot{u}$ is tangent to $\mathrm{Ad}_{K}H_{d}$, i.e. $\dot{u}=\left[Y,\mathrm{Ad}_{k}H_{d}\right]$
for some $Y\in \mathfrak{k}$ and $k\in K$, the last condition is
equivalent to \[
0=\left\langle \left[Y,\mathrm{Ad}_{k}H_{d}\right],Z\right\rangle _{\mathbf{1}}\]
for all $Z\in \mathfrak{k}$. This is satisfied, because\begin{eqnarray*}
\left\langle \left[Y,\mathrm{Ad}_{k}H_{d}\right],Z\right\rangle _{\mathbf{1}} & = & \left\langle \mathrm{Ad}_{k}\left[\mathrm{Ad}_{k^{-1}}Y,H_{d}\right],Z\right\rangle _{\mathbf{1}}\\
 & = & \left\langle \left[\mathrm{Ad}_{k^{-1}}Y,H_{d}\right],\mathrm{Ad}_{k^{-1}}Z\right\rangle _{\mathbf{1}}\\
 & = & -\left\langle H_{d},\left[\mathrm{Ad}_{k^{-1}}Y,\mathrm{Ad}_{k^{-1}}Z\right]\right\rangle _{\mathbf{1}}\\
 & = & 0
\end{eqnarray*}
holds as a consequence of the $\mathrm{Ad}$-invariance of the inner
product $\left\langle \cdot ,\cdot \right\rangle _{\mathbf{1}}$ and
of our assumption $H_{d}\in \mathfrak{k}^{\perp }$. 
\end{proof}
\begin{summary}
>From Theorem 2.4.6 we obtain a whole family of extremal trajectories
associated with the control system (2.4.9). These are parametrized
by real numbers $A$ and $C$, their role being the following. The
parameter $A=\dot{g}(0)$ determines the direction of the trajectory
$t\mapsto g(t)$ at its starting point, while the parameters $A$
and $C$ jointly fix the direction at $t=0$ of the component $x(t)$
of the Hamiltonian lift of $g(t)$, as $\dot{x}(0)=-\dot{u}(0)=\left[C,A\right]$.
Thus in a subsequent step one would have to determine those pairs
$(A,C)$ which actually give rise to a time-optimal trajectory.\\
For the special class of adjoint systems that we shall consider in
the following section, it turns out that only the choice $C=0$ can
lead to time-optimal trajectories. So from this example one can see
that the set of extremal trajectories in the sense of PMP will in
general be considerably larger than that of actually time-optimal
trajectories.
\end{summary}

\section{Time-Optimal Torus Theorem }

In the following an explicit solution to the control problem as described
in Section 2.2 and reformulated in the Equivalence Theorem 2.3.6 will
be discussed under the additional assumption that the homogeneous
space $G/K$ in that \\
theorem gives rise to a symmetric Lie algebra pair $\left(\mathfrak{g},\mathfrak{k}\right)$.\\
Thus in the following we fix a compact, simply connected, semisimple
Lie group $G$ together with a closed subgroup $K$ such that their
Lie algebras $\mathfrak{g}$ and $\mathfrak{k}$ form a symmetric
Lie algebra pair $\left(\mathfrak{g},\mathfrak{k}\right)$. Let $\mathfrak{g}=\mathfrak{k}\oplus \mathfrak{p}$
the corresponding Cartan-like decomposition, and take $\mathfrak{h}\subseteq \mathfrak{p}$
to be a maximal abelian subalgebra of $\mathfrak{p}$. Denote by $A$
the torus in $G$ with Lie algebra $\mathfrak{h}$.

\begin{thm}
\emph{(Controllability).} Let $G$ be a compact Lie group with simple
Lie algebra $\mathfrak{g}$. On $G$ consider the affine right-invariant
control system (2.3.1) of Section 2.3 with the Hamiltonian\begin{equation}
H(v_{1},...,v_{m})=H_{d}+\sum _{j=1}^{m}v_{j}H_{j}.\label{eq:}\end{equation}
Let $\mathfrak{k}$ the Lie algebra generated by $H_{1},...,H_{m}$
and assume $(\mathfrak{g},\mathfrak{k})$ to be a symmetric Lie algebra
pair. Denote by $\theta $ its Cartan involution and let $\mathfrak{h}$
be a maximal abelian subalgebra of $\mathfrak{p}$ that contains the
projection $H_{0}$ of $H_{d}$ on $\mathfrak{p}$ (such exists in
view of Lemma 1.3.8). Assume furthermore that $H_{0}$ is generic
in the sense that it is not contained in any root hyperplane (of a
root space decomposition of $\mathfrak{g}_{\mathbb{C}}$ with respect
to $\mathfrak{h}$). \\
Then system (2.3.1) has reachable set $\mathbf{R}(\mathbf{1})=G$,
i.e. it is controllable. The same holds for the respective reduced
system on the symmetric space $G/K$.
\end{thm}
\begin{proof}
The result follows from Theorem 1.4.5 if we can show that the Lie
algebra generated by $\mathfrak{k}$ and $H_{d}$ is equal to $\mathfrak{g}$.
Extend $\mathfrak{h}$ to a maximal abelian subalgebra $\mathfrak{t}$
of $\mathfrak{g}$ and let\[
\mathfrak{g}_{\mathbb{C}}=\mathfrak{t}_{\mathbb{C}}\oplus \bigoplus _{\alpha \in \Sigma }\mathfrak{g}_{\alpha }\]
be the root space decomposition of $\mathfrak{g}_{\mathbb{C}}$ with
respect to $\mathfrak{t}$. Choose as in (1.3.7) a subset $\Sigma ^{+}\subseteq \Sigma $
of so-called positive roots. Since $\mathfrak{g}$ is simple, it follows
from \\
Theorem 1.3.6 (iii) that the root spaces $\mathfrak{g}_{\alpha }$
are all $1$-dimensional. We can therefore write\begin{equation}
\mathfrak{g}_{\mathbb{C}}=\mathfrak{t}_{\mathbb{C}}\oplus \bigoplus _{\alpha \in \Sigma }\mathbb{C}X_{\alpha }\label{eq:}\end{equation}
with $X_{\alpha }\in \mathfrak{g}_{\alpha }\setminus \left\{ 0\right\} $
arbitrary. The root space decomposition (2.5.2) is related to the
$-1$-eigenspace $\mathfrak{p}$ of the Cartan-like decomposition
$\mathfrak{g}=\mathfrak{k}\oplus \mathfrak{p}$ in the following way:\[
\mathfrak{p}=\mathfrak{h}\oplus \mathfrak{n},\]
where\[
\mathfrak{n}:=\sum _{\alpha \in \Sigma ^{+}}\mathfrak{p}\cap \mathbb{C}\left(X_{\alpha }-\theta X_{\alpha }\right),\]
cf. \cite{key-6}, p. 335-336. Furthermore, the following sum is direct:\[
\sum _{\alpha \in \Sigma ^{+}}\mathfrak{k}\cap \mathbb{C}\left(X_{\alpha }+\theta X_{\alpha }\right)\subseteq \mathfrak{k}.\]
So for any $\alpha \in \Sigma ^{+}$ we can choose $c\in \mathbb{C}$
such that $c\left(X_{\alpha }+\theta X_{\alpha }\right)\in \mathfrak{k}\setminus \left\{ 0\right\} $.
We then find that\begin{eqnarray*}
\left[H_{0},c\left(X_{\alpha }+\theta X_{\alpha }\right)\right] & = & c\alpha (H_{0})X_{\alpha }+\theta \left[\theta H_{0},cX_{\alpha }\right]\\
 & = & c\alpha (H_{0})X_{\alpha }-c\theta \left[H_{0},X_{\alpha }\right]\\
 & = & c\alpha (H_{0})\left(X_{\alpha }-\theta X_{\alpha }\right).
\end{eqnarray*}
The commutator relation $\left[\mathfrak{p},\mathfrak{k}\right]\subseteq \mathfrak{p}$
together with our assumption $\alpha (H_{0})\neq 0$ for all roots
$\alpha $ now implies that\[
\left[H_{0},\mathfrak{k}\right]=\mathfrak{n}.\]
Set $\mathfrak{i}:=\left\langle \mathfrak{k},\mathfrak{n},H_{d}\right\rangle _{\mathrm{Lie}}=\left\langle \mathfrak{k},\mathfrak{n},H_{0}\right\rangle _{\mathrm{Lie}}$.
It remains to show that $\mathfrak{h}\subseteq \mathfrak{i}$. Assume
$X\in \mathfrak{h}\setminus \mathfrak{i}$. Let $Z\in \mathfrak{i}$
arbitrary and write \[
Z=Z_{1}+Z_{2}+Z_{3}\textrm{ with }Z_{1}\in \mathfrak{k},Z_{2}\in \mathfrak{n},Z_{3}\in \mathfrak{h}.\]
Repeating the calculation before yields $\left[X,Z_{1}\right]\in \mathfrak{n}$.
Furthermore, $\left[\mathfrak{h},\mathfrak{h}\right]=0$ and $\left[\mathfrak{h},\mathfrak{n}\right]\subseteq \left[\mathfrak{p},\mathfrak{p}\right]\subseteq \mathfrak{k}$,
so that \[
\left[X,Z\right]=\left[X,Z_{1}\right]+\left[X,Z_{2}\right]\in \mathfrak{n}+\mathfrak{k}\subseteq \mathfrak{i}.\]
This shows that $\mathfrak{i}$ is an ideal of the simple Lie algebra
$\mathfrak{g}$, which is not possible unless $\mathfrak{i}=\mathfrak{g}$.\\
The statement on the reduced system follows immediately from Theorem
2.3.6 (ii).
\end{proof}
\begin{rem}
A different proof of Theorem 2.5.1, without the assumption that $H_{0}$
be not contained in any root hyperplane, can be found in \cite{key-41}. 
\end{rem}
We now turn to a discussion of time-optimal control. To fix ideas
we initially consider the simple example of a single-particle system.
Here the underlying Lie group is $G=SU(2)$, the drift operator can
be chosen to be \[
H_{d}=\left(\begin{array}{cc}
 \mathrm{i} & 0\\
 0 & -\mathrm{i}\end{array}
\right)\in \mathfrak{su}(2),\]
and the free Hamiltonian to be \[
H_{1}=\left(\begin{array}{cc}
 0 & 1\\
 -1 & 0\end{array}
\right)\in \mathfrak{su}(2).\]
Denote by $\mathfrak{k}$ the $1$-dimensional Lie algebra spanned
by $H_{1}$. It generates the compact Lie subgroup\[
K=\left\{ \left.\left(\begin{array}{cc}
 \cos t & \sin t\\
 -\sin t & \cos t\end{array}
\right)\right|t\in \mathbb{R}\right\} \]
of $G$. We first observe that any element $U_{F}=\left(\begin{array}{cc}
 a & b\\
 -\bar{b} & \bar{a}\end{array}
\right)$, $a\bar{a}+b\bar{b}=1$, of $G$ can be decomposed as \begin{equation}
U_{F}=U_{1}k\label{eq:}\end{equation}
with $k=\left(\begin{array}{cc}
 \cos t & \sin t\\
 -\sin t & \cos t\end{array}
\right)\in K$ and $U_{1}=\left(\begin{array}{cc}
 r & s\\
 s & \bar{r}\end{array}
\right)$ symmetric. Namely, a calculation shows that we can choose\begin{equation}
\tan (t)=\frac{\textrm{Re}\, b}{\textrm{Re}\, a},\quad r=a\cos t+b\sin t,\quad s=-a\sin t+b\cos t.\label{eq:}\end{equation}
 Now $U_{F}$ and $U_{1}$, being contained in the same coset of $K$,
are for all $t\geq 0$ either both contained in the approximately
reachable set $S_{1}(\mathbf{1},t)$ or both not contained in $S_{1}(\mathbf{1},t)$,
see Proposition 2.3.11. Therefore, in the time-optimal control problem
for the unreduced system, $U_{F}$ can be replaced by $U_{1}$. Time-optimal
trajectories from $\mathbf{1}$ to $U_{1}$ for the adjoint system\begin{equation}
\dot{U}=XU,\quad X\in \textrm{Ad}_{K}H_{d}\label{eq:}\end{equation}
may now be obtained as follows. The symmetric matrix $U_{1}$ admits
a decomposition as \begin{equation}
U_{1}=k\left(\exp t_{0}H_{d}\right)k^{-1}\label{eq:}\end{equation}
where $k\in SU(2)$ is some orthogonal matrix, i.e. $kk^{T}=\mathbf{1}$,
and $t_{0}\in \mathbb{R}$. A calculation shows that one can choose
$k\in K$. Furthermore, we choose $t_{0}$ such that $\left|t_{0}\right|$
is as small as possible. Since the set of controls for the adjoint
system is invariant under conjugation by elements of $K$, $U_{1}$
can be replaced by the new terminal point $U_{2}=\exp t_{0}H_{d}$.
The key observation is then that a time-optimal trajectory between
$\mathbf{1}$ and $U_{2}$ is given by geodesics $t\mapsto \exp tH_{d}$
of a maximal torus $A$ of $G$ that contains $H_{d}$. This trajectory
is generated by a constant control function $u(t)=H_{d}$ or $u(t)=-H_{d}$,
the sign depending on that of $t_{0}$. \\
The main argument in the proof of this observation will be that a
suitable projection of an arbitrary trajectory $t\mapsto U(t)$ between
$\mathbf{1}$ and $U_{2}$ into the maximal torus $A$ leads to another
trajectory $t\mapsto V(t)$ for the adjoint system, which reaches
$U_{2}$ in the same time as $U(t)$.\\
This projection process reduces the original (adjoint) control system
to a system on the torus $A$, which can be solved easily because
the admissible vector fields are now pairwise commuting. In our particular
example, those admissible vector fields are the right-invariant extensions
of $H_{d}$ and $-H_{d}$. In the higher-dimensional cases, this set
of vector fields will become the Weyl orbit $W\cdot H_{d}$ of $H_{d}$
as defined in Section 1.3.1. Optimal trajectories will then again
be generated by choosing piecewise constant controls within the set
$W\cdot H_{d}$, and therefore will be geodesic arcs on a maximal
torus $A$ of $G$.\\
We next consider the general case of a compact semisimple Lie group
$G$ and a symmetric Lie algebra pair $\left(\mathfrak{g},\mathfrak{k}\right)$.
Our observations made so far motivate the following theorem.

\begin{thm}
\emph{(Time-optimal torus theorem\index{time-optimal torus theorem})}.
\emph{}Let $G$ be a compact, simply connected, semisimple Lie group,
and $K\subseteq G$ a closed subgroup. Assume that their respective
Lie algebras form a symmetric pair $\left(\mathfrak{g},\mathfrak{k}\right)$,
its Cartan-like decomposition being $\mathfrak{g}=\mathfrak{k}\oplus \mathfrak{p}$.
Let $\mathfrak{h}$ be a maximal abelian subalgebra of $\mathfrak{p}$,
$H_{d}\in \mathfrak{h}$ a generic point in the sense of the previous
theorem, and $A\subseteq G$ the maximal torus with Lie algebra $\mathfrak{h}$.
Furthermore, denote by $Y_{1},...,Y_{l}$ the elements of the Weyl
orbit $W\cdot H_{d}$. Set \begin{equation}
\Theta :=\mathrm{Ad}_{K}\bar{\Omega }=\left\{ \left.kak^{-1}\right|a\in \bar{\Omega },k\in K\right\} ,\label{eq:}\end{equation}
 and let $U_{F}\in \Theta $ arbitrary. Here $\Omega \subseteq A$
is a certain domain in $A$, which will be specified in Lemma 2.5.7.\\
Then the minimal time $t_{\mathrm{mi}\mathsf{n}}\left(U_{F}\right)$
for steering the adjoint system\begin{equation}
\dot{U}=XU,\quad U(0)=\mathbf{1},\quad X\in \mathrm{Ad}_{K}H_{d}\label{eq:}\end{equation}
to $U_{F}$ is equal to $\alpha ^{*}\left(U_{F}\right)$, which we
define to be the smallest non-negative value of $\alpha $ such that
the equation\begin{equation}
U_{F}=k\exp \left(\alpha \sum _{i=1}^{l}\beta _{i}Y_{i}\right)k^{-1}\label{eq:}\end{equation}
can be solved with $k\in K$, $\beta _{i}\in \mathbb{R}$, and $\sum _{i=1}^{l}\beta _{i}=1$.
Moreover, a time-optimal trajectory to $U_{F}$ is given by\begin{equation}
U:t\longmapsto \left\{ \begin{array}{cc}
 \exp \left(tkY_{1}k^{-1}\right), & t\in \left[0,\alpha \beta _{1}\right],\\
 \vdots  & \vdots \\
 \exp \left(tkY_{l}k^{-1}\right)\prod _{i=1}^{l-1}\exp \left(\alpha \beta _{i}kY_{i}k^{-1}\right), & t\in \left[\alpha \sum _{i=1}^{l-1}\beta _{i},\alpha \right].\end{array}
\right.\label{eq:}\end{equation}

\end{thm}
The proof of this theorem needs some preparation. 

To start with, we introduce the root space decomposition\index{root space decomposition}\begin{equation}
\mathfrak{g}_{\mathbb{C}}=\mathfrak{g}_{0}\oplus \bigoplus _{\alpha \in \Sigma }\mathfrak{g}_{\alpha }\label{eq:}\end{equation}
of $\mathfrak{g}_{\mathbb{C}}$ with respect to $\mathfrak{h}$, cf.
Theorem 1.3.2. Note that this is not quite the usual root-space decomposition,
since $\mathfrak{h}$ was only assumed to be maximal abelian in $\mathfrak{p}$,
but not in $\mathfrak{g}$. We will nevertheless call the spaces $\mathfrak{g}_{\alpha }$,
$\alpha \in \Sigma $, root-spaces, and the linear forms $\alpha :\mathfrak{h}\rightarrow \mathbb{R}$
roots. The subset of roots which do not vanish identically on $\mathfrak{p}$
is denoted $\Sigma _{\mathfrak{p}}$. We define the Weyl group\index{Weyl group}
in the same manner as before to be the quotient\begin{equation}
W:=N(\mathfrak{h})/\mathrm{S}\mathsf{t}\mathrm{ab}(\mathfrak{h}).\label{eq:}\end{equation}
In contrast to the previous definition, the isomorphism type of $W$
now depends on the pair $\left(\mathfrak{g},\mathfrak{k}\right)$,
not on $\mathfrak{g}$ alone.\\
Having fixed our notation in this way, we make now the following definition.

\begin{defn}
The following subset of $\mathfrak{h}$ is called the \emph{diagram\index{diagram}}
$D$ associated with the symmetric Lie algebra pair $\left(\mathfrak{g},\mathfrak{k}\right)$:\begin{equation}
D:=\left\{ \left.X\in \mathfrak{h}\right|\alpha (X)\in \mathrm{i}\pi \mathbb{Z}\textrm{ for some }\alpha \in \Sigma _{\mathfrak{p}}\right\} .\label{eq:}\end{equation}

\end{defn}
The diagram $D$ is the union of finitely many families of equidistant
hypersurfaces of $\mathfrak{h}$. The connected components of $\mathfrak{h}\setminus D$
are called \emph{cells\index{cell}}. Those are pairwise isometric,
open polytopes in $\mathfrak{h}$. The set $D$ contains the root
hyperplanes $\left\{ \left.X\in \mathfrak{h}\right|\alpha (X)=0\textrm{ for some }\alpha \in \Sigma \right\} $,
so that the arrangement of Weyl chambers in $\mathfrak{h}$ is now
further subdivided into the system $\mathfrak{h}\setminus D$ of cells.
\\
The set of reflexions on the hyperplanes that constitute the diagram
$D$ gene-\\
rates a group $W_{\mathrm{aff}}$, the so-called \emph{affine Weyl
group}\index{affine Weyl group}. This group contains $W$ as a subgroup.
More precisely, $W_{\mathrm{aff}}$ is isomorphic to the semidirect
product $W\rtimes T$ of $W$ with the group $T$ of translations
in $\mathfrak{h}$ that map $D$ onto itself.\\
The affine Weyl group acts transitively on the set of cells. Furthermore,
if $\Delta $ is a cell such that $0\in \bar{\Delta }$, then any
orbit $W_{\mathrm{aff}}\cdot X$, $X\in \mathfrak{h}$, intersects
$\bar{\Delta }$ in exactly one point.\\
For further details on the affine Weyl group and the diagram, see
\cite{key-6} and \cite{key-3}.\\
The relevance of the affine Weyl group for the proof of the time-optimal
torus theorem becomes clear through the following decomposition lemma.

\begin{lem}
\emph{($KAK$-decomposition\index{KAK-decomposition}).} Let $G$
be a compact, simply connected, semisimple Lie group. Then each $g\in G$
yields a decomposition\begin{equation}
g=k_{1}ak_{2},\label{eq:}\end{equation}
with $k_{1},k_{2}\in K$ and $a=\exp X\in A$. The factor $a=\exp X$
in this decomposition is unique up to an action of the affine Weyl
group. So if \begin{equation}
k_{1}ak_{2}=k'_{1}a'k'_{2}\label{eq:}\end{equation}
are two decompositions of the above type with $a=\exp X$ and $a'=\exp X'$,
then there exists $w\in W_{\mathrm{aff}}$ such that $X'=w\cdot X$.
The factor $a$ becomes determined uniquely, if in addition the requirement
$X\in \bar{\Delta }$ is imposed. Here $\Delta $ denotes as before
a cell whose closure contains $0$. 
\end{lem}
\begin{proof}
\cite{key-6}, p. 321-323. 
\end{proof}
\begin{rem}
Our initial assumption on $G$ to be simply connected is only needed
for the proof of the last lemma. We do not know a version of the $KAK$-decomposition
lemma without this assumption.
\end{rem}
\begin{lem}
Let $\Delta \in \mathfrak{h}$ a cell as in Lemma 2.5.5, and set $\Omega :=\exp \Delta \subseteq A$.
Define the map\begin{equation}
\Phi :K\times A\times K\longrightarrow G,\quad (k_{1},a,k_{2})\longmapsto k_{1}ak_{2},\label{eq:}\end{equation}
and let \begin{equation}
\pi _{2}:K\times A\times K\longrightarrow A\label{eq:}\end{equation}
be the projection map onto $A$. Then the following holds:\\
(i) For each $g\in G$, the set \begin{equation}
\left(\pi _{2}\circ \Phi ^{-1}(g)\right)\cap \bar{\Omega }\label{eq:}\end{equation}
consists of a single element $\pi _{A}(g)$. The map \begin{equation}
\pi _{A}=:G\longrightarrow \bar{\Omega },\quad g\longmapsto \pi _{A}(g)\label{eq:}\end{equation}
is continuous. Moreover, \begin{equation}
\left.\pi _{A}\right|_{\bar{\Omega }}=\mathrm{id}.\label{eq:}\end{equation}

\begin{flushleft}(ii) Set $G^{\mathrm{reg}}:=\pi _{A}^{-1}(\Omega )$.
Then the restriction of $\pi _{A}$ to the set $G^{\mathrm{reg}}$
yields a dif-\\
ferentiable map. Its differential is for all $g=\Phi (k_{1},a,k_{2})\in G^{\mathrm{reg}}$
and for all $X\in \mathfrak{g}$ given by the following formula: \begin{equation}
D_{\mathbf{1}}\left(\pi _{A}\circ R_{g}\right)(X)=D_{\mathbf{1}}R_{\pi _{A}(g)}\left(\Gamma \left(k_{1}^{-1}Xk_{1}\right)\right),\label{eq:}\end{equation}
where $\Gamma :\mathfrak{g}\rightarrow \mathfrak{h}$ denotes orthogonal
projection. In particular, as a consequence of Kostant's Convexity
Theorem 1.3.9,\begin{equation}
D_{\mathbf{1}}\left(\pi _{A}\circ R_{g}\right)(\mathrm{Ad}_{K}X)=D_{\mathbf{1}}R_{\pi _{A}(g)}\left(\mathfrak{c}\left(W\cdot X\right)\right)\label{eq:}\end{equation}
holds for all $X\in \mathfrak{h}$.\end{flushleft}
\end{lem}
\begin{proof}
(i) Using Lemma 2.5.5 on the $KAK$-decomposition, we can describe
the set $\left(\pi _{2}\circ \Phi ^{-1}\right)(g)$ as \[
\left(\pi _{2}\circ \Phi ^{-1}\right)(g)=\exp \left(W_{\mathrm{aff}}\cdot X\right),\]
where $X\in \bar{\Delta }$ is the unique element such that\[
g=k_{1}\exp Xk_{2}\]
holds for some $k_{1},k_{2}\in K$. We now claim that\[
\left(\pi _{2}\circ \Phi ^{-1}\right)(g)\cap \bar{\Omega }=\left\{ \exp X\right\} .\]
So suppose that $\exp X'\in \left(\pi _{2}\circ \Phi ^{-1}\right)(g)\cap \bar{\Omega }$.
Hence $X'=w\cdot X$ for some $w\in W_{\mathrm{aff}}$, and $\exp X'=\exp Z$
for some $Z\in \bar{\Delta }$. Thus $\exp Z=\exp \left(w\cdot X\right)$.
It follows that\[
Z=w\cdot X+V\]
for some $V\in \exp ^{-1}\left(\mathbf{1}\right)$. \\
Now $V$ can be regarded as an element of the subgroup $T$ of translations
of $W_{\mathrm{aff}}$, and we can further write\[
Z=w'\cdot X\]
for some $w'\in W_{\mathrm{aff}}$. This implies that\[
Z\in W\cdot X\cap \bar{\Delta }=\left\{ X\right\} ,\]
and finally $\exp X'=\exp Z=\exp X$, as claimed. This shows in particular
that the map $\pi _{A}$ is well-defined.\\
For the proof of continuity we define the quotient space $A/_{\sim _{W_{\mathrm{aff}}}}$
by\[
a\sim _{W_{\mathrm{aff}}}a'\quad \Longleftrightarrow \quad a'=w_{1}a\exp (d)w_{1}^{-1}\textrm{ for some }\left(w_{1},d\right)\in W_{\mathrm{aff}},\]
and notice that $\bar{\Omega }$ and $A/_{\sim _{W_{\mathrm{aff}}}}$
can be identified as topological spaces. Indeed, for any $X\in \mathfrak{h}$
there exists a unique $w\in W_{\mathrm{aff}}$, which we may write
as $w=\left(w_{1},d\right)\in W\times T$, such that\[
w\cdot X=w_{1}(X+d)w_{1}^{-1}\in \bar{\Delta }\]
holds. Hence $\exp w\cdot X=w_{1}\left(\exp X\exp d\right)w_{1}^{-1}$,
and $\exp X\sim \exp w\cdot X\in \bar{\Omega }$. Now let $a,a'\in \bar{\Omega }$
with $a\sim a'$. Then $a'=w_{1}a\exp (d)w_{1}^{-1}\textrm{ for some }\left(\textrm{w}_{\textrm{1}},d\right)\in W\times T$,
and both sides of the last equation define a $KAK$-decomposition
for $a'$. By Lemma 2.5.5 it follows that $a'=a$. This shows that
$\bar{\Omega }$ is a complete set of representatives for the equivalence
relation $\sim _{W_{\mathrm{aff}}}$. So the projection map\[
\textrm{pr}:\bar{\Omega }\longrightarrow A/_{\sim _{W_{\mathrm{aff}}}},\quad a\longmapsto \left[a\right]\]
is bijective. It is also continuous, because the action of $W_{\mathrm{aff}}$
on $A$ is continuous. Now that $A/_{\sim _{W_{\mathrm{aff}}}}$ is
compact, we see that the map $\textrm{pr}$ is in fact a homeomorphism.
Next consider an arbitrary open subset $U$ of $\bar{\Omega }$. Denote
the projection of $A$ onto $A/_{\sim _{W_{\mathrm{aff}}}}$ by $\pi _{\sim }$.
Then the preimage of $U$ under $\pi _{A}$ can be described as\[
\pi _{A}^{-1}(U)=\left(\pi _{2}\circ \Phi ^{-1}\right)^{-1}(U)=\Phi \left(K,\pi _{\sim }^{-1}(U),K\right).\]
As a consequence of the identification of $\bar{\Omega }$ with $A/_{\sim _{W_{\mathrm{aff}}}}$
it follows that the set $\pi _{\sim }^{-1}(U)$ is open in $A$. This
implies that $\Phi \left(K,\pi _{\sim }^{-1}(U),K\right)$ is open
in $G$ (cf. \cite{key-3} where such a statement is proved for general
actions of compact Lie groups on manifolds), and finally proves the
continuity of the map $\pi _{A}$.\\
For any $a\in A$ one has the $KAK$-decompositions $a=\mathbf{1}\cdot a\cdot \mathbf{1}$
and $a=k_{1}\pi _{A}(a)k_{2}$ with suitable factors $k_{1},k_{2}\in K$.
Now the middle factor of a $KAK$-decomposition is uniquely determined
if it is in addition required to lie in $\bar{\Omega }$. So for $a\in \bar{\Omega }$,
the elements $a$ and $\pi _{A}(a)$ have to coincide. This proves
the identity (2.5.20).\\
(ii) We first consider the action $\gamma $ of $K\times K$ on $G$
given by \[
\left(k_{1},k_{2}\right)\cdot g=k_{1}gk_{2}^{-1}.\]
A calculation shows that the orbit $\mathcal{O}_{\gamma }(a)$ of
$a\in A$ intersects the torus $A$ perpendicularly. Namely, for all
$X_{1},X_{2}\in \mathfrak{k}$ and $Z\in \mathfrak{h}$ the following
holds:

\begin{eqnarray*}
\left\langle \frac{d}{dt}\left.\left(\exp \left(tX_{1}\right)\cdot a\cdot \exp \left(tX_{2}\right)\right)\right|_{t=0},\frac{d}{dt}\left.\left(\exp \left(tZ\right)\cdot a\right)\right|_{t=0}\right\rangle _{a} & = & 
\end{eqnarray*}
\begin{eqnarray*}
 & = & \left\langle X_{1}a+aX_{2},Za\right\rangle _{a}\\
 & = & \left\langle X_{1}a,Za\right\rangle _{a}+\left\langle aX_{2}a^{-1}a,Za\right\rangle _{a}\\
 & = & \left\langle X_{1},Z\right\rangle _{\mathbf{1}}+\left\langle aX_{2}a^{-1},Z\right\rangle _{\mathbf{1}}\\
 & = & \left\langle X_{1},Z\right\rangle _{\mathbf{1}}+\left\langle X_{2},a^{-1}Za\right\rangle _{\mathbf{1}}\\
 & = & 0.
\end{eqnarray*}
Here we used the bi-invariance of the metric together with the fact
that the sum $\mathfrak{g}=\mathfrak{k}\oplus \mathfrak{p}$ is orthogonal.
>From this there follows the direct sum decomposition

\begin{equation}
T_{a}G=T_{a}A\oplus T_{a}\mathcal{O}_{\gamma }(a),\label{eq:}\end{equation}
if we can show that the dimension of $\mathcal{O}_{\gamma }(a)$ is
complementary to that of $A$ in $G$. This is in fact the case, provided
that $a$ is contained in $\Omega $, as will be proven now.\\
To this aim we make use of the following result, cf. \cite{key-6},
p. 294-295. Denote by $M$ the centralizer of $\mathfrak{h}$ in $K$,
i.e.\[
M:=\left\{ \left.k\in K\right|kXk^{-1}=X\: \forall X\in \mathfrak{h}\right\} .\]
Then the map\[
\varphi :K/M\times \mathfrak{h}\longrightarrow G/K,\quad \left(\left[k\right],X\right)\longmapsto \left[\exp \left(\textrm{Ad}_{k}X\right)\right]\]
is surjective. It is moreover regular on $K/M\times \left(\mathfrak{h}\setminus D\right)$,
i.e. in particular regular on $K/M\times \Delta $.\\
Now let $U\subseteq K/M\times \Delta $ be a sufficiently small open
subset such that the map \[
\left.\varphi \right|_{U}:U\longrightarrow \varphi (U)\]
is a diffeomorphism, and the set $W:=\pi ^{-1}(V)\subseteq G$ admits
a trivialization\[
\psi :W\longrightarrow V\times K\]
with\[
\psi \left(\exp \left(\textrm{Ad}_{k}X\right)k'\right)=\left(\left[\exp \left(\textrm{Ad}_{k}X\right)\right],k'^{-1}\right)\]
for all $\left(\left[k\right],X\right)\in U$ and $k'\in K$. Then,
by construction, the map\[
\sigma :U\times K\longrightarrow G,\quad \left(\left[k\right],X,k'\right)\longmapsto \psi ^{-1}\left(\varphi \left(\left[k\right],X\right),k'\right)\]
is regular on $U\times K$. Furthermore, the map $\gamma $ is related
to $\sigma $ as follows:\begin{equation}
\gamma \left(k,k'k,\exp X\right)=\sigma \left(\left[k\right],X,k'\right).\label{eq:}\end{equation}
The regularity of $\sigma $ implies now that for all $X\in \textrm{pr}_{2}(U)\subseteq \Delta $
the differential of the map $\sigma \left(\cdot ,X,\cdot \right)$
at the point $\left(\left[k\right],k'\right)\in \textrm{pr}_{1}(U)\times K\subseteq K\times K$
has maximal rank $r=\dim G-\dim \Delta =\dim G-\dim A$. From equation
(2.5.24) it follows that the same holds for the map $\gamma (\cdot ,\cdot ,a)$,
where $a=\exp X\in \exp \left(\textrm{pr}_{2}(U)\right)\subseteq \Omega $.
This finally proves equation (2.5.23).\\
We now proceed in the proof of formula (2.5.21). By part (i) of the
proof we have for all $a\in \Omega $, $Z\in \mathfrak{h}$ and $t$
sufficiently small the identity\[
\pi _{A}\left(\left(\exp tZ\right)a\right)=\left(\exp tZ\right)a,\]
and therefore\begin{equation}
D_{a}\pi _{A}\left(Za\right)=D_{\mathbf{1}}R_{a}(Z).\label{eq:}\end{equation}
 On the other hand, for all $tX\in T_{a}\mathcal{O}_{\gamma }(a)$
it follows that\begin{equation}
\pi _{A}\left(\left(\exp tX\right)a\right)=\pi _{A}\left(a\right)\label{eq:}\end{equation}
since $\pi _{A}$ is constant on the orbits of $\gamma $. Equations
(2.5.23), (2.5.25), and (2.5.26) now imply that\begin{equation}
D_{a}\pi _{A}\left(D_{\mathbf{1}}R_{a}(X)\right)=D_{\mathbf{1}}R_{a}\left(\Gamma (X)\right)\label{eq:}\end{equation}
holds for all $X\in \mathfrak{g}$.\\
Now let $g\in G^{\mathrm{reg}}$ arbitrary. Then $g=k_{1}ak_{2}$
for $a=\pi _{A}(g)\in \Omega $ and some $k_{1},k_{2}\in K$. Then
for all $X\in \mathfrak{g}$,\begin{eqnarray*}
\pi _{A}\left(\left(\exp tX\right)g\right) & = & \pi _{A}\left(\left(\exp tX\right)k_{1}ak_{2}\right)\\
 & = & \pi _{A}\left(k_{1}\left(\exp tk_{1}^{-1}Xk_{1}\right)ak_{2}\right)\\
 & = & \pi _{A}\left(\left(\exp tk_{1}^{-1}Xk_{1}\right)a\right).
\end{eqnarray*}
Differentiating both sides of this equation with respect to $t$ at
$t=0$ and applying equation (2.5.27) yields\begin{eqnarray}
D_{g}\pi _{A}\left(D_{\mathbf{1}}R_{g}(X)\right) & = & D_{a}\pi _{A}\left(D_{\mathbf{1}}R_{a}\left(k_{1}^{-1}Xk_{1}\right)\right)\label{eq:}\\
 & = & D_{\mathbf{1}}R_{a}\left(\Gamma \left(k_{1}^{-1}Xk_{1}\right)\right),\nonumber 
\end{eqnarray}
as claimed. \\
The identity (2.5.22) is now immediate from Kostant's Convexity Theorem
1.3.9, which for fixed $k_{1}\in K$ and $X\in \mathfrak{k}$ implies
that \[
\Gamma \left(k_{1}^{-1}\left(\textrm{Ad}_{K}X\right)k_{1}\right)=\Gamma \left(\textrm{Ad}_{K}X\right)=\mathfrak{c}\left(W\cdot X\right)\]
holds. Combining this with identity (2.5.21) we obtain (2.5.22).\\
We finally note that the map $\pi _{A}$ is differenciable in $g\in G^{\mathrm{reg}}{}$,
since equation (2.5.28) implies that the partial derivatives of $\pi _{A}$
in all directions $D_{\mathbf{1}}R_{g}(X)$ of $T_{g}G$ exist, with
continuous dependency on $g$.
\end{proof}
We now turn to the proof of Theorem 2.5.3.

\begin{proof}
First consider the special case $U_{F}\in \bar{\Omega }$. For such
an element $U_{F}$ there always exists an $\alpha $ which satisfies
equation (2.5.9) with $k=\mathbf{1}$. This is evident in the case
where $\mathfrak{g}$ is a simple Lie algebra, since then $W$ acts
irreducibly on $\mathfrak{h}$ (cf. \cite{key-8}, p. 53), so that
the Weyl orbit $W\cdot H_{d}$ of any $H_{d}\neq 0$ spans $\mathfrak{h}$.
In the semisimple case $\mathfrak{h}$ decomposes as a direct sum
$\mathfrak{h}=\bigoplus _{j}\mathfrak{h}_{j}$ of abelian subalgebras,
and $W$still acts irreducibly on each summand $\mathfrak{h}_{j}$.
By our choice of $H_{d}$, the component of $H_{d}$ in $\mathfrak{h}_{j}$
is non-zero, such that as in the simple case the set $W\cdot H_{d}=\left\{ Y_{1},...,Y_{l}\right\} $
spans $\mathfrak{h}$.\\
Moreover, for any tuple $\left(k,\alpha ,\beta _{1},...,\beta _{l}\right)$
that satisfies (2.5.9) it is clearly possible to steer system (2.5.8)
to $U_{F}$ at time $t_{F}=\alpha $ along the trajectory as specified
in (2.5.10). Let $\alpha ^{*}(U_{F})$ be the smallest non-negative
number such that equation (2.5.9) can be solved with $\alpha =\alpha ^{*}(U_{F})$.\\
It remains to show that any trajectory of system (2.5.8) with $U(t_{F})=U_{F}$
necessarily satisfies $t_{F}\geq \alpha ^{*}(U_{F})$. In order to
prove this we use the projection map $\pi _{A}$ of Lemma 2.5.6 to
replace $t\mapsto U(t)$ by another trajectory of the system which
joins $U(0)=\mathbf{1}$ to $U(t_{F})=U_{F}$, but is completely contained
in $A$. Since the tangents $\dot{U}(t)$ of $U(t)$ are by definition
right translates of the set $\mathrm{Ad}_{K}H_{d}\subseteq \mathfrak{g}$
we can apply Lemma 2.5.6 (ii) to conclude that the path\[
t\longmapsto \pi _{A}\left(U(t)\right)=:V(t)\]
has tangents $\dot{V}(t)\in D_{\mathbf{1}\textrm{ }}R_{V(t)}\left(\mathfrak{c}\left(W\cdot H_{d}\right)\right)$.
By part (i) of the same lemma, \[
V(t_{F})=\pi _{A}(U_{F})=U_{F}.\]
However, $t\mapsto V(t)$ is in general not a trajectory of the control
system under consideration. But for all $t\in \left[0,t_{F}\right]$,
its tangents are of the form\[
\dot{V}(t)=D_{\mathbf{1}\textrm{ }}R_{V(t)}\left(\sum _{i=1}^{l}\beta _{i}(t)Y_{i}\right)\]
 with $\sum _{i=1}^{l}\beta _{i}(t)=1$ and $Y_{i}$ as stated. Therefore,
since $\left[Y_{i},Y_{j}\right]=0$ for all $i,j=1,...,l$,\[
V(t)=\prod _{i=1}^{l}\left(\exp \left(\int _{0}^{t}\beta _{i}(\tau )\, d\tau \right)Y_{i}\right).\]
So steering the adjoint system according to the control\[
t\longmapsto \left\{ \begin{array}{cc}
 Y_{1}, & t\in \left[0,\gamma _{1}\right],\\
 \vdots  & \vdots \\
 Y_{l}, & t\in \left[\sum _{i=1}^{l-1}\gamma _{l-1},\sum _{i=1}^{l}\gamma _{l}\right],\end{array}
\right.\]
where \[
\gamma _{i}:=\int _{0}^{t_{F}}\beta _{i}(\tau )\, d\tau ,\]
results in a trajectory $t\mapsto W(t)$ on $A$ which satisfies \[
W\left(\sum _{i=1}^{l}\gamma _{l}\right)=W(t_{F})=V(t_{F})=U_{F}.\]
This is again immediate from the pairwise commutativity of the vector
fields $Y_{i}$, $i=1,...,l$. Moreover, it follows from the definition
of $\alpha ^{*}(U_{F})$ that $\alpha ^{*}(U_{F})\leq t_{F}$, as
we have claimed. Finally, a trajectory from $U(0)$ to $U\left(\alpha ^{*}(U_{F})\right)=U_{F}$,
and therefore (by construction) a time-optimal trajectory, is given
by $t\mapsto W(t)$, and this has the form as stated.\\
Now let $U_{F}\in \Theta $ be arbitrary. By definition of $\Theta $
we can choose $\tilde{k}\in K$ such that $\tilde{U}_{F}:=\tilde{k}U_{F}\tilde{k}^{-1}\in \bar{\Omega }$.
The trajectories joining the system from $U(0)=\mathbf{1}$ to $U_{F}$,
respectively to $\tilde{U}_{F}$ at fixed time $t_{F}\geq 0$ are
in bijective correspondence to each other, see the proof of Proposition
2.3.7 (i). Namely, if $t\mapsto X(t)$, $t\in \left[0,t_{F}\right]$,
is a control which leads to a trajectory $t\mapsto U(t)$ with $U(t_{F})=U_{F}$,
then $t\mapsto \tilde{k}X(t)\tilde{k}^{-1}$, $t\in \left[0,t_{F}\right]$,
is a control (with values again in $\mathrm{Ad}_{K}H_{d}$) which
leads to a trajectory with endpoint $\tilde{U}_{F}$, and vice versa.\\
Hence $\alpha ^{*}\left(U_{F}\right)=\alpha ^{*}\left(\tilde{U}_{F}\right)$,
which is, by the first part of the proof, the smallest non-negative
value of $\alpha $ such that\[
\tilde{U}_{F}=k\exp \left(\alpha \sum _{i=1}^{l}\beta _{i}Y_{i}\right)k^{-1}\]
can be solved for some $k\in K$. At the same time, this $\alpha $
is the smallest one possible such that\[
U_{F}=\tilde{k}^{-1}\tilde{U}_{F}\tilde{k}=\tilde{k}^{-1}k\exp \left(\alpha \sum _{i=1}^{l}\beta _{i}Y_{i}\right)k^{-1}\tilde{k}\]
can be solved with $k\in K$, which means that the value of $\alpha ^{*}(U_{F})$
is as stated. \\
The statement on the time-optimality of the trajectory $t\mapsto U(t)$
with $U(t_{F})=U_{F}$ follows from that on time-optimal trajectories
in the special case $U_{F}\in \bar{\Omega }$ by using again the correspondence
of trajectories with $K$-conjugated endpoints as formulated before.
\end{proof}
\begin{cor}
Assume the Lie groups $G$ and $K$ to satisfy the prerequisities
of Theorem 2.5.3, and let $P_{F}\in G/K$ arbitrary. Then the set
\begin{equation}
X_{F}:=\pi ^{-1}(P_{F})\cap \Theta \subseteq G\label{eq:}\end{equation}
 is non-empty. Furthermore, the canonical projection of any trajectory
of type (2.5.10) with endpoint in $X_{F}$ yields a time-optimal trajectory
between $\pi (\mathbf{1})=K$ and $P_{F}$ for the reduced system
$3$.
\end{cor}
\begin{proof}
Let $P_{F}=gK$ for some $g\in G$. Then $g$ yields a $KAK$-decomposition
of the form $g=k_{1}ak_{1}^{-1}k_{2}$ with $k_{1},k_{2}\in K$ and
$a\in \bar{\Omega }$. Hence $k_{1}ak_{1}^{-1}\in \pi ^{-1}(P_{F})\cap \Theta =X_{F}$.\\
Now let $t\mapsto u(t)\in \mathrm{Ad}_{K}H_{d}$, $t\in \left[0,t_{F}\right]$,
be a control for system $3$, such that the corresponding trajectory
$t\mapsto P(t)$ has terminal point $P(t_{F})=P_{F}$. Let system
$2$ evolve according to the same control function $u$, and denote
the resulting trajectory by $t\mapsto U(t)$. From the proof of Proposition
2.3.7 (ii) it follows that $P(t)=\pi \circ U(t)$ holds on $\left[0,t_{F}\right]$.
In particular, $U_{F}:=U(t_{F})$ satisfies $\pi (U_{F})=P_{F}$.
\\
In complete analogy to the proof of the previous Theorem 2.5.3 we
can apply the projection map $\pi _{A}$ together with a conjugation
to the trajectory $U$ in order to obtain a trajectory $t\mapsto W(t)$,
$t\in \left[0,t_{F}\right]$, which is of the form (2.5.10) and satisfies
$W(t_{F})=U_{F}k'$ for some $k'\in K$. Namely, if \[
U_{F}=k\pi _{A}(U_{F})k^{-1}k'^{-1}\]
is a $KAK$-decomposition of $U_{F}$, then we will set $V(t):=k\pi _{A}(U(t))k^{-1}$
for $t\in \left[0,t_{F}\right]$. As in the proof of Theorem 2.5.3
we then obtain from the path $t\mapsto V(t)$ a trajectory $t\mapsto W(t)$
of system $2$ which is of the special form (2.5.10) and has endpoint
$W(t_{F})=V(t_{F})=U_{F}k'$. \\
We denote the control function which generates this trajectory by
$v$. Let system $3$ evolve according to the same control function
$v$, and denote the corresponding trajectory by $t\mapsto P_{1}(t)$.
It follows as before that \[
P_{1}(t)=\pi \circ W(t)\]
holds on $\left[0,t_{F}\right]$. In particular, \[
P_{1}(t_{F})=\pi (U_{F}k')=\pi (U_{F})=P_{F}.\]
This shows that any trajectory $t\mapsto P(t)$ which reaches $P_{F}$
at time $t_{F}$ can be replaced by a trajectory $t\mapsto P_{1}(t)$
which also satisfies $P_{1}(t_{F})=P_{F}$, and has the special form
as stated.\\
It remains to show that any trajectory of that type is indeed time-optimal
for system $3$. So let $W'_{F}\in X_{F}$ be arbitrary, and denote
by $t\mapsto W'(t)$, $t\in \left[0,t'_{F}\right]$, a trajectory
of type (2.5.10) which reaches $W'_{F}$ at time $t'_{F}$. The statement
of the time-optimal torus theorem is that such a trajectory is time-optimal
for system $2$. Since $\pi (W'_{F})=\pi (W_{F})=P_{F}$, we see that
$W_{F}=W'_{F}k'$ holds for some $k'\in K$. Thus $\pi _{A}(W'_{F})=\pi _{A}(W_{F})=:a\in \bar{\Omega }$.
Furthermore, as $W_{F},W'_{F}\in \Theta $, it follows that both $W_{F}$
and $W'_{F}$ are conjugated within $K$ to $a$. Hence $W_{F}=k''W'_{F}k''^{-1}$
for some $k''\in K$. Finally, as a consequence of Proposition 2.3.7
(i), it follows that $t_{F}=t'_{F}$, as otherwise $t\mapsto W(t)$
and $t\mapsto W'(t)$ could not both be time-optimal trajectories
for system $2$. \\
This shows that the projection of $t\mapsto W'(t)$ under $\pi $
again yields a time-optimal trajectory for system $3$ between $K$
and $P_{F}$.
\end{proof}
We finally describe how the combination of the Equivalence Theorem
2.3.6 with the previous Corollary 2.5.8 can be used to solve the time-optimal
control problem for the unreduced system (2.3.1) we have originally
been interested in. We therefore keep all the assumptions made in
the time-optimal torus theorem, and let $U_{F}\in G$ arbitrary. From
Corollary 2.3.12 it follows that $t_{F}:=t_{\textrm{inf},1}(U_{F})$
equals $t_{\textrm{inf},3}(U_{F}K)$. Corollary 2.5.8 can now be used
as follows to construct a trajectory $t\mapsto U(t)$ for system $1$,
which satisfies the time-optimality condition $U(t_{F})=U_{F}$.

\begin{enumerate}
\item Decompose $U_{F}$ as $U_{F}=k_{1}ak_{1}^{-1}k_{2}$ with $k_{1},k_{2}\in K$
and $a\in \bar{\Omega }$. This is just a $KAK$-decomposition as
introduced in Lemma 2.5.5. By definition of $\bar{\Omega }$, the
factor $a$ in that decomposition is uniquely determined.
\item Set $V_{F}:=k_{1}ak_{1}^{-1}\in \Theta $, where $\Theta \subseteq G$
is as defined in Theorem 2.5.3. Let $t\mapsto V(t)$, $t\in \left[0,t_{F}\right]$,
be a trajectory for system $2$ of type (2.5.10) such that $V(t_{F})=V_{F}$
holds. By Corollary 2.5.8, such a trajectory exists and is time-optimal.
\item By construction, the trajectory $t\mapsto V(t)$ consists of a finite
number of geodesic arcs of the form\[
t\longmapsto k'\exp \left(tY_{j+1}\right)k'^{-1}V(t_{j}),\quad t\in \left[t_{j},t_{j+1}\right],\]
where $k'\in K$, $Y_{j}\in W\cdot H_{d}$, and $\left[t_{j},t_{j+1}\right]$
is a subinterval of $\left[0,t_{F}\right]$ as specified in Theorem
2.5.3. Write $Y_{j}$ as $k''H_{d}k''^{-1}$ for some $k''\in K$.
\item System $1$ can be steered from $V(t_{j})$ to \[
V(t_{j+1})=k'\exp \left(t_{j+1}Y_{j+1}\right)k'^{-1}V(t_{j})\]
 within time $t_{j+1}-t_{j}$ by first producing the element $\left(k'k''\right)^{-1}V(t_{j})$
within zero infimizing time. Evolution under the influence of the
drift operator $H_{d}$ for time $t_{j+1}-t_{j}$ transfers the system
in a second step from $\left(k'k''\right)^{-1}V(t_{j})$ to $\exp \left(t_{j+1}H_{d}\right)\left(k'k''\right)^{-1}V(t_{j})$.
The point $V(t_{j+1})$ is finally reached from $\exp \left(t_{j+1}H_{d}\right)\left(k'k''\right)^{-1}V(t_{j})$
within zero infimizing time.
\item The iteration of such so-called pulse-drift-pulse sequences transfers
the unreduced system within infimizing time $t_{F}$ from $V(0)=\mathbf{1}$
to $V_{F}$. The point $U_{F}=V_{F}k_{2}$ differs from $U_{F}$ by
only an element of $K$ and thus can also be reached within infimizing
time $t_{F}$.
\end{enumerate}

\chapter{Discussion of some Explicit Spin Systems}

\section{General Considerations}

In this chapter we shall apply the results obtained so far to a number
of concrete examples. While the Equivalence Theorem 2.3.6 and the
Time-Optimal Torus Theorem 2.5.3 have been formulated for right-invariant
control systems on arbitrary compact Lie groups (with additional assumptions
such as semisimplicity to be satisfied in 2.5.3), we are now focussing
on the particular case $G_{n}=SU\left(2^{n}\right)$ and the time-optimal
control of certain $n$-particle spin systems. As a typical example
of that kind of control-system we want to discuss system (2.2.2) of
Section 2.2, where the subgroup $K_{n}$ generated by the control
Hamiltonians was $K_{n}=SU(2)^{\otimes n}$. The equivalence theorem
allows us to replace this control system by the corresponding reduced
system\begin{equation}
\dot{P}=XP,\quad P(0)=K,\quad X\in \textrm{Ad}_{K_{n}}H_{d}\label{eq:}\end{equation}
on the homogeneous space $G_{n}/K_{n}$. Let $\mathfrak{g}_{n}:=\mathfrak{su}\left(2^{n}\right)$
and $\mathfrak{k}_{n}:=\mathfrak{su}(2)^{\otimes n}$ be the Lie algebras
of $G_{n}$, respectively of $K_{n}$.\\
To proceed in our discussion of time-optimal control, we first of
all determine those examples in the family (3.1.1) of control systems
which meet the requirements of Theorem 2.5.3, respectively its Corollary
2.5.8. The following theorem shows that these are satisfied for $n=1$
and $n=2$ only. 

\begin{thm}
The pair $\left(\mathfrak{g}_{n},\mathfrak{k}_{n}\right)$ is a symmetric
Lie algebra pair if and only if $n\leq 2$.
\end{thm}
\begin{proof}
In course of proving this theorem we need to work in the tensor-product
basis of $\mathfrak{g}_{n}=\mathfrak{su}\left(2^{n}\right)$ as introduced
in Example 1.6.4. This basis comprises the $n$-fold products of elements
$\left\{ \mathbf{1},I_{x},I_{y},I_{z}\right\} $ with at least one
factor being different from $\mathbf{1}$.\\
Assume $\left(\mathfrak{g}_{n},\mathfrak{k}_{n}\right)=\left(\mathfrak{su}\left(2^{n}\right),\mathfrak{su}(2)^{\otimes n}\right)$,
$n\in \mathbb{N}$, to be a symmetric Lie algebra pair with Cartan-like
decomposition $\mathfrak{g}_{n}=\mathfrak{k}_{n}\oplus \mathfrak{p}_{n}$.
We claim that the set\[
X:=\bigcup _{j\geq 2}X_{j}\]
with $X_{j}$ as defined in Example 1.6.4 is a basis of $\mathfrak{p}_{n}$.
Because $\mathfrak{k}_{n}$ is spanned by $X_{1}$, the $\mathbb{R}$-linear
span of $X$ is seen to be complementary to $\mathfrak{k}_{n}$ in
$\mathfrak{g}_{n}$. It remains to show that $X$ is orthogonal to
$\mathfrak{k}_{n}$ (cf. Lemma 1.2.6, and use that $\mathfrak{g}_{n}$
is semisimple). Therefore, let \[
A:=\mathrm{i}^{\varepsilon _{j}}\mathbf{1}\otimes ...\otimes I_{\alpha _{1}}\otimes ...\otimes I_{\alpha _{j}}\otimes ...\otimes \mathbf{1}\in X_{j},\]
$j\geq 2$, and \[
B:=\mathbf{1}\otimes ...\otimes I_{\beta }\otimes ...\otimes \mathbf{1}\in X_{1}.\]
Let $k\in \left\{ 1,...,j\right\} $ be an index such that the position
of $I_{\alpha _{k}}$ in $A$ differs from that of $I_{\beta }$ in
$B$. Choose $\gamma ,\delta \in \left\{ x,y,z\right\} $ such that
$\left[I_{\gamma },I_{\delta }\right]=I_{\alpha _{j}}$ and set \[
C:=\mathbf{1}\otimes ...\otimes I_{\gamma }\otimes ...\otimes \mathbf{1}\in X_{1},\]
$I_{\gamma }$ at the same position as $I_{\alpha _{j}}$, and \[
D:=\mathrm{i}^{\varepsilon _{j}}\mathbf{1}\otimes ...\otimes I_{\alpha _{1}}\otimes ...\otimes I_{\alpha _{j-1}}\otimes ...\otimes I_{\delta }\otimes ...\otimes \mathbf{1}\in X_{j},\]
$I_{\delta }$ at the same position as $I_{\alpha _{j}}$ and all
other positions coinciding with those of $A$. Using Lemma 1.6.3 (iv)
we find that $A=\left[C,D\right]$ and $\left[B,C\right]=0$. From
the $\mathrm{ad}$-invariance of the Killing-form $\kappa $ it now
follows that\[
\kappa \left(A,B\right)=\kappa \left(\left[C,D\right],B\right)=\kappa \left(D,\left[B,C\right]\right)=\kappa \left(D,0\right)=0.\]
This shows $A\in X_{1}^{\perp }$. So \[
\left\langle X\right\rangle \subseteq \left\langle X_{1}\right\rangle ^{\perp }=\mathfrak{k}_{n}^{\perp }=\mathfrak{p}_{n},\]
and finally, for dimensional reasons, $\left\langle X\right\rangle =\mathfrak{p}_{n}$.\\
Now let $n\geq 3$ and consider the elements \[
A_{1}:=I_{x}\otimes I_{y}\otimes I_{z}\otimes \mathbf{1}\otimes ...\otimes \mathbf{1}\in \mathfrak{p}_{n}\]
and\[
A_{2}:=\mathrm{i}I_{y}\otimes I_{y}\otimes \mathbf{1}\otimes ...\otimes \mathbf{1}\in \mathfrak{p}_{n}.\]
We use Lemma 1.6.3 (iv) to calculate\begin{eqnarray*}
-\mathrm{i}\left[A_{1},A_{2}\right] & = & \left[I_{x}\otimes I_{y}\otimes I_{z}\otimes \mathbf{1}\otimes ...\otimes \mathbf{1},I_{y}\otimes I_{y}\otimes \mathbf{1}\otimes ...\otimes \mathbf{1}\right]\\
 & = & \left[I_{x},I_{y}\right]\otimes \left(I_{y}\otimes I_{z}\otimes \mathbf{1}\otimes ...\otimes \mathbf{1}\right)\left(I_{y}\otimes \mathbf{1}\otimes ...\otimes \mathbf{1}\right)\\
 &  & +I_{y}I_{x}\otimes \left[I_{y}\otimes I_{z}\otimes \mathbf{1}\otimes ...\otimes \mathbf{1},I_{y}\otimes \mathbf{1}\otimes ...\otimes \mathbf{1}\right]\\
 & = & I_{z}\otimes I_{y}^{2}\otimes I_{z}\otimes \mathbf{1}\otimes ...\otimes \mathbf{1}\\
 &  & -I_{z}\otimes \left(\left[I_{y}\otimes I_{y}\right]\otimes \left(I_{z}\otimes \mathbf{1}\otimes ...\otimes \mathbf{1}\right)\left(\mathbf{1}\otimes ...\otimes \mathbf{1}\right)\right)\\
 &  & -I_{z}\otimes \left(I_{y}^{2}\otimes \left[I_{z}\otimes \mathbf{1}\otimes ...\otimes \mathbf{1},\mathbf{1}\otimes ...\otimes \mathbf{1}\right]\right)\\
 & = & -I_{z}\otimes \mathbf{1}\otimes I_{z}\otimes \mathbf{1}\otimes ...\otimes \mathbf{1}\\
 &  & -I_{z}\otimes \left(0\otimes \left(I_{z}\otimes \mathbf{1}\otimes ...\otimes \mathbf{1}\right)\left(\mathbf{1}\otimes ...\otimes \mathbf{1}\right)+I_{y}^{2}\otimes 0\right)\\
 & = & -I_{z}\otimes \mathbf{1}\otimes I_{z}\otimes \mathbf{1}\otimes ...\otimes \mathbf{1}\\
 & \notin  & \mathfrak{k}_{n}.
\end{eqnarray*}
 So $\left[\mathfrak{p}_{n},\mathfrak{p}_{n}\right]$ is not contained
in $\mathfrak{k}_{n}$, and $\left(\mathfrak{g}_{n},\mathfrak{k}_{n}\right)$
cannot be a symmetric Lie algebra pair.\\
In the trivial case $n=1$ the assertion clearly holds.\\
Now let $n=2$. We have to check that $\mathfrak{g}_{2}$ admits an
involutive Lie algebra automorphism $\theta $ with $\mathfrak{k}_{2}$
as its $1$-eigenspace. On the basis $X_{1}\cup X_{2}$ as introduced
before define\[
\theta (X)=\left\{ \begin{array}{ll}
 X, & \textrm{if }X\in X_{1},\\
 -X, & \textrm{if }X\in X_{2},\end{array}
\right.\]
and extend $\theta $ to a linear map on $\mathfrak{g}_{2}$. So $\theta $
is by definition an involution. A calculation now shows that the commutator
$\left[Y_{1},Y_{2}\right]$ of any two elements $Y_{1},Y_{2}\in X_{2}$
is contained in $\mathfrak{k}_{2}$, while for all $Z\in \mathfrak{k}_{2}$
the commutator $\left[Y_{1},Z\right]$ is in the linear span of $X_{2}$.
So \[
\theta \left[Y_{1},Y_{2}\right]=\left[Y_{1},Y_{2}\right]=\left[\theta (Y_{1}),\theta (Y_{2})\right]\]
and\[
\theta \left[Y_{1},Z\right]=-\left[Y_{1},Z\right]=\left[\theta (Y_{1}),\theta (Z)\right],\]
i.e. $\theta $ is also a Lie algebra automorphism.
\end{proof}
\begin{summary}
Our discussion so far lead to the result that the problem of time-optimal
control of an $n$-particle spin system with Hamiltonian $H=H_{d}+\sum _{j=1}^{m}\left(v_{jx}I_{jx}+v_{jy}I_{jy}+v_{jz}I_{jz}\right)\in \mathfrak{su}\left(2^{n}\right)$
can be solved by applying the results of Section 2.5 if and only if
$n\in \left\{ 1,2\right\} $. This will be carried out in detail in
the subsequent two sections.\\
Although the number of cases where Theorem 2.5.3 on time-optimal control
applies is quite limited as long as we are only interested in Hamiltonians
of the special form above, one nevertheless could imagine other interesting
right-invariant control systems on $G=SU(n)$ or any other compact
semisimple Lie group $G$ that allow for the application of that theorem.
\end{summary}

\section{Single Particle Systems}

We turn to a discussion of control system (3.1.1) in the case $n=1$.
Here we assume the exterior magnetic field to excite rapidly the $x$-component
$I_{x}$ of the spin $I=\left(I_{x},I_{y},I_{z}\right)$ and consider
$I_{z}$ to be the drift Hamiltonian. This leads to the unreduced
control system\begin{equation}
\dot{U}=\left(I_{z}+vI_{x}\right)U,\quad U(0)=\mathbf{1},\quad v\in \mathbb{R}\label{eq:}\end{equation}
on $G=SU(2)$. $I_{x}$ generates the Lie subgroup \begin{equation}
K=\left\{ \left.\left(\begin{array}{cc}
 \cos t & -\sin t\\
 \sin t & \cos t\end{array}
\right)\right|t\in \mathbb{R}\right\} ,\label{eq:}\end{equation}
 which is isomorphic to $U(1)$. So the corresponding reduced system
is\begin{equation}
\dot{P}=PX,\quad P(0)=K,\quad X\in \textrm{Ad}_{K}I_{z}\label{eq:}\end{equation}
 on the two-dimensional homogeneous space $G/K=SU(2)/U(1)$. This
space is diffeomorphic to the projective plane $\mathbb{RP}^{2}$,
as will become clear later. The set $\textrm{Ad}_{K}I_{z}$ of control
variables is a circle around zero, cf. Example 1.3.10.\\
The pair $\left(\mathfrak{g},\mathfrak{k}\right)=\left(\mathfrak{su}(2),\mathfrak{u}(1)\right)$
is symmetric with Cartan involution $\theta $ defined by\begin{equation}
\theta \left(\alpha I_{x}+\beta I_{y}+\gamma I_{z}\right)=\alpha I_{x}-\beta I_{y}-\gamma I_{z}\label{eq:}\end{equation}
Hence Theorems 2.5.1 and 2.5.3 apply to system (3.2.3). So this system
is in particular controllable.\\
The orthogonal complement of $\mathfrak{k}=\mathbb{R}I_{x}$ with
respect to the Killing form on $\mathfrak{su}(2)$ is \begin{equation}
\mathfrak{p}=\mathbb{R}I_{y}+\mathbb{R}I_{z}=\left\{ \left.\left(\begin{array}{cc}
 \mathrm{i}\alpha  & \mathrm{i}\beta \\
 \mathrm{i}\beta  & -\mathrm{i}\alpha \end{array}
\right)\right|\alpha ,\beta \in \mathbb{R}\right\} ,\label{eq:}\end{equation}
which leads to the Cartan-like decomposition $\mathfrak{g}=\mathfrak{k}\oplus \mathfrak{p}$.
One can identify $\mathfrak{p}$ with the tangent space $T_{K}(G/K)$
of $G/K$ and then argue that the geodesics emanating from $K\in G/K$
are of the form \begin{equation}
t\longmapsto \left(\exp tX\right)K,\quad X\in \mathfrak{p},\label{eq:}\end{equation}
cf. \cite{key-6}, p. 212. Since $G/K$ is compact, the Hopf-Rinow
theorem implies that any point $gK\in G/K$ is of the form $gK=\left(\exp X\right)K$
for some $X\in \mathfrak{p}$. Because $\exp X$ is a symmetric matrix
if $X$ is symmetric, we see that the points of $G/K$ can be represented
by the symmetric unitary $(2\times 2)$-matrices. These comprise the
set \begin{equation}
\mathrm{Sym}(2):=\left\{ \left.\left(\begin{array}{cc}
 \cos \psi e^{\mathrm{i}\varphi } & \mathrm{i}\sin \psi \\
 \mathrm{i}\sin \psi  & \cos \psi e^{-\mathrm{i}\varphi }\end{array}
\right)\right|\varphi ,\psi \in \left[0,2\pi \right]\right\} ,\label{eq:}\end{equation}
which is a $2$-dimensional submanifold of $G$. Moreover, the map\begin{equation}
\psi :\mathrm{Sym}(2)\longrightarrow \mathbb{S}^{2}\subseteq \mathbb{R}^{3},\label{eq:}\end{equation}
\[
\left(\begin{array}{cc}
 \cos \psi e^{\mathrm{i}\varphi } & \mathrm{i}\sin \psi \\
 \mathrm{i}\sin \psi  & \cos \psi e^{-\mathrm{i}\varphi }\end{array}
\right)\longmapsto \left(\cos \varphi \cos \psi ,\sin \varphi \cos \psi ,\sin \psi \right)\]
is a diffeomorphism between $\mathrm{Sym}(2)$ and the $2$-sphere
$\mathbb{S}^{2}$. A calculation now shows that two elements $g_{1},g_{2}\in \mathrm{Sym}(2)$
are in the same coset modulo $K$ if and only if $g_{1}=g_{2}$ or
$g_{1}=-g_{2}$. Hence \begin{equation}
G/K\cong \mathrm{Sym}(2)/_{g\sim -g}\cong \mathbb{S}^{2}/_{g\sim -g}\cong \mathbb{RP}^{2}.\label{eq:}\end{equation}
In order to solve the time-optimal control problem related to the
unreduced system (3.2.1) we shall procede as outlined at the end of
Section 2.5. To this aim we fix the maximal abelian subalgebra $\mathfrak{h}:=\mathbb{R}I_{z}$
of $\mathfrak{p}$ and determine the Weyl orbit $W\cdot H_{d}$ of
$H_{d}$, the action of $W_{\mathrm{aff}}$ on $\mathfrak{h}$, and
the sets $\Omega ,\Theta \subseteq G$.\\
The action of the affine Weyl group $W_{\mathrm{aff}}$ on $\mathfrak{h}$
is generated by reflexions\begin{equation}
X\longmapsto -X\label{eq:}\end{equation}
and translations\begin{equation}
X\longmapsto X+\frac{\pi n}{2}I_{z},\quad n\in \mathbb{Z},\label{eq:}\end{equation}
as is seen from the root space decomposition of $\mathfrak{g}$ with
respect to $\mathfrak{h}$, cf. Example 1.3.7. The cells in $\mathfrak{h}$
are then the sets\begin{equation}
L_{n}:=\left\{ \left.\frac{\pi \nu }{2}I_{z}\right|\nu \in (n,n+1)\right\} ,\quad n\in \mathbb{Z}.\label{eq:}\end{equation}
We fix the cell $\Delta :=L_{0}$ and set \begin{equation}
\Omega :=\exp \Delta =\left\{ \left.\left(\begin{array}{cc}
 e^{\mathrm{i}\varphi } & 0\\
 0 & e^{-\mathrm{i}\varphi }\end{array}
\right)\right|\varphi \in \left(0,\frac{\pi }{2}\right)\right\} \subseteq A.\label{eq:}\end{equation}
Here $A$ denotes as usual the maximal torus of $G$ with Lie algebra
$\mathfrak{h}$. The set $\Theta =\mathrm{Ad}_{K}\bar{\Omega }$ of
Theorem 2.5.3 is in this situation the following:\begin{equation}
\Theta =\left\{ \left.\left(\begin{array}{cc}
 \cos \psi e^{\mathrm{i}\varphi } & \mathrm{i}\sin \psi \\
 \mathrm{i}\sin \psi  & \cos \psi e^{-\mathrm{i}\varphi }\end{array}
\right)\right|\varphi ,\psi \in \left[-\frac{\pi }{2},\frac{\pi }{2}\right]\right\} .\label{eq:}\end{equation}
Note that $\mathrm{Sym}(2)=\Theta \cup -\Theta $ so $\Theta $ can
be thought of as a hemissphere in $\mathrm{Sym}(2)\cong \mathbb{S}^{2}$.\\
Given a terminal point $U_{F}\in G$, a time-optimal trajectory between
$U(0)=\mathbf{1}$ and $U_{F}$ is now obtained in the following manner.

\begin{enumerate}
\item Decompose $U_{F}$ as $U_{F}=U_{1}k_{1}$ with $U_{1}\in \Theta $
and $k_{1}\in K$. This can be accomplished by making use for instance
of (2.5.4). 
\item The general form of a matrix $U_{1}\subseteq \Theta $ is \[
U_{1}=\left(\begin{array}{cc}
 \cos \psi e^{\mathrm{i}\varphi } & \mathrm{i}\sin \psi \\
 \mathrm{i}\sin \psi  & \cos \psi e^{-\mathrm{i}\varphi }\end{array}
\right),\quad \varphi ,\psi \in \left[-\frac{\pi }{2},\frac{\pi }{2}\right].\]
Calculate the parameters $\varphi $ and $\psi $ of the matrix $U_{1}$
determined in step (1) and set \[
a:=\exp \left(\alpha I_{z}\right)\in \bar{\Omega },\]
with\[
\alpha :=\arg \left(\cos \psi \cos \varphi +\mathrm{i}\sqrt{1-\cos ^{2}\psi \cos ^{2}\varphi }\right)\in \left[0,\frac{\pi }{2}\right].\]
By construction, $U_{1}=kak^{-1}$ holds for some $k\in K$. 
\item Calculate $k\in K$ such that $U_{1}=kA_{1}k^{-1}$.
\item A time-optimal control sequence to generate $U_{F}$ is \[
\mathbf{1}\longrightarrow k^{-1}k_{1}\longrightarrow ak^{-1}k_{1}\longrightarrow kak^{-1}k_{1}=U_{F}.\]
Here the first and the last arrow mean synthesizing $k^{-1}k_{1}$
and $k$ by so-called hard pulses (the infimizing time for accomplishing
this being equal to zero), while the middle arrow denotes evolution
of the system under the influence of the drift Hamiltonian $I_{z}$
for time $\alpha $.
\end{enumerate}

\section{Two-Particle Systems}

In this section we discuss control system (3.1.1) for the special
case of $n=2$ spin-particles, and assume that the $x$- and the $y$-component
of each of the spins may be excited individually. The problem of controlling
the spin of such a system then reads\begin{equation}
\dot{U}=\left(H_{d}+\sum _{j=1}^{4}v_{j}H_{j}\right)U,\quad U(0)=\mathbf{1},\quad v_{j}\in \mathbb{R}\label{eq:}\end{equation}
with\begin{equation}
H_{1}=I_{1x},\quad H_{2}=I_{1y},\quad H_{3}=I_{2x},\quad H_{4}=I_{2y}.\label{eq:}\end{equation}
The drift operator $H_{d}$ needs to be chosen within a maximal abelian
subalgebra $\mathfrak{h}\subseteq \mathfrak{su}(4)$ (which will be
specified later) subject to the restriction that it is not contained
in any root hyperplane. The elements $H_{j}$, $j=1,...,4$, generate
a subalgebra $\mathfrak{k}$ isomorphic to $\mathfrak{su}(2)\otimes \mathfrak{su}(2)$,
and we already know by Theorem 3.1.1 that the pair $\left(\mathfrak{su}(4),\mathfrak{su}(2)\otimes \mathfrak{su}(2)\right)$
is a symmetric Lie algebra pair. This allows us to argue along the
lines of the previous section. \\
We denote by $K=SU(2)\otimes SU(2)$ the connected subgroup of $G=SU(4)$
with Lie algebra $\mathfrak{k}$. So the resulting reduced system
is\begin{equation}
\dot{P}=PX,\quad P(0)=K,\quad X\in \textrm{Ad}_{K}H_{d},\label{eq:}\end{equation}
on the $9$-dimensional homogeneous space $G/K$. In order to describe
this space more succintly, we first of all observe that the Lie algebras
$\mathfrak{k}=\mathfrak{su}(2)\otimes \mathfrak{su}(2)\subseteq \mathfrak{su}(4)$
and $\mathfrak{so}(4)=\left\{ \left.X\in \mathfrak{su}(4)\right|X+X^{T}=0\right\} $
are isomorphic. An isomorphism $\varphi :\mathfrak{k}\rightarrow \mathfrak{so}(4)$
is given by conjugation with the unitary matrix\begin{equation}
U:=\frac{1}{2}\left(\begin{array}{cccc}
 1 & 0 & 0 & 1\\
 0 & 1 & -1 & 0\\
 \mathrm{i} & 0 & 0 & -\mathrm{i}\\
 0 & \mathrm{i} & \mathrm{i} & 0\end{array}
\right).\label{eq:}\end{equation}
This is for instance seen by using Lemma 1.6.3 (vi) in order to represent
the elements of a basis of $\mathfrak{k}$ by $(4\times 4)$-matrices,
such as \[
\begin{array}{ll}
 I_{x}\otimes \mathbf{1}=\left(\begin{array}{cccc}
 0 & 0 & 1 & 0\\
 0 & 0 & 0 & 1\\
 -1 & 0 & 0 & 0\\
 0 & -1 & 0 & 0\end{array}
\right), & \mathbf{1}\otimes I_{x}=\left(\begin{array}{cccc}
 0 & 1 & 0 & 0\\
 -1 & 0 & 0 & 0\\
 0 & 0 & 0 & 1\\
 0 & 0 & -1 & 0\end{array}
\right),\\
 I_{y}\otimes \mathbf{1}=\left(\begin{array}{cccc}
 0 & 0 & \mathrm{i} & 0\\
 0 & 0 & 0 & \mathrm{i}\\
 \mathrm{i} & 0 & 0 & 0\\
 0 & \mathrm{i} & 0 & 0\end{array}
\right), & \mathbf{1}\otimes I_{y}=\left(\begin{array}{cccc}
 0 & \mathrm{i} & 0 & 0\\
 \mathrm{i} & 0 & 0 & 0\\
 0 & 0 & 0 & \mathrm{i}\\
 0 & 0 & \mathrm{i} & 0\end{array}
\right),\\
 I_{z}\otimes \mathbf{1}=\left(\begin{array}{cccc}
 \mathrm{i} & 0 & 0 & 0\\
 0 & \mathrm{i} & 0 & 0\\
 0 & 0 & -\mathrm{i} & 0\\
 0 & 0 & 0 & -\mathrm{i}\end{array}
\right), & \mathbf{1}\otimes I_{z}=\left(\begin{array}{cccc}
 \mathrm{i} & 0 & 0 & 0\\
 0 & -\mathrm{i} & 0 & 0\\
 0 & 0 & \mathrm{i} & 0\\
 0 & 0 & 0 & -\mathrm{i}\end{array}
\right),\end{array}
\]
 and then by checking that the map $\varphi $ sends this basis to
a basis of $\mathfrak{so}(4)$. The Lie algebra isomorphism $\varphi $
can be integrated to a Lie group isomorphism $\Phi :K\rightarrow SO(4)=:\tilde{K}$
which likewise is given by conjugation with $U$. This also shows
that the homogeneous spaces $SU(4)/\left(SU(2)\otimes SU(2)\right)$
and $SU(4)/SO(4)$ are diffeomorphic as $SU(4)$-homogeneous spaces.
Namely, an equivariant diffeomorphism is given by the map\begin{equation}
\psi :SU(4)/\left(SU(2)\otimes SU(2)\right)\longrightarrow SU(4)/SO(4),\quad gK\longmapsto UgU^{-1}\tilde{K}.\label{eq:}\end{equation}
The map $\psi $ is well-defined: \begin{eqnarray*}
\psi (gkK) & = & UkgU^{-1}\tilde{K}\\
 & = & UgU^{-1}UkU^{-1}\tilde{K}\\
 & = & UgU^{-1}\tilde{K}\\
 & = & \psi (gK)
\end{eqnarray*}
 holds for all $k\in K$.

We remark that the space $SU(4)/SO(4)$ appears as a symmetric space
of type A I in Example 1.2.7. It also can be shown to be diffeomorphic
to the Grassmannian manifold of $3$-dimensional subspaces in $\mathbb{R}^{6}$,
cf. \cite{key-40}, p. 322.\\
An argument analogous to that in the previous section shows that the
elements of $G/K\cong SU(4)/SO(4)$ can be represented (again not
uniquely) by those of the space $\mathrm{Sym}(4)$ of symmetric unitary
$(4\times 4)$-matrices. In this case a calculation yields\begin{equation}
G/K\cong \mathrm{Sym}(4)/_{\sim },\label{eq:}\end{equation}
where\begin{equation}
g_{1}\sim g_{2}\quad \Longleftrightarrow \quad g_{1}=g_{2}d,\quad d\in D,\label{eq:}\end{equation}
with\begin{equation}
D:=\left\{ \pm \mathbf{1},\pm \mathrm{diag}(1,1,-1,-1),\pm \mathrm{diag}(1,-1,1,-1),\pm \mathrm{diag}(1,-1,-1,1)\right\} .\label{eq:}\end{equation}
We now turn to a discussion of the time-optimal control problem as
formulated in Section 2.3 in the here relevant case of $\left(\mathfrak{g},\mathfrak{k}\right)=\left(\mathfrak{su}(4),\mathfrak{so}(4)\right)$
constituting a symmetric Lie algebra pair. The Cartan-like decomposition
of the Lie algebra $\mathfrak{g}$ is\begin{equation}
\mathfrak{g}=\mathfrak{k}\oplus \mathrm{sym}(4),\label{eq:}\end{equation}
where $\mathrm{sym}(4)=\left\{ \left.X\in \mathfrak{g}\right|X^{T}=X\right\} $.
Indeed, $\mathfrak{p}:=\mathrm{sym}(4)$ is the orthogonal complement
of $\mathfrak{k}$, because it is complementary as a vector space,
and for all $X\in \mathfrak{p}$, $Y\in \mathfrak{k}$ we compute
(making use of equation (1.1.16)) that\begin{eqnarray*}
\kappa \left(X,Y\right) & = & 8\mathrm{tr}\left(XY\right)\\
 & = & 8\sum _{j=1}^{4}\sum _{k=1}^{4}X_{jk}Y_{kj}\\
 & = & 8\sum _{j\leq k}X_{jk}Y_{kj}-8\sum _{j>k}X_{jk}Y_{kj}\\
 & = & 8\sum _{j<k}X_{jk}Y_{kj}-8\sum _{j>k}X_{jk}Y_{kj}\\
 & = & 8\sum _{j<k}X_{kj}Y_{jk}-8\sum _{j>k}X_{jk}Y_{kj}\\
 & = & 8\sum _{j>k}X_{jk}Y_{kj}-8\sum _{j>k}X_{jk}Y_{kj}\\
 & = & 0,
\end{eqnarray*}
as $X_{kj}=X_{jk}$ and $Y_{kj}=-Y_{jk}$.\\
The next step is to determine a root space decomposition of $\mathfrak{g}$.
Choose $\mathfrak{h}:=\mathbb{R}H_{1}+\mathbb{R}H_{2}+\mathbb{R}H_{3}\subseteq \mathfrak{p}$
with\begin{equation}
H_{1}:=\mathrm{diag}\left(\mathrm{i},\mathrm{-i},0,0\right),\quad H_{2}:=\mathrm{diag}\left(\mathrm{i},0,\mathrm{-i},0\right),\quad H_{3}:=\mathrm{diag}\left(\mathrm{i},0,0,\mathrm{-i}\right)\label{eq:}\end{equation}
to serve as a maximal abelian subalgebra of $\mathfrak{g}$. Let $A\in G$
the maximal torus with Lie algebra $\mathfrak{h}$. The root space
decomposition of $\mathfrak{g}_{\mathbb{C}}\cong \mathfrak{sl}_{4}\mathbb{C}$
with respect to $\mathfrak{h}$ is then given by\begin{equation}
\mathfrak{g}_{\mathbb{C}}=\mathfrak{h}_{\mathbb{C}}\oplus \bigoplus _{i\neq j}\mathfrak{g}_{ij},\label{eq:}\end{equation}
$\mathfrak{g}_{ij}$ as in Example 1.3.7. The coroots have also been
determined before; they are the following\begin{equation}
Y_{1}=H_{1},\: Y_{2}=H_{2},\: Y_{3}=H_{3},\: Y_{4}=H_{2}-H_{1},\: Y_{5}=H_{3}-H_{1},\: Y_{6}=H_{3}-H_{2},\label{eq:}\end{equation}
together with \begin{equation}
Y_{j+6}:=-Y_{j},j=1,...,6.\label{eq:}\end{equation}
It is easily checked that reflexion in $\mathfrak{h}$ on the hyperplane
perpendicular to $Y\in \mathfrak{h}$ is given by\begin{equation}
X\longmapsto X-2\frac{\left\langle X,Y\right\rangle }{\left\Vert Y\right\Vert ^{2}}Y.\label{eq:}\end{equation}
>From this we obtain the Weyl orbit $W\cdot H_{d}$ of the element
$H_{d}=\sum _{i=1}^{3}a_{i}Y_{i}\in \mathfrak{h}$ by reflexion on
the root hyperplanes $Y_{j}^{\perp }$, $j=1,...,6$, which is, in
coordinates with respect to the ordered basis $\left(Y_{1},Y_{2},Y_{3}\right)$
of $\mathfrak{h}$, the set\begin{eqnarray*}
W\cdot H_{d} & = & \left\{ \left(a_{1},a_{2},a_{3}\right),\left(a_{1},a_{3},a_{2}\right),\left(a_{2},a_{1},a_{3}\right),\left(a_{2},a_{3},a_{1}\right),\left(a_{3},a_{1},a_{2}\right),\right.\\
 &  & \left(a_{3},a_{2},a_{1}\right),\left(-a_{1}-a_{2}-a_{3},a_{2},a_{3}\right),...,\left(a_{3},a_{2},-a_{1}-a_{2}-a_{3}\right),\\
 &  & \left(a_{1},-a_{1}-a_{2}-a_{3},a_{3}\right),...,\left(a_{3},a_{1}-a_{2}-a_{3},a_{1}\right),\\
 &  & \left.\left(a_{1},a_{2},-a_{1}-a_{2}-a_{3}\right),...,\left(a_{1}-a_{2}-a_{3},a_{2},a_{1}\right)\right\} .
\end{eqnarray*}
In the same manner as in 3.2 we determine the cell $\Delta \subseteq \mathfrak{h}$
to be the convex hull of $0$, $\frac{\pi }{2}H_{1}$, $\frac{\pi }{2}H_{2}$,
and $\frac{\pi }{2}H_{3}$, and set $\Omega :=\exp \Delta \subseteq A$.
It is easily checked that any $a\in A$ permits a decomposition $a=a_{1}d$
with $a_{1}\in \bar{\Omega }$ and $d\in D$, where the group $D$
is as defined in (3.3.8).\\
By proceding as described at the end of Section 2.5 we obtain a time-optimal
trajectory between the identity and any given point $U_{F}\in G$
in the following way.

\begin{enumerate}
\item Perform a polar decomposition of $U_{F}$ to obtain $k_{1}\in K$
and $U_{1}\in \mathrm{Sym}(4)$ with $U_{F}=U_{1}k_{1}$. If $U_{1}\notin \Theta $
then replace $U_{1}$ suitably by $U_{1}d$, and $k_{1}$ by $d^{-1}k_{1}$,
where $d\in D$.
\item Diagonalize $U_{1}$ as $U_{1}=kak^{-1}$ with $k\in K$ and $a\in \bar{\Omega }$.
Write $a$ in the form \[
a=\prod _{j=1}^{24}\exp \left(\alpha \beta _{j}Z_{j}\right)\]
with $\sum _{k=1}^{24}\beta _{k}=1$, and $Z_{1},...,Z_{24}$ the
elements of the Weyl orbit $W\cdot H_{d}$. Choose the parameter $\alpha \geq 0$
to be the smallest one possible. This $\alpha $ then satisfies the
minimality condition of Theorem 2.5.3.
\item Steer system (3.3.1) as depicted below:\begin{eqnarray*}
\mathbf{1} & \longrightarrow  & k^{-1}k_{1}\\
 & \longrightarrow  & \exp \left(\alpha \beta _{24}Z_{24}\right)k^{-1}k_{1}\\
 & \longrightarrow  & \exp \left(\alpha \beta _{23}Z_{23}\right)\exp \left(\alpha \beta _{24}YZ_{24}\right)k^{-1}k_{1}
\end{eqnarray*}
\begin{eqnarray*}
 & \vdots  & \vdots \\
 & \longrightarrow  & \left(\prod _{j=1}^{24}\exp \left(\alpha \beta _{j}Z_{j}\right)\right)k^{-1}k_{1}\\
 & \longrightarrow  & k\left(\prod _{j=1}^{24}\exp \left(\alpha \beta _{j}Z_{j}\right)\right)k^{-1}k_{1}=U_{F}.
\end{eqnarray*}
Here again the first and the last arrow means producing movement within
the subgroup $K$ of $G$ and is realized by performing a so-called
hard pulse, while the middle arrows symbolize evolution of the system
in direction of $Z_{1}$,..., $Z_{24}$ for times $\alpha \beta _{1}$,...,
$\alpha \beta _{24}$, respectively. 
\end{enumerate}

\printindex{}

%\newpage

%\thispagestyle{empty}

%\

%\normalsize

%\newpage

%\thispagestyle{empty}

%\

%\vfill

%\begin{center}

%\Huge \textbf{Erkl\"arung}

%\end{center}

%\vspace{25mm}

%\newpage

%\thispagestyle{empty}

%\begin{center}

%\huge \textbf{Erkl\"arung}

%\end{center}

%\vfill

%\Large

%Hiermit best\"atige ich, diese Diplomarbeit selbst\"andig und ohne weitere Hilfe ausgearbeitet und angefertigt zu haben. Verwendete Literatur kann dem entsprechenden Verzeichnis entnommen werden.

%\vspace{50mm}

%\begin{flushright}

%Miltenberg, den 15. Mai 2005

%\end{flushright}

%\vfill

\end{document}